\documentclass[fleqn,usenatbib]{mnras}

\usepackage{newtxtext,newtxmath}

\usepackage[T1]{fontenc}

\DeclareRobustCommand{\VAN}[3]{#2}
\let\VANthebibliography\thebibliography
\def\thebibliography{\DeclareRobustCommand{\VAN}[3]{##3}\VANthebibliography}

\usepackage{graphicx}	
\usepackage{amsmath}	
\usepackage{amstext,amscd,bm}
\usepackage{array}
\usepackage{multirow}
\usepackage[table,svgnames]{xcolor}

\newcommand\an{Astron. Nachr}
\newcommand\chaa{Chinese Astron. Astrophys}

\newcolumntype{R}{>{$}r<{$}}
\newcolumntype{L}{>{$}l<{$}}
\newcolumntype{A}{R@{${}\pm{}$}L}
\newcolumntype{E}{R@{${}-{}$}L}
\newcolumntype{B}{R@{${}\,/\,{}$}L}
\newcommand{\mcl}[1]{\multicolumn{2}{c}{#1}}
\newcommand{\mcll}[1]{\multicolumn{2}{c|}{#1}}
\newcommand{\mcc}[1]{\multicolumn{1}{c}{#1}}

\newcommand{\mrw}[1]{\multirow{4}{*}{#1}}
\newcommand{\rot}[1]{\rotatebox[origin=c]{90}{#1}}

\title[Study of the Open Clusters in Kepler Field]{Study of the Open Clusters in Kepler Prime Field}

\author[Y\"uksel Karata\c{s} et al.]{
Y\"uksel Karata\c{s}$^1\thanks{E-mail: karatas@istanbul.edu.tr}$,
Hikmet {\c C}akmak$^1$,
{\.I}nci Akkaya Oralhan$^2$,
Charles Bonatto $^3$,
Ra\'ul Michel$^4$\newauthor
and Martin Netopil$^5$
\\
$^{1}$Department of Astronomy and Space Sciences, Science Faculty, {\.I}stanbul University, 34116, \"Universite-Istanbul, T\"urkiye\\
$^{2}$Department of Astronomy and Space Sciences, Faculty of Arts and Sciences, Erciyes University, Talas Yolu, 38039, Kayseri, T\"urkiye\\
$^3$Universidade Federal do Rio Grande do Su, Departamento de Astronomia, CP\,15051, RS, Porto Alegre 91501-970, Brazil\\
$^4$Observatorio Astron\'omico Nacional, Universidad Nacional Aut\'onoma de M\'exico, Apartado Postal 877, C.P. 22800, Ensenada, B.C., M\'exico\\
$^5$Kuffner Observatory, Johann-Staud-Stra{\ss}e 10, A-1160 Wien, Austria
}

\date{Accepted XXX. Received YYY; in original form ZZZ}

\pubyear{2022}

\begin{document}
\label{firstpage}
\pagerange{\pageref{firstpage}--\pageref{lastpage}}
\maketitle

\begin{abstract}
We present a detailed study of NGC~6791, NGC~6811, NGC~6819 and NGC~6866, the four open clusters that are located in the Kepler prime field. We use new CCD~$UBV(RI)_{KC}$ photometry, which was combined with Gaia EDR3 photometric/astrometric data, to derive the astrophysical parameters with two independent methods - one of them the \textit{fitCMD} algorithm. Furthermore, we provide among others estimates of the mass and mass function, the cluster structure, derive the cluster orbits, and discuss the cluster dynamics. All objects belong to the older open cluster population ($\sim1-7$\,Gyr), are in an advanced dynamical stage with signs of mass segregation, and are located close to the solar circle, but show a large range in respect of radii, member stars or observed cluster mass ($\sim100-2000\,M_\odot$). For the three younger objects we were also able to provide photometric metallicity estimates, which confirms their status as clusters with a roughly solar metallicity. The most outstanding object is clearly NGC~6791, a very old cluster with a high metallicity at a distance of about 4.5\,kpc from the Sun. We estimate a probable radial migration by about 7\,kpc, resulting in a birth position close to the Galactic center.

\end{abstract}




\section{Introduction}
In this paper, we present new CCD photometry of the open clusters (OCs) NGC~6791, NGC~6811, NGC~6819, and NGC~6866, which was combined with Gaia EDR3 photometric and astrometric data \citep{gaia3}. These objects are Kepler asteroseismic targets, so a detailed understanding of the properties of these open clusters is important also for studies on correlations with variable stars. 

\begin{figure}
	\centering{\includegraphics[width=0.98\columnwidth]{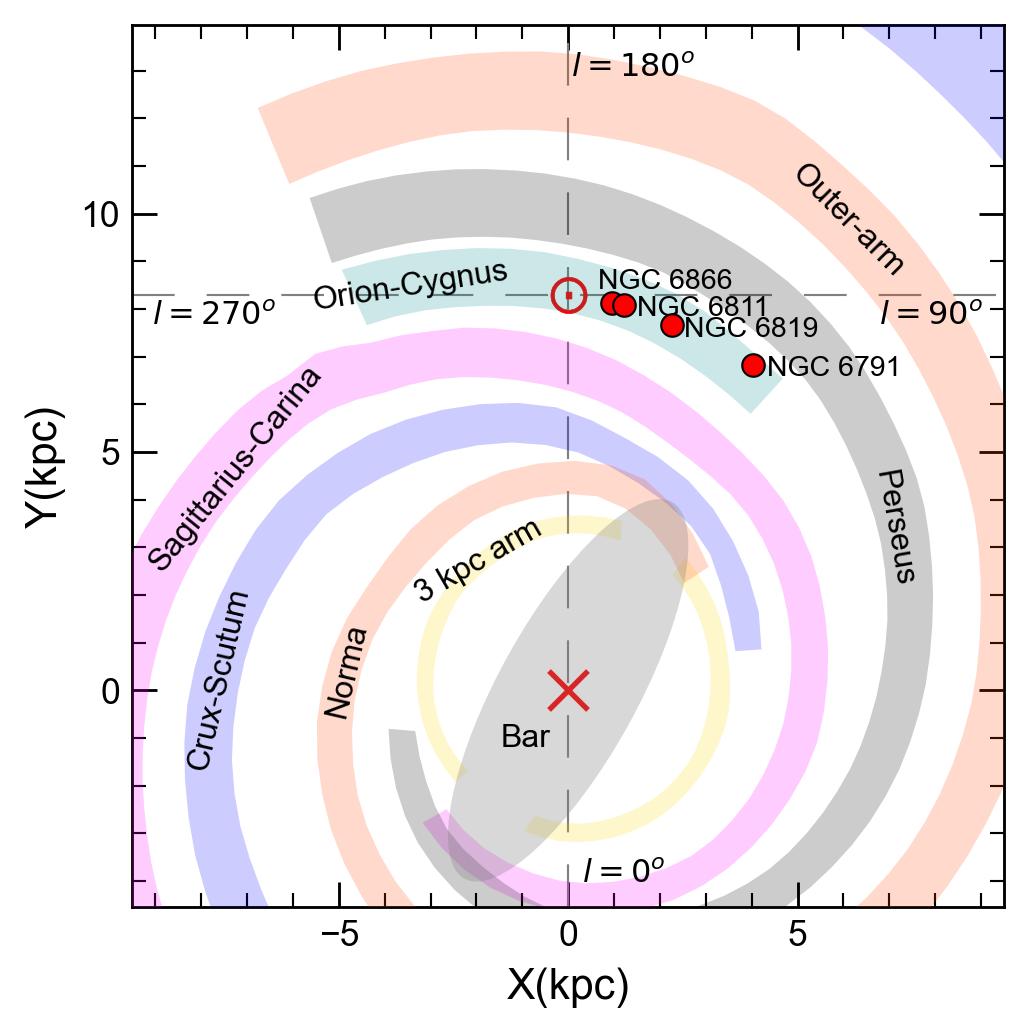}}\vspace*{-2ex}
	\caption{Spatial distribution of the four OCs (filled red circles) in Galactocentric cartesian coordinates. The schematic projection of the Galaxy with its spiral arms is seen from the North pole. The Sun is located at 8.2 kpc. The figure is adapted from the fig.~10 by \citet{rei19}.}
\label{f1_galpos}	
\end{figure}

These four old-aged OCs represent objects of the first Galactic quadrant (see Fig.~\ref{f1_galpos} and Table~\ref{t1_summary}). Their location criteria are important due to the survival rate of the OC population. As discussed e.g. by \cite{Bonatto2007} and \cite{Gunes2017}, the majority of OCs older than 1~Gyr lie outside the solar circle. On the other hand, the OC population gets rare in direction to the Galactic centre, because of the effects of strong absorption, crowding or dissolution by Giant Molecular Clouds (GMCs).

A detailed understanding of the dynamical evolution of the four OCs depends on the knowledge of the astrophysical parameters (reddening, distance, age), structural parameters (core, cluster and tidal radii), overall masses, mass function (MF), relaxation times and evolutionary parameters. The members of the OCs undergo internal and external perturbations such as stellar evolution, two-body relaxation, mass segregation, tidal interactions with the Galactic disc and bulge, spiral arm shocks, Galactic tidal field, and collisions with GMCs  \citep{Lamers2006,Gieles2007}.

\cite{heg2003} theoretically interpreted the proportional relations between the half-mass, core, and tidal radii.  Later, there have been attempts to explain the dynamic evolution from the observations of star clusters e.g. by \cite{bau2010}, \cite{Angelo2018, Angelo2020, Angelo2021} - hereafter A18, A20, A21, or \cite{pia09}. According to these studies, as a star cluster expands to the point of being tidally filling, it is exposed to internal dynamical evolution in its core region due to two-body relaxation, mass segregation, and core-collapse. Due to mass segregation and core collapse, the cores contract whereas the half mass radius remains almost constant. The binaries and possible stellar black-holes in the central parts of the clusters may be responsible for their expansion. This expansion is accompanied by mass losses from the outer parts. By the effect of tidal interaction, an OC heats and its stars gain kinetic energy, which leads to an increase in the evaporation rate. Finally, they are dissolved in the Galactic field.

In this paper we will investigate the role and degree of internal and external dynamic effects of the four OCs. The astrophysical open cluster parameters (the colour excess, the distance, and the age) are determined from CCD $UBV(RI)_{KC}$ and  Gaia EDR3 photometric data. For this, we adopt spectroscopic metal abundances and employ the approach \textit{fitCMD}, presented by \cite{Bonatto2019}. Furthermore, Gaia EDR3 data are used to obtain structural information, the mass and mass function, dynamical evolution parameters and kinematics of the objects.

Our programme objects are covered by several individual studies, but are certainly also included in larger scale surveys. For example, \cite{cantat2020} derived the distance, age and reddening for 1867 OCs using Gaia DR2 data, a sample similar in size was also investigated by \cite{dia21} using the same data. 
\cite{Tarr2022} adopt Gaia EDR3 data to study the structural parameters of 389 OCs. They note that older OCs have on average smaller core radii and that mass segregation operates more efficiently in older OCs. Such catalogue provide also an unique opportunity for detailed comparisons of the results for our sample OCs. 

Furthermore, to understand that these objects move away from their birth places (or migrate radially), 
their birth radii and radial migration distances are also estimated. According to \cite{and07}, non- or inward migrating OCs may be more prone to disruption, leading to an appearance of metal-rich OCs (e.g. NGC~6791) in the solar vicinity.   

This paper is organized as follows. The CCD $UBV(RI)_{KC}$ photometry of the four OCs is presented in Section~\ref{sect:2}. The cluster membership technique is discussed in Section~\ref{sect:3}. The derivation of the astrophysical parameters based on \textit{fitCMD} and the differential grid technique is presented in Section~\ref{sect:4}. The obtained cluster dimensions, masses/mass function slopes, dynamical parameters together with their indicators, kinematics, and orbital parameters are given in Sections~\ref{sect:5}--\ref{sect:7}. A discussion/conclusion about the above topics is finally presented in the last Section together with a comparison with the literature and investigation of the dynamical evolution. 

\section{Observation and Data Reduction}
\label{sect:2}
\begin{figure}
	\centering{\includegraphics[width=0.48\columnwidth]{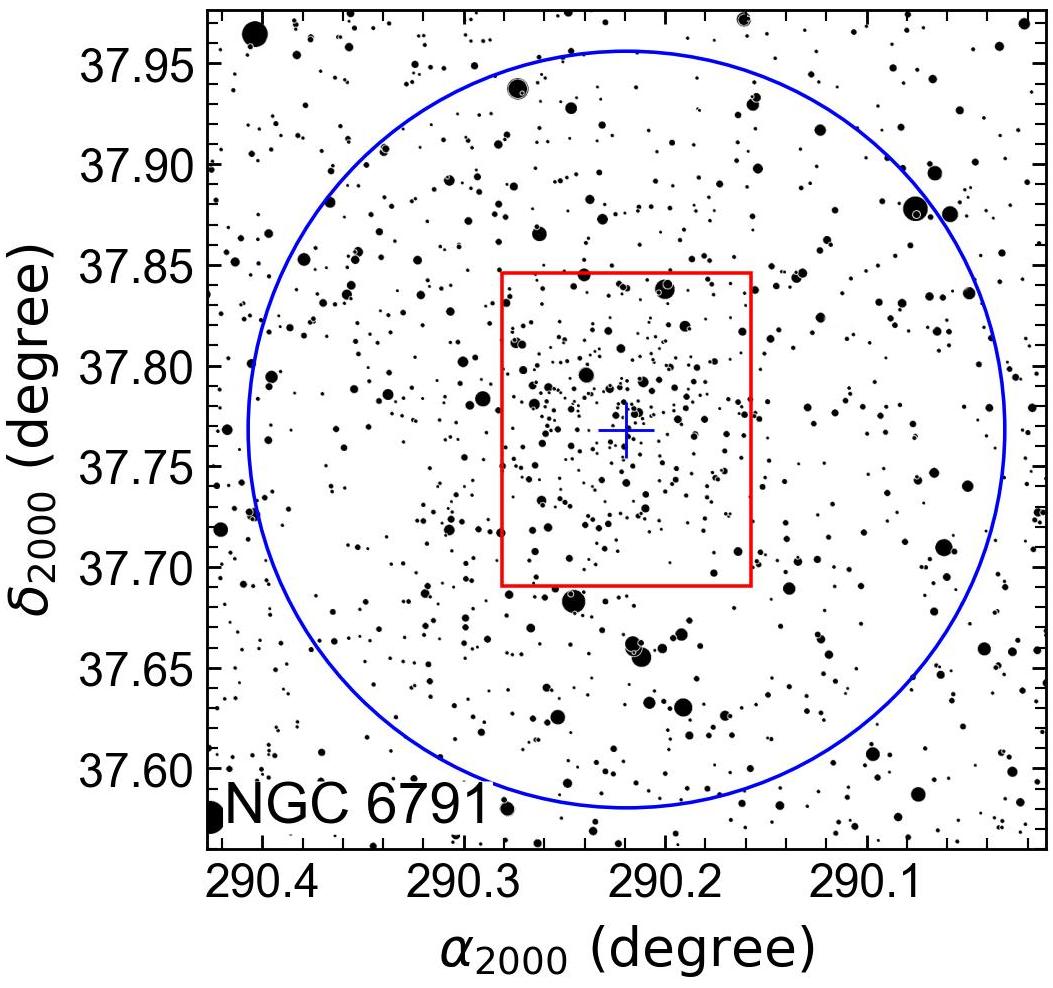} \hspace*{1ex}
		\includegraphics[width=0.48\columnwidth]{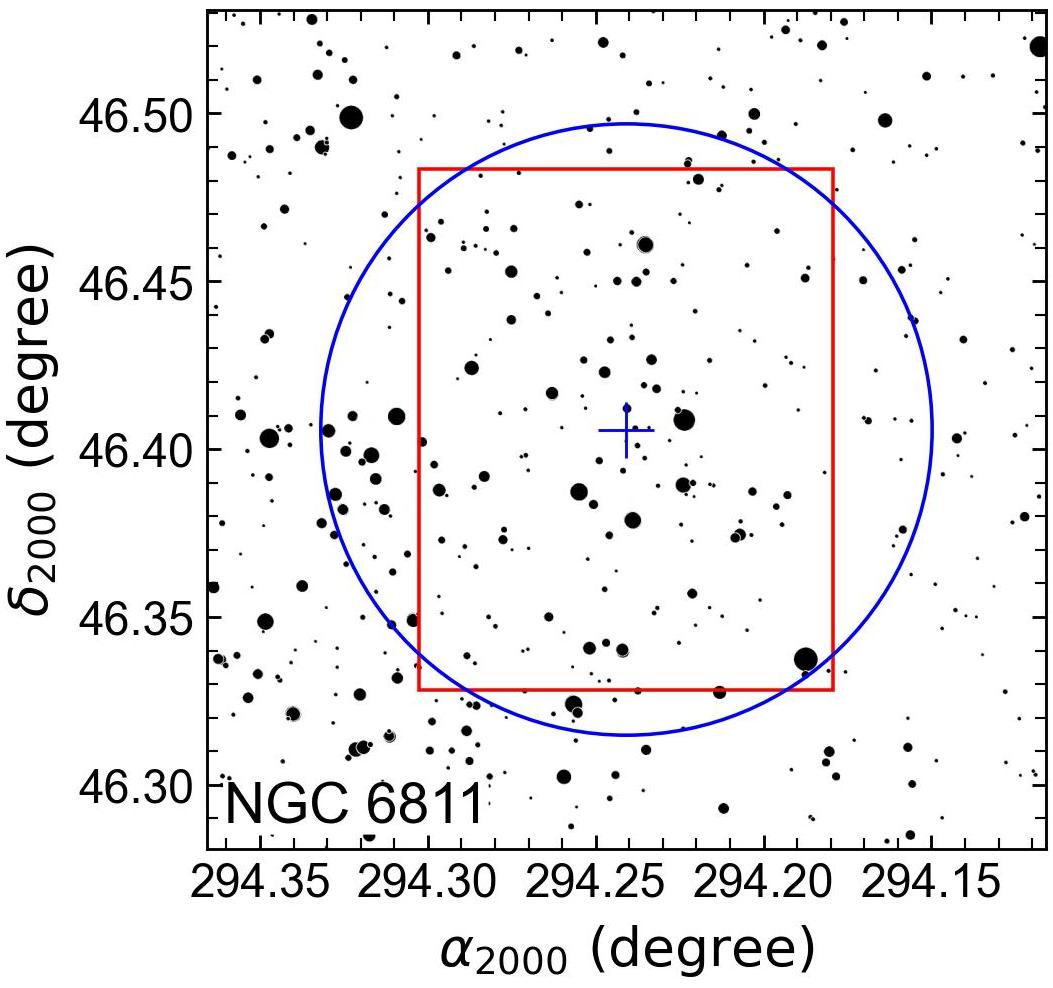}}\\ [1ex] 
	\centering{\includegraphics[width=0.48\columnwidth]{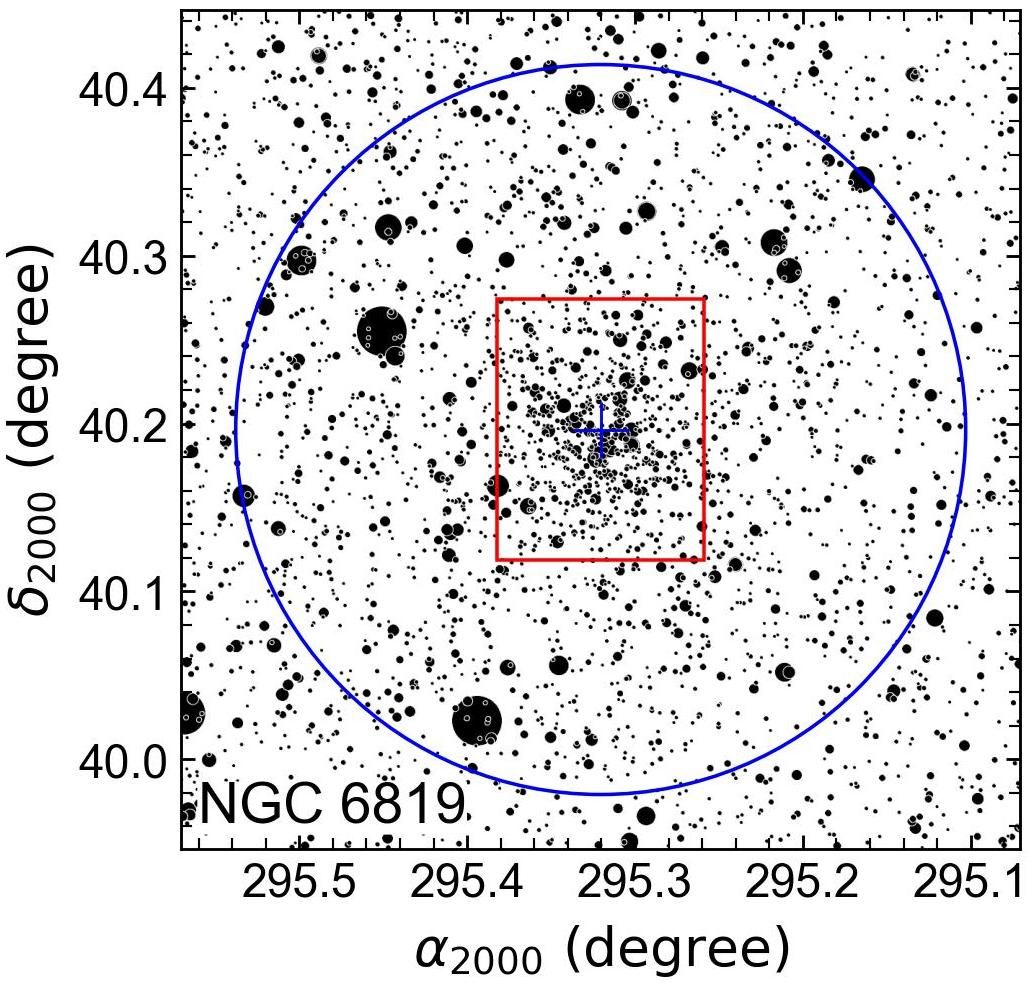} \hspace*{1ex}
		\includegraphics[width=0.48\columnwidth]{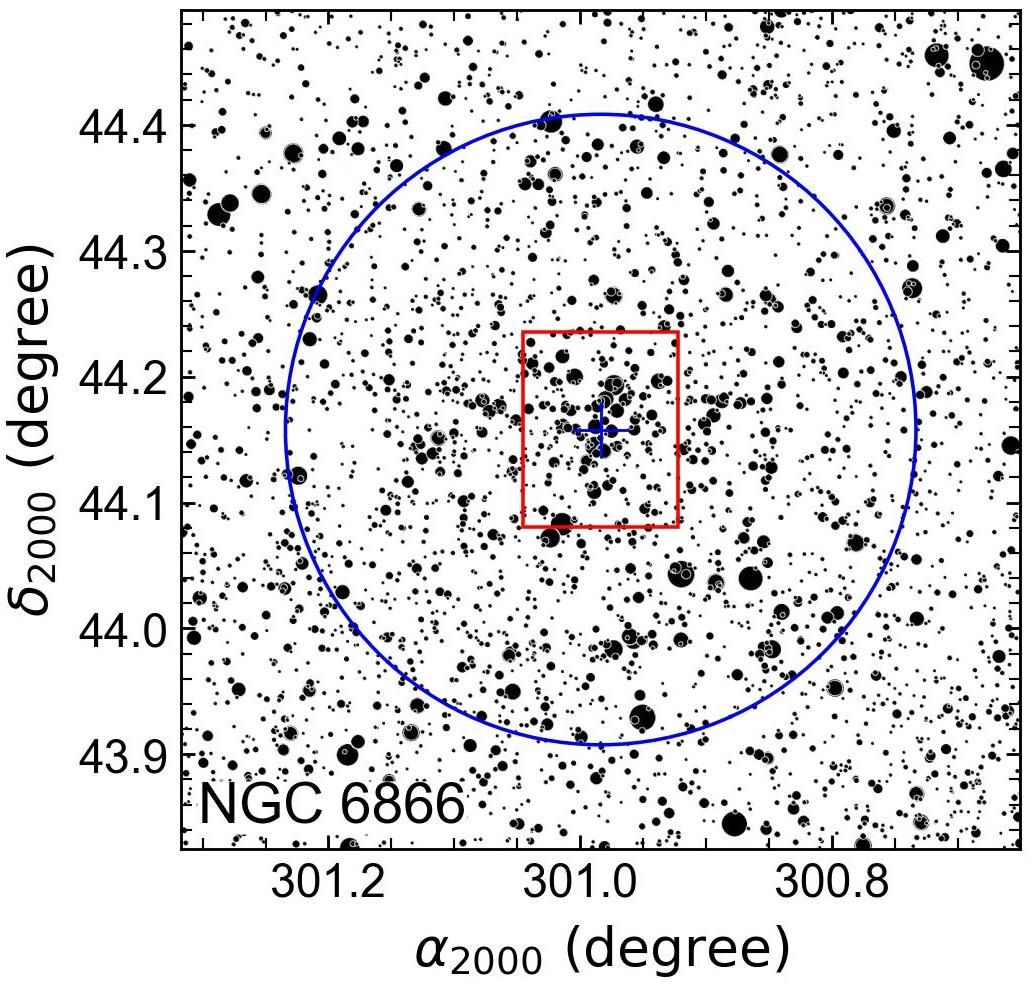}}\vspace*{-1ex}
	\caption{The star charts of the four OCs are produced using the chart tool at https://www.aavso.org/apps/vsp/. The field of view of the SPM detector is shown by the red rectangle ($7.4^{\prime}$ E-W $\times$ $9.3^{\prime}$ N-S), the blue circles represent the radii obtained from the radial density profile ($R_{RDP}$) listed in Table~\ref{t5_struct}. Big plus symbols show the central equatorial coordinates.
	}
\label{f2_chart}	
\end{figure}

The observations of NGC~6791, NGC~6811, NGC~6819 and NGC~6866 were carried out at the San Pedro Martir Observatory (SPMO) during photometric nights (7-10 June 2013) with very good seeing (0\farcs6 in long V exposures) using the 0.84-m (f/15) Ritchey-Chretien telescope equipped with the Mexman filter wheel and the ESOPO CCD detector. The ESOPO detector, a 2048$\times$4608 13.5-$\mu m$ square pixels E2V CCD42-40, has a gain of 1.7 e$^-$/ADU and a readout noise 3.8 e$^-$ at 2$\times$2 binning. The combination of telescope and detector ensures an unvignetted field of view of 7.4$\times$9.3 arcmin$^2$. The star charts of the OC areas are shown in Fig.~\ref{f2_chart}. It is noticeable that our observations are unfortunately restricted to the core of most objects and do not cover cluster members in the outskirts, thus e.g. a structural analysis of the objects (see Sect.~\ref{sect:5}) has to be based on Gaia data.

Each OC was observed through the Johnson's $UBV$ and the Kron-Cousins' $RI$ filters with short and long exposure times 
in order to properly cover both, bright and faint stars in the region. Standard star fields \citep{lan09} were observed at the meridian and at about two airmasses to determine the atmospheric extinction coefficients. 

The log of the observations is shown in Table \ref{t1_summary}. It includes the object names, centre coordinates of the observed fields, air mass range during the observations, and exposure times in each band. The flat fields were taken at the beginning and end of each night, and bias images were obtained between cluster observations. Data reduction was carried out by Raul Michel with the IRAF/DAOPHOT\footnote {IRAF is distributed by the National Optical Observatories, operated by the Association of Universities for Research in Astronomy, Inc., under cooperative agreement with the National Science Foundation.} package \citep{stet87}. The standard magnitude in a given filter $\lambda$ is obtained using the following relation:
\begin{equation}
	M_{\lambda} = m_{\lambda} - [k_{1\lambda} -k_{2\lambda}C)] X + \eta_{\lambda} C + \zeta_{\lambda}
\end{equation} 
where $m_{\lambda}$, $k_{1\lambda}$, $k_{2\lambda}$, $C$, and $X$  are the observed instrumental magnitude, primary/secondary extinction coefficients, colour index and air mass, respectively. $M_{\lambda}$, $\eta_{\lambda}$, $\zeta_{\lambda}$ are standard magnitude, transformation coefficient and photometric zero point, respectively.  More details about the data reduction, the extinction coefficients and zero points for the $UBVRI$ filters can be found in the papers by \cite{akk10}, \cite{akk15} and \cite{akk19}. The photometric errors in $V$ and the colours $(R$--$I)$, $(V$--$I)$, $(B$--$V)$, $(U$--$B)$ of the four OCs are presented in Fig.~\ref{f1_app_pherr} and the mean errors in $V$-mag intervals are listed in Table~\ref{t1_app_err} in Appendix.

\renewcommand{\tabcolsep}{1.9mm}
\renewcommand{\arraystretch}{1.3}
\begin{table*}
	\footnotesize
	\begin{center}
		\caption{Equatorial/Galactic coordinates and observation summary of the four OCs.}
		\label{t1_summary}
		\begin{tabular}{lrrrrcccccc}
			\hline
			Cluster   &$\alpha(2000)$&$\delta(2000)$&$\ell$&$b$&Airmass&U&B&V&R&I\\
			&(h\,m\,s)  &$(^{\circ}\,^{\prime}\,^{\prime\prime})$&$(^{\circ})$&$(^{\circ})$&&Exp.Time (s)& Exp.Time (s)&  Exp.Time (s)& Exp.Time (s)& Exp.Time (s)\\
			\hline
			NGC\,6791 &  19 20 52.6  & 37 46 05.6  & 69.95 & 10.90&1.012-1.050&100,1200&20,30,800&25,50,500&10,30,300&15,50,300\\
			NGC\,6811 &  19 37 17.0  & 46 23 18.0  & 79.20 & 12.07&1.039-1.054& 90,1200&   20,500& 6,10,200&    6,150&    6,150\\
			NGC\,6819 &  19 41 16.9  & 40 11 47.3  & 73.98 &  8.49&1.016-1.035& 90, 900&   20,300& 5,50,200& 4,30,120&   20,120\\
			NGC\,6866 &  20 03 56.1  & 44 09 28.7  & 79.58 &  6.84&1.045-1.077& 30, 900&   20,500&   15,300&   15,200&   15,200\\
			\hline
		\end{tabular}
	\end{center}
\end{table*} 

\section{Membership selection}
\label{sect:3}
In order to identify the cluster members of NGC~6791, NGC~6811, NGC~6819 and NGC~6866, we have obtained Gaia EDR3 astrometric/photometric data \citep{gaia3} from VizieR\footnote{http://vizier.u-strasbg.fr/viz-bin/VizieR?-source=II/246.} for a large area of 40-70 arcmin. 

We applied the Gaussian Mixture Model (GMM) and the scikit-learn package \citep{ped11} to determine the membership probabilities $P$(\%)\footnote{$P$ is defined by $\Phi_c$ /$\Phi$.  Here $\Phi = \Phi_c + \Phi_f$ is the total probability distribution. \textit{c} and \textit{f} are subscripts for cluster and field parameters, respectively. Parameters for the estimation of $\Phi_c$ and $\Phi_f$ are $\mu_{\alpha}$, $\mu_{\delta}$, $\varpi$, $\sigma_{\mu\alpha}$, $\sigma_{\mu\delta}$, $\sigma_\varpi$.} of the cluster stars. The GMM model considers that the distribution of proper motions of the stars in a cluster region can be represented by two elliptical bivariate Gaussians. The used expressions can be found in the papers by \cite{bal98}, \cite{wu02}, \cite{sar12}, \cite{dia18}, or \cite{Cakmak2021}. 

\begin{figure}
	\centering{\includegraphics[width=0.47\columnwidth]{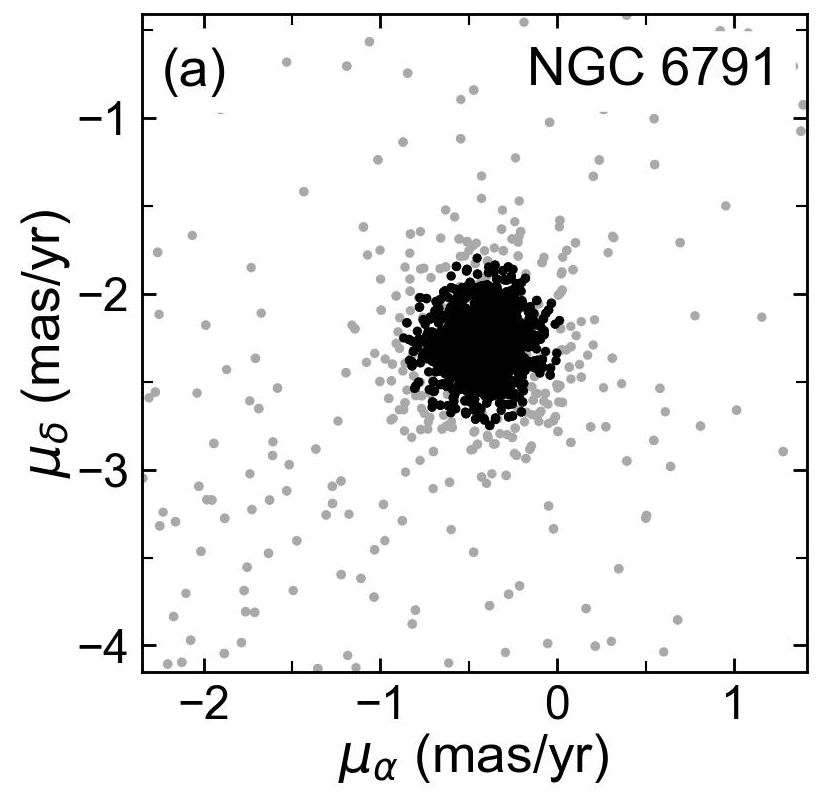} \hspace*{2ex}
		\includegraphics[width=0.475\columnwidth]{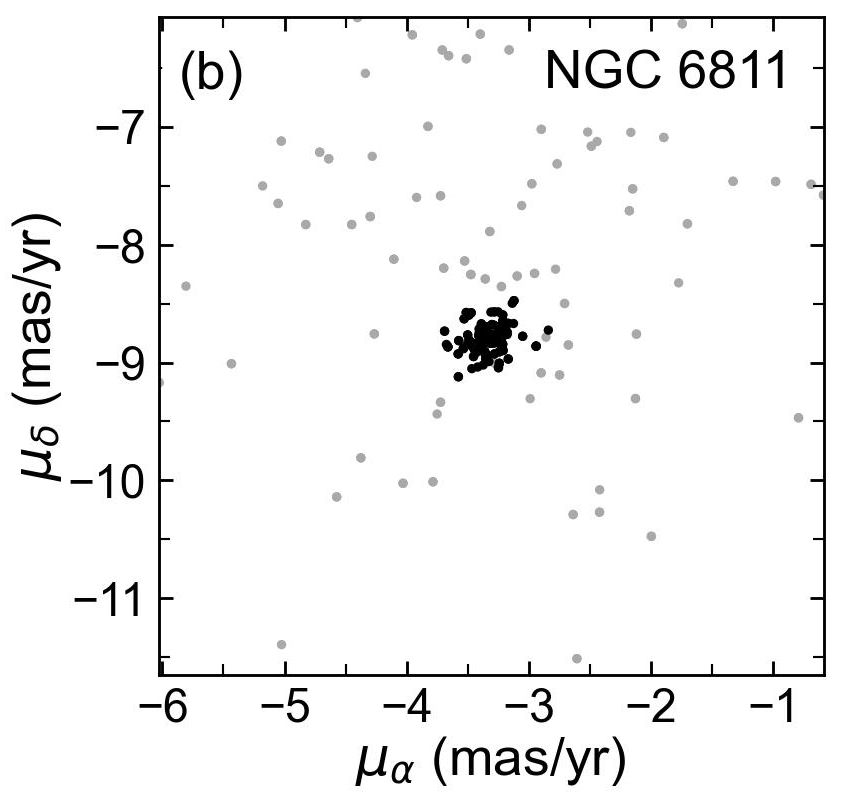}\\[2ex]
		\includegraphics[width=0.47\columnwidth]{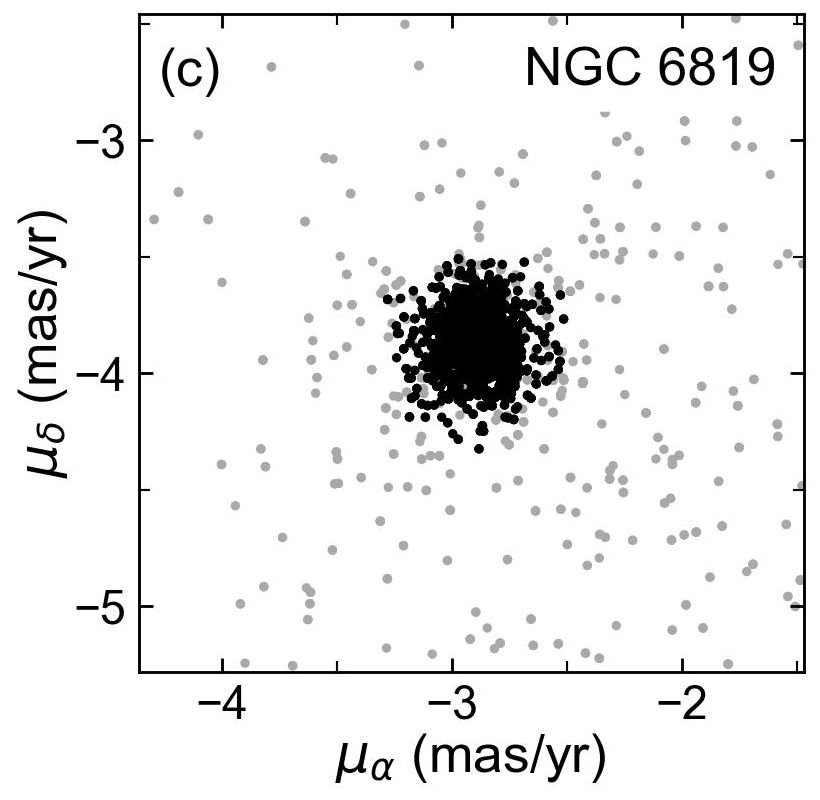} \hspace*{2ex}
		\includegraphics[width=0.47\columnwidth]{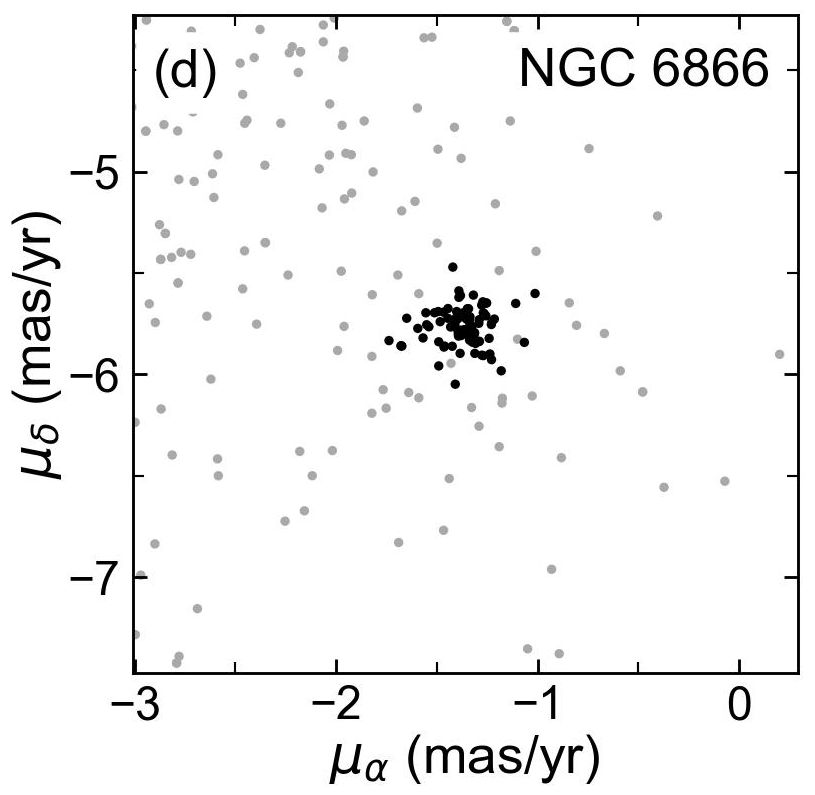}}\vspace*{-1ex}
	\caption{The ($\mu_{\alpha}$,~$\mu_{\delta}$) diagrams of the four OCs. Black and gray dots represent the members and field stars, respectively.}
\label{f3_membrs}	
\end{figure}

\begin{figure}
	\centering{\includegraphics[width=0.8\columnwidth]{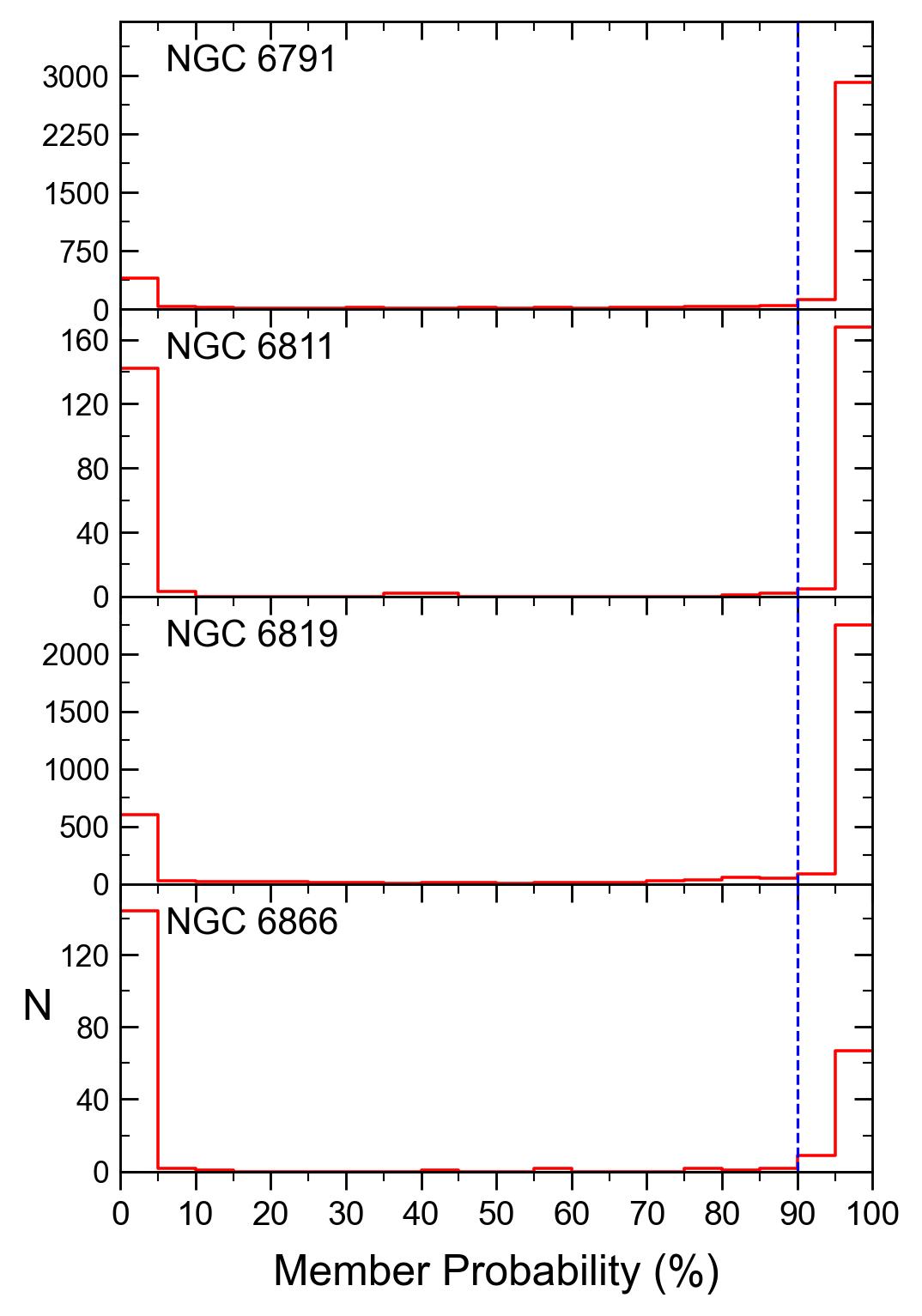}}\vspace{-2ex}
	\caption{The distributions of membership probabilities for the  four OCs according to GMM. The vertical dashed blue line shows the selected probability limit.}
\label{f4_proba}	
\end{figure}

Figures~\ref{f3_membrs} and \ref{f4_proba} show the proper motion and membership distributions of the cluster stars. Here we adopt the first significant rise in the distribution of the membership probabilities ($P>90\%$) as the membership percentage limit. In Fig.~\ref{f3_membrs}, the potential cluster members are clearly standing out compared to the scatter caused by field stars.

The distances of the four OCs based on Gaia-EDR3 parallaxes are obtained from the posterior probability density functions  \citep{Bailer2018,Bailer2021}. For this, we use the global zero point of $-$0.017 mas \citep{lin21}. The median equatorial coordinates, proper motion components, the median parallaxes and the distances of the four OCs are listed in Table~\ref{t2_proper}. 

\cite{cantat2020} and \cite{dia21} use UPMASK and maximum likelihood methods, respectively, for the membership determination based on Gaia DR2 data. These authors consider a membership probability higher than $50\%$ as limit and their number of members is almost close to each other (see Table~\ref{t2_proper}). Our membership determination, on the other hand, is based on the GMM technique and Gaia EDR3 data using P$>90\%$. This certainly leads to some discrepancies in the number of the members. In particular for the more populous clusters (NGC~6791 and NGC~6819), Gaia EDR3 data apparently reveal a much higher number of members. However, our derived astrometric median values are within the errors compatible with the results by these authors.


\renewcommand{\tabcolsep}{1.6mm}
\renewcommand{\arraystretch}{1.4}
\begin{table}
	\centering
	\caption{The median proper motion components and parallaxes/distances of the likely cluster members. }
	\label{t2_proper}
	\begin{tabular}{@{}lcccccp{0.5em}@{}}
		\hline
		Cluster & $\mu_{\alpha}$ & $\mu_{\delta}$ & $\varpi$ & $d$ & N&\\  
		&   (mas/yr)     &   (mas/yr)     &   (mas)  &  (kpc)  & & \\  
		\hline
		NGC\,6791  & -0.42$\pm$0.08  & -2.28$\pm$0.09 & 0.21$\pm$0.08 & 4.37$\pm$0.02 &2923& \\
		NGC\,6811  & -3.33$\pm$0.12  & -8.80$\pm$0.11 & 0.87$\pm$0.05 & 1.05$\pm$0.04 &166&\\
		NGC\,6819  & -2.90$\pm$0.03  & -3.88$\pm$0.04 & 0.37$\pm$0.03 & 2.37$\pm$0.20 &2430&\\
		NGC\,6866  & -1.38$\pm$0.02  & -5.79$\pm$0.02 & 0.69$\pm$0.04 & 1.30$\pm$0.03 & 67&\\
		\hline
		NGC\,6791  & -0.42$\pm$0.16  & -2.27$\pm$0.19 & 0.19$\pm$0.09 &      4.23     & 1629&\mrw{\rot{Cantat (2020)}}\\
		NGC\,6811  & -3.40$\pm$0.12  & -8.81$\pm$0.12 & 0.87$\pm$0.04 &      1.16     &  296\\
		NGC\,6819  & -2.92$\pm$0.13  & -3.86$\pm$0.14 & 0.36$\pm$0.05 &      2.76     & 1527\\
		NGC\,6866  & -1.36$\pm$0.08  & -5.74$\pm$0.09 & 0.69$\pm$0.03 &      1.41     &   72\\
		\hline
		NGC\,6791  & -0.43$\pm$0.18  & -2.27$\pm$0.22 & 0.19$\pm$0.09 &      4.23     & 1520&\mrw{\rot{Dias (2021)}}\\
		NGC\,6811  & -3.40$\pm$0.15  & -8.80$\pm$0.16 & 0.87$\pm$0.04 &      1.16     &  302\\
		NGC\,6819  & -2.92$\pm$0.13  & -3.86$\pm$0.16 & 0.36$\pm$0.05 &      2.76     & 1535\\
		NGC\,6866  & -1.36$\pm$0.09  & -5.73$\pm$0.09 & 0.69$\pm$0.03 &      1.41     &  104\\
		\hline	
	\end{tabular}
\flushleft
Notes: The top rows list our results, the lower rows the results by \citet{cantat2018,cantat2020} and \citet{dia21}.	
\end{table}

\section{Determination of the Reddening, Distance, Age and Mass}
\label{sect:4}
We first apply \textit{fitCMD}, an algorithm improved by \cite{Bonatto2019}, on the CCD~$UBV(RI)_{KC}$ and Gaia EDR3 $(G, G_{BP}-G_{RP})$ photometry of the probable members in the four OCs.

The algorithm transposes theoretical initial mass function (IMF) properties for the isochrones of given age and metallicity to their observational colour-magnitude diagrams (CMDs) \citep[see e.g.][]{Bonatto2019,Cakmak2021}. 
Based on the IMF properties of the B12 PARSEC isochrones \citep{bre12} \footnote{http://stev.oapd.inaf.it/cgi-bin/cmd.},  \textit{fitCMD} searches for the values of the cluster stellar mass $(m_{cl})$, Age, global metallicity ($Z$), foreground reddening, distance modulus $(m-M)$, and magnitude-dependent photometric completeness that produce the artificial and observational CMDs.

The observed CMD is converted into a Hess diagram, representing the density of stars in a magnitude range. The mass in each Hess cell is then  computed based on the mass range of the cells' magnitude range read from the best-fitting isochrone. The theoretical IMF is used only for the purpose of estimating the completeness-corrected mass by using the difference between the number of actually detected stars at a given magnitude to the expected ones. Therefore, all the other parameters are unaffected by this procedure. The \textit{fitCMD} algorithm was already succesfully applied to several OCs \citep[see e.g.][]{Cakmak2021}. 

\begin{figure}
	\centering{\includegraphics[width=0.47\columnwidth]{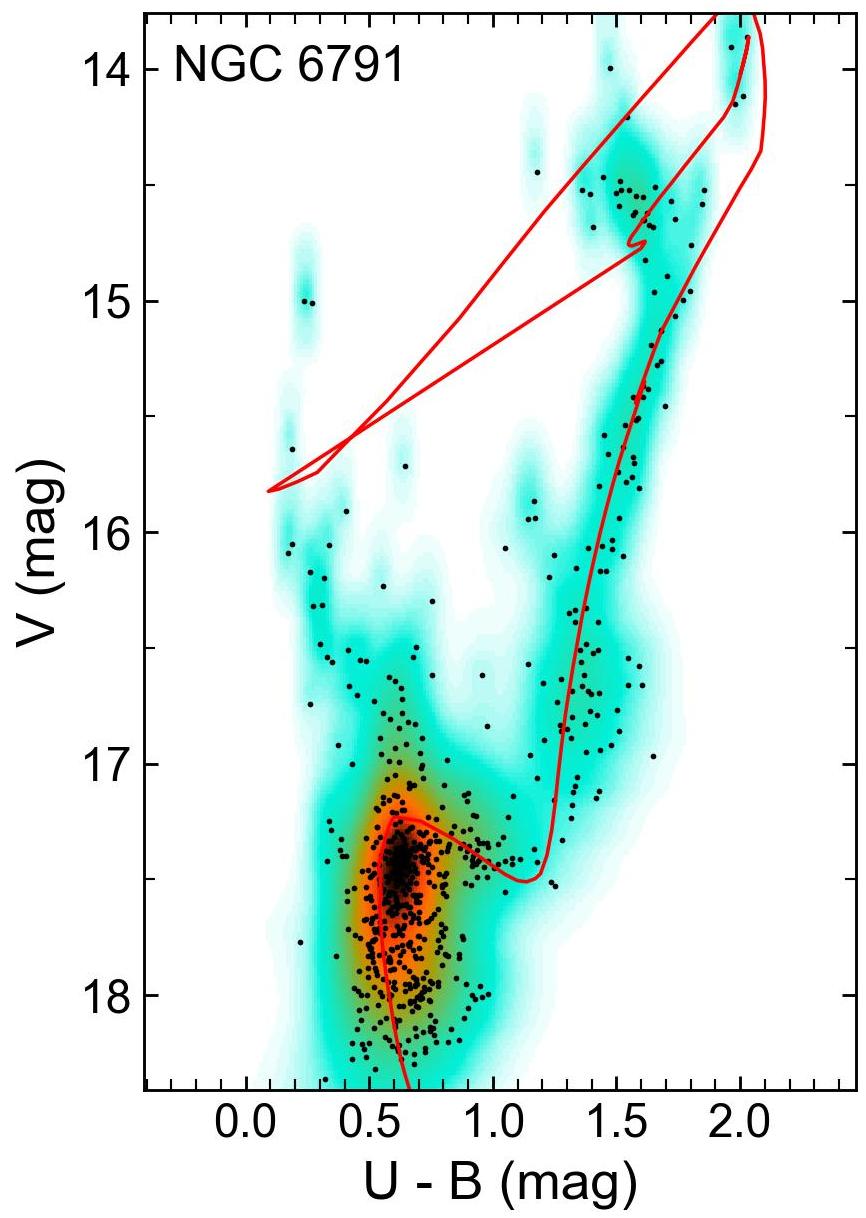} \hspace*{2ex} 
		\includegraphics[width=0.47\columnwidth]{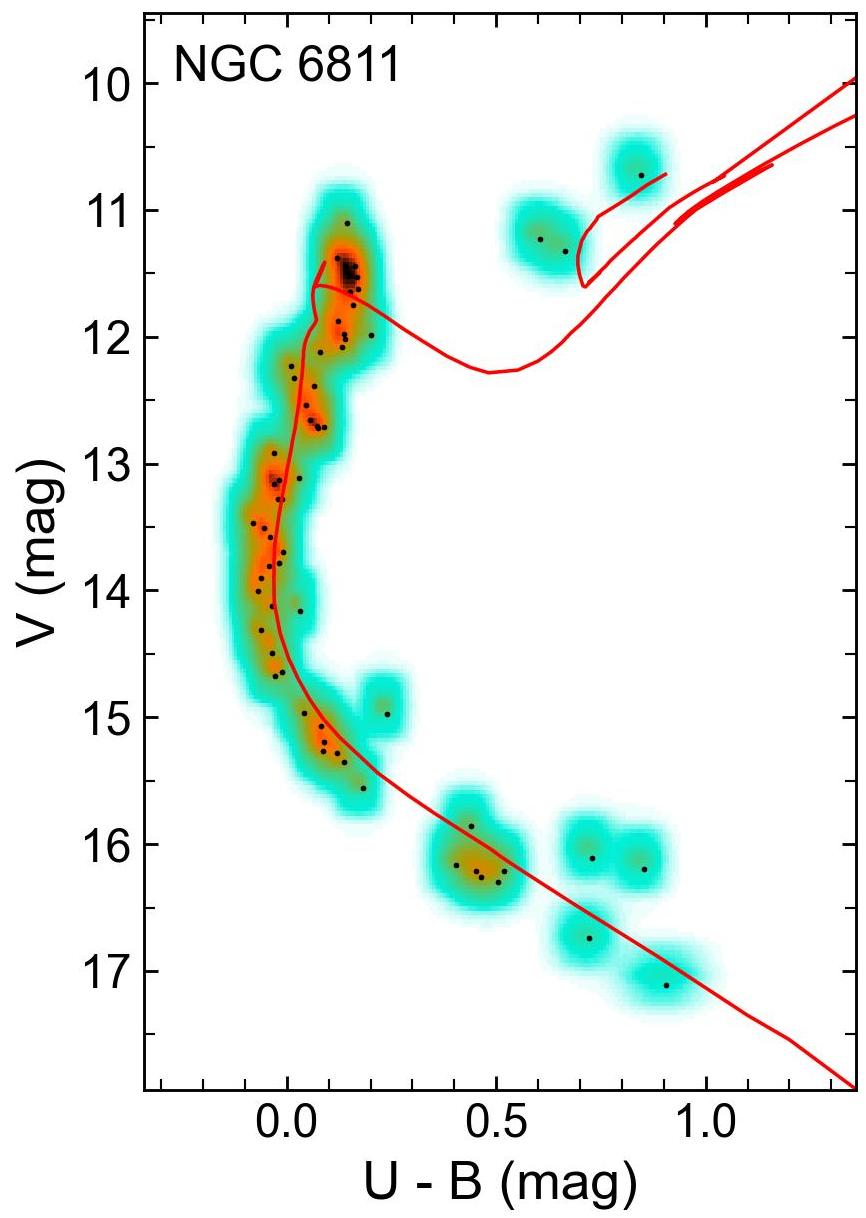}\\[1ex]
		\includegraphics[width=0.47\columnwidth]{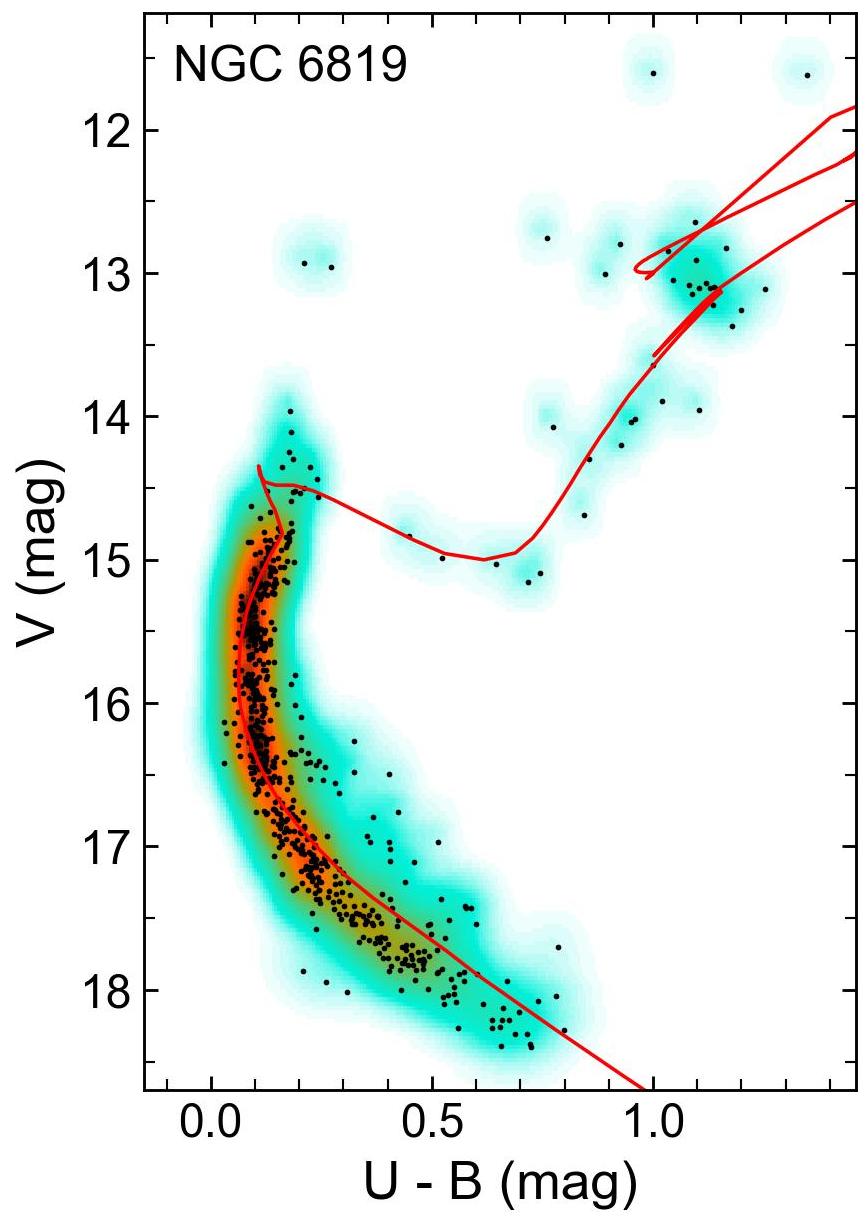} \hspace{2ex}
		\includegraphics[width=0.47\columnwidth]{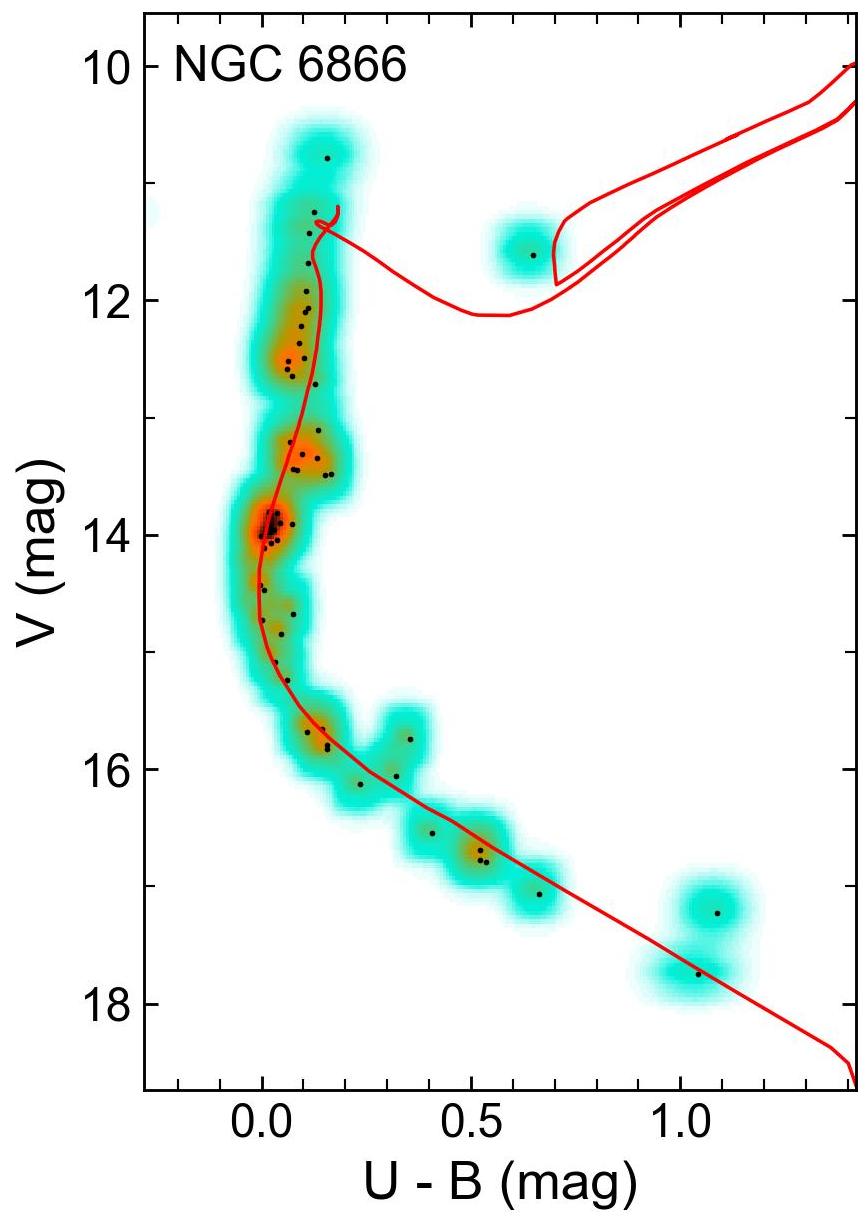}}\vspace{-2ex}
	\caption{V-(U-B) CMDs of the four OCs. The solid red line represents the best-fit PARSEC isochrones, and the coloured areas the Hess diagram.}
	\label{f5_cmdvub}
\end{figure}

\begin{figure*}
	\centering{
		\includegraphics[width=0.455\textwidth]{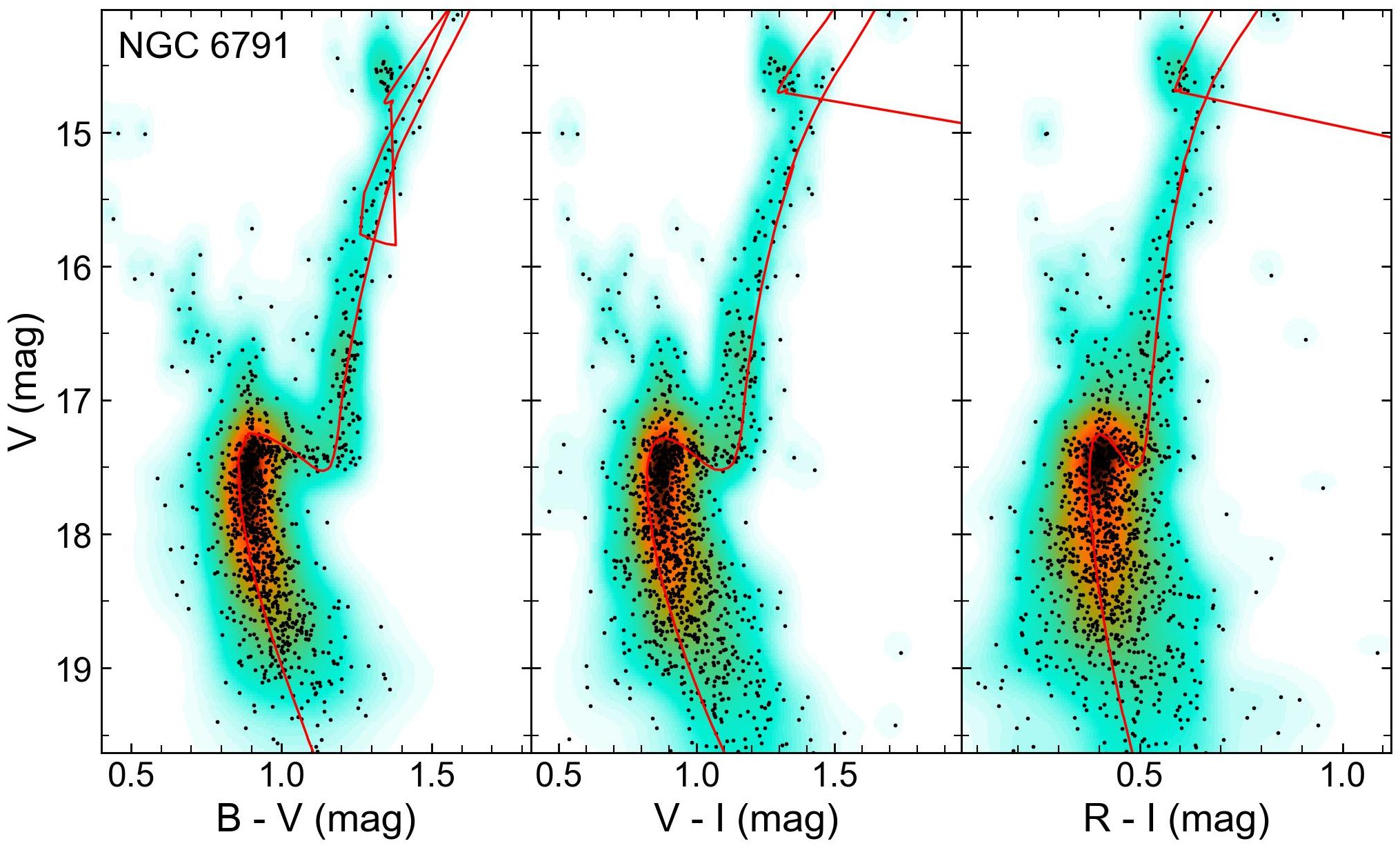}\hspace*{2ex}
		\includegraphics[width=0.455\textwidth]{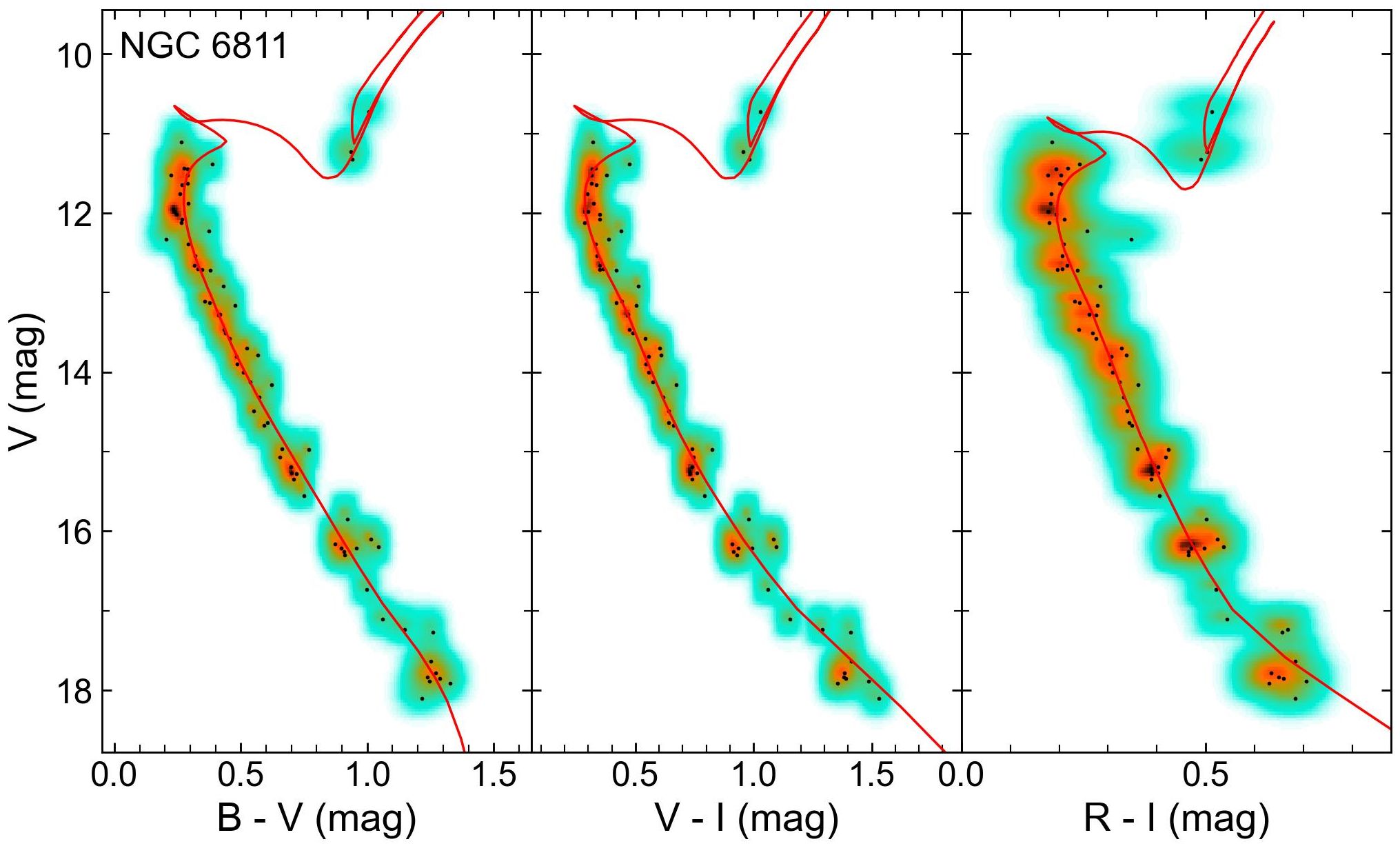}\\[2ex]
		\includegraphics[width=0.455\textwidth]{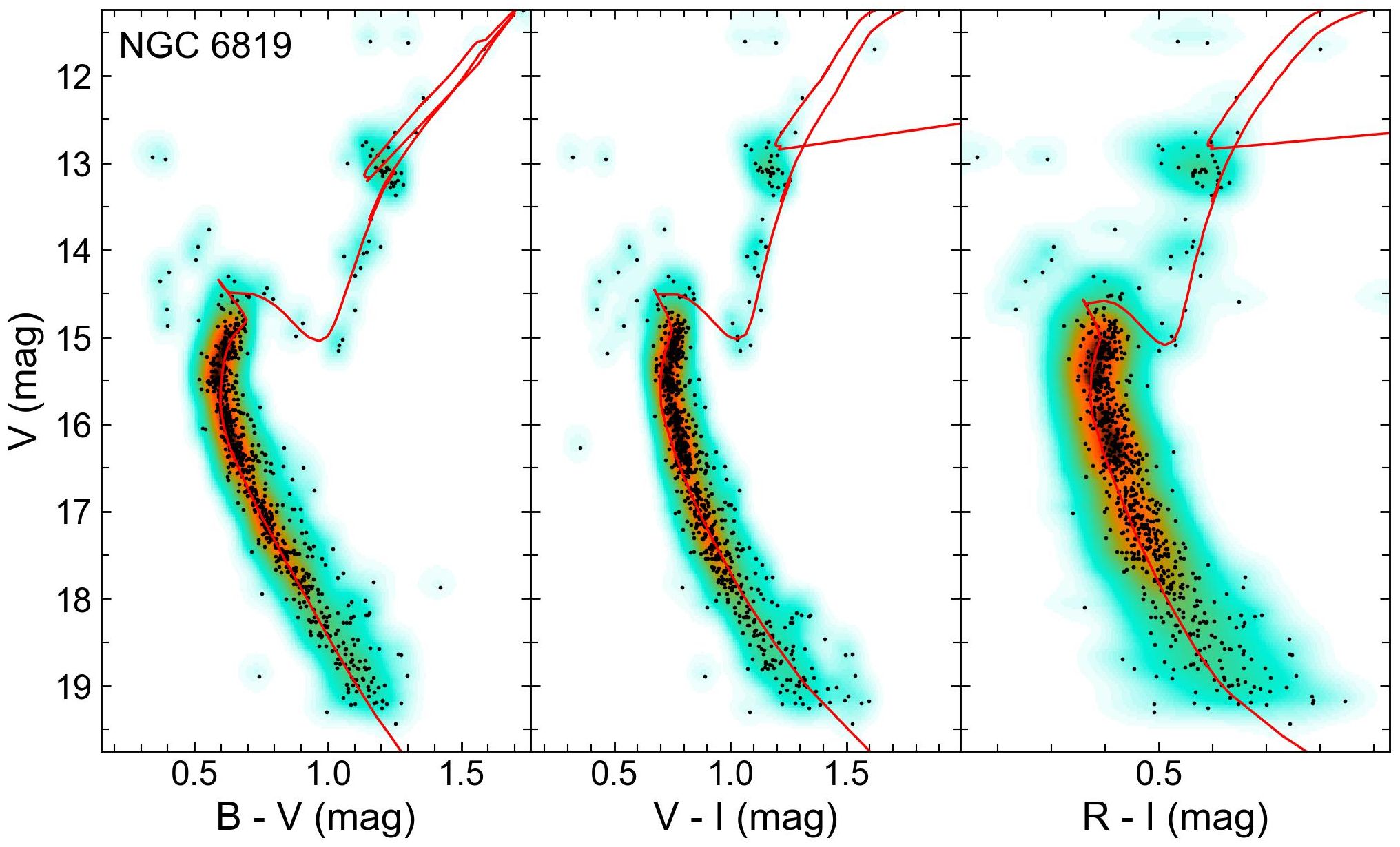}\hspace*{2ex}
		\includegraphics[width=0.455\textwidth]{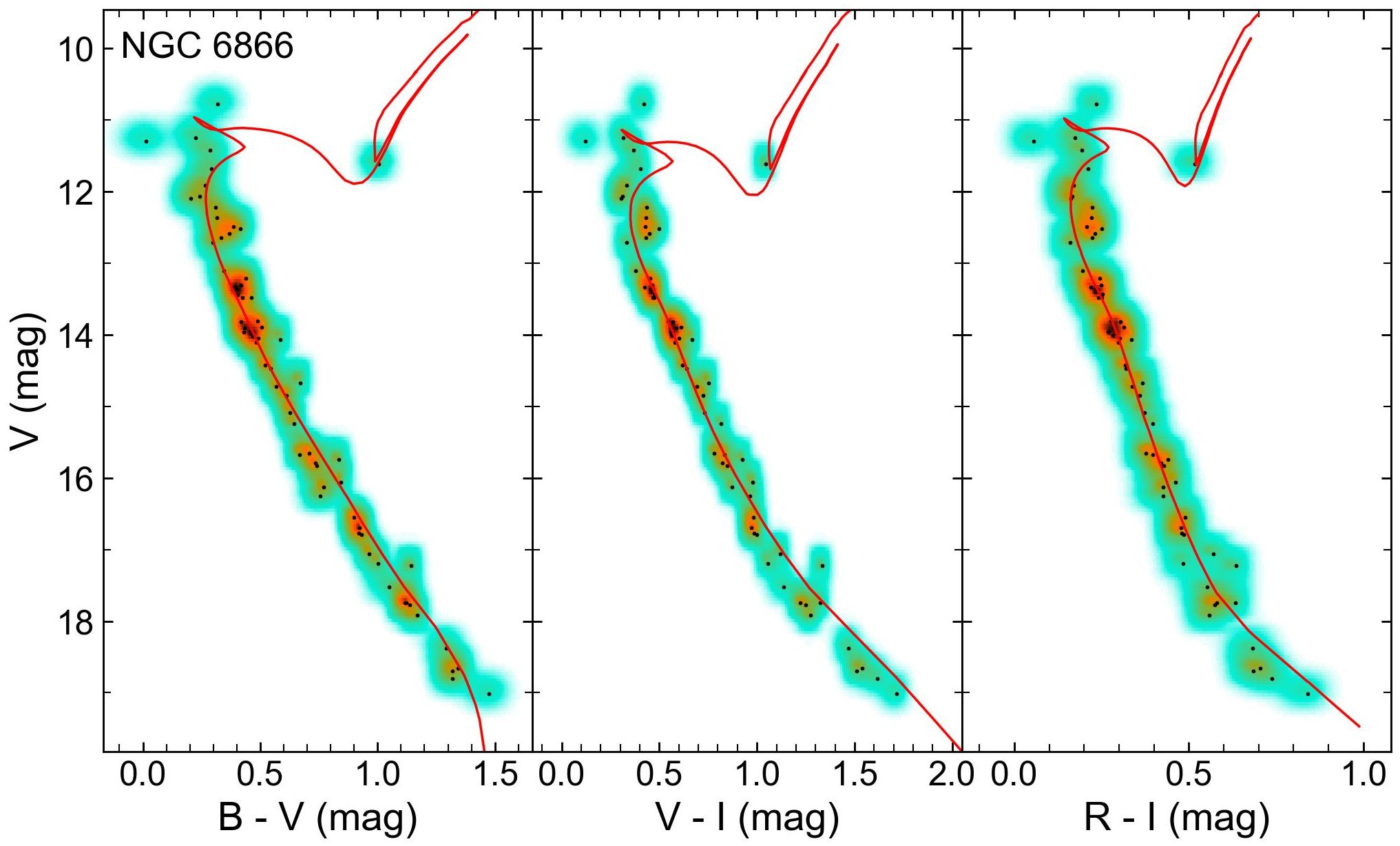}}\vspace*{-1ex}
	\caption{The V-(B-V), V-(V-I) and V-(R-I) CMDs of the four OCs. The symbols are the same as in Fig.~\ref{f5_cmdvub}.}
\label{f6_cmdubvri}
\end{figure*}

\begin{figure}
	\centering{\includegraphics[width=0.455\columnwidth]{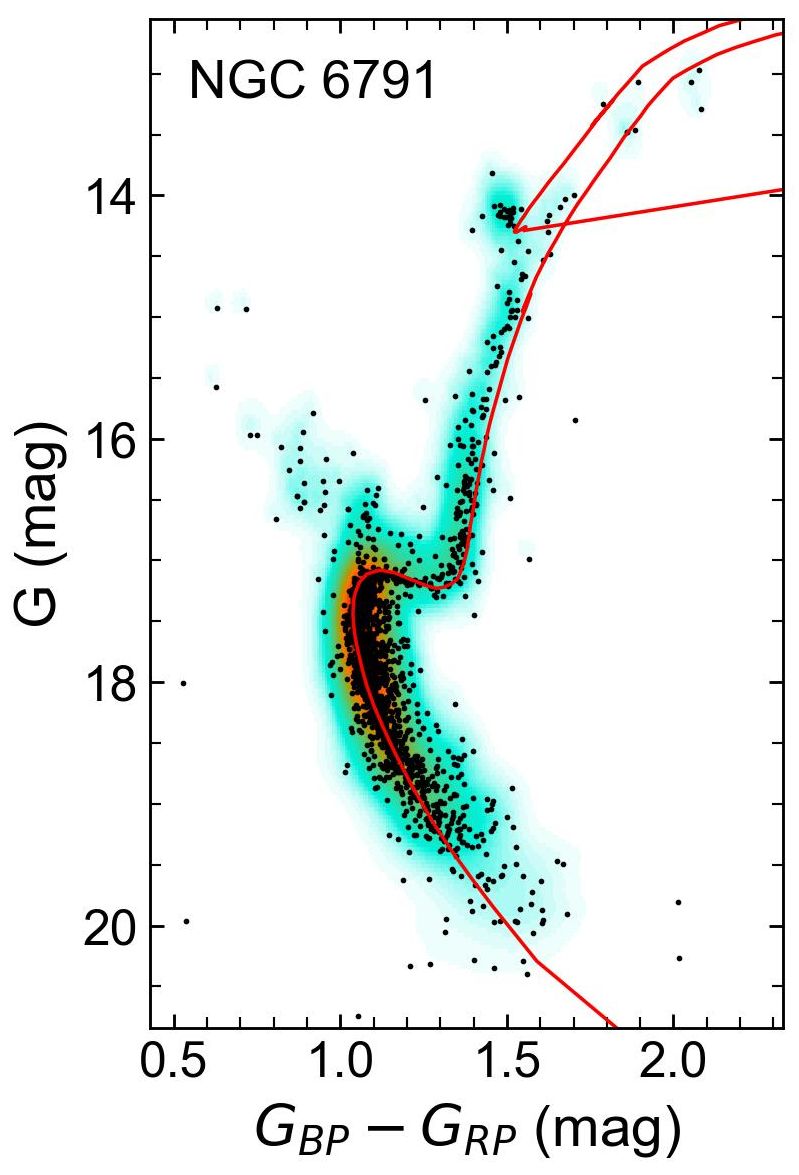} \hspace*{2ex} 
		\includegraphics[width=0.455\columnwidth]{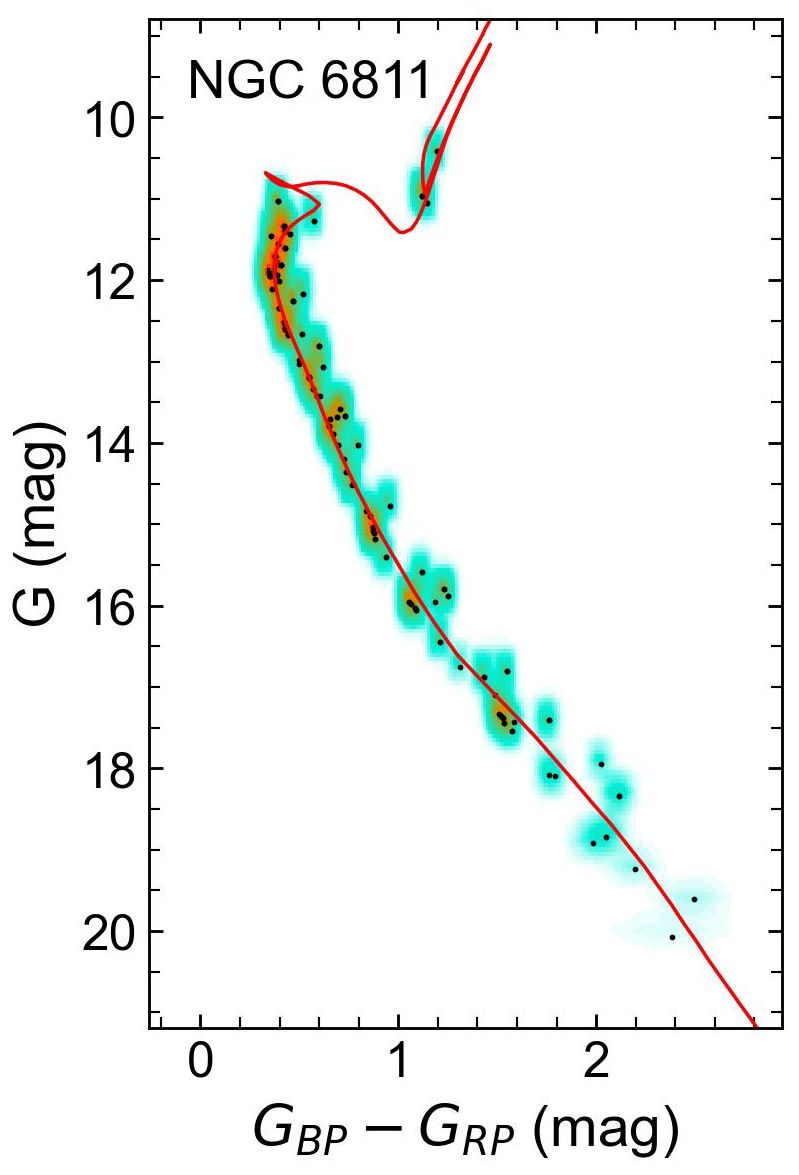}\\[2ex]
		\includegraphics[width=0.455\columnwidth]{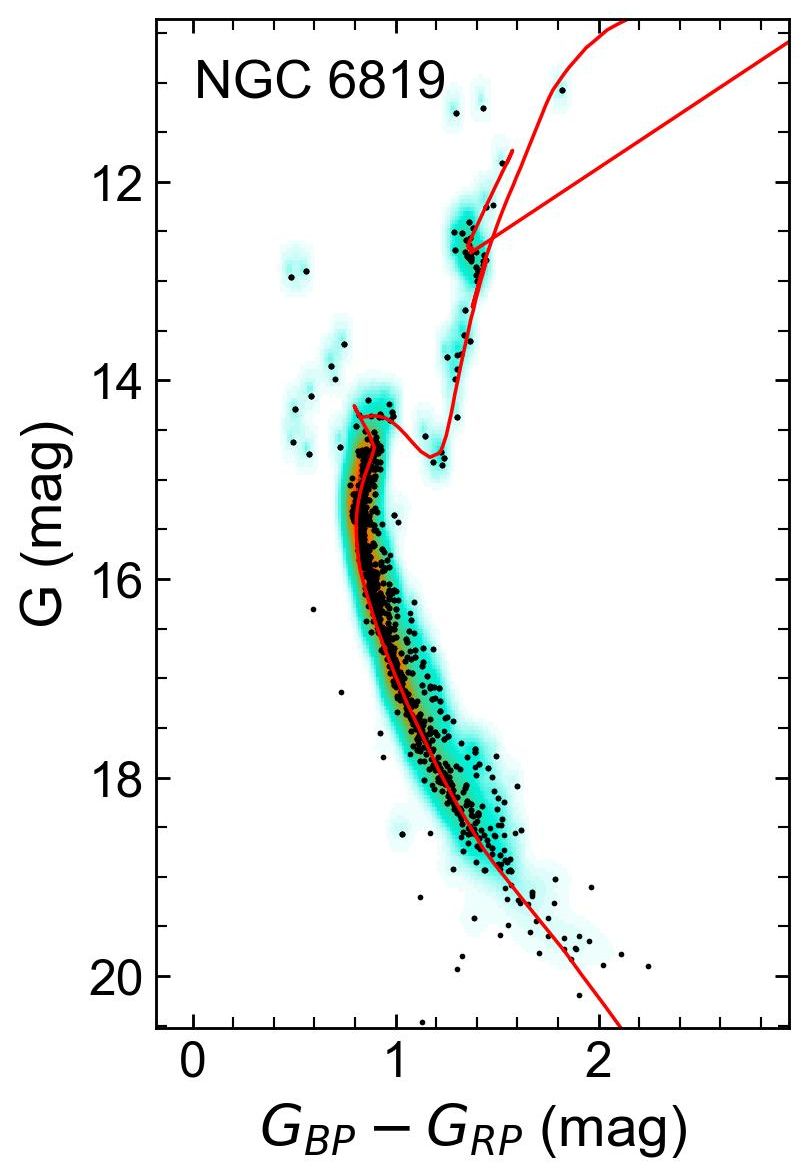} \hspace*{2ex}
		\includegraphics[width=0.455\columnwidth]{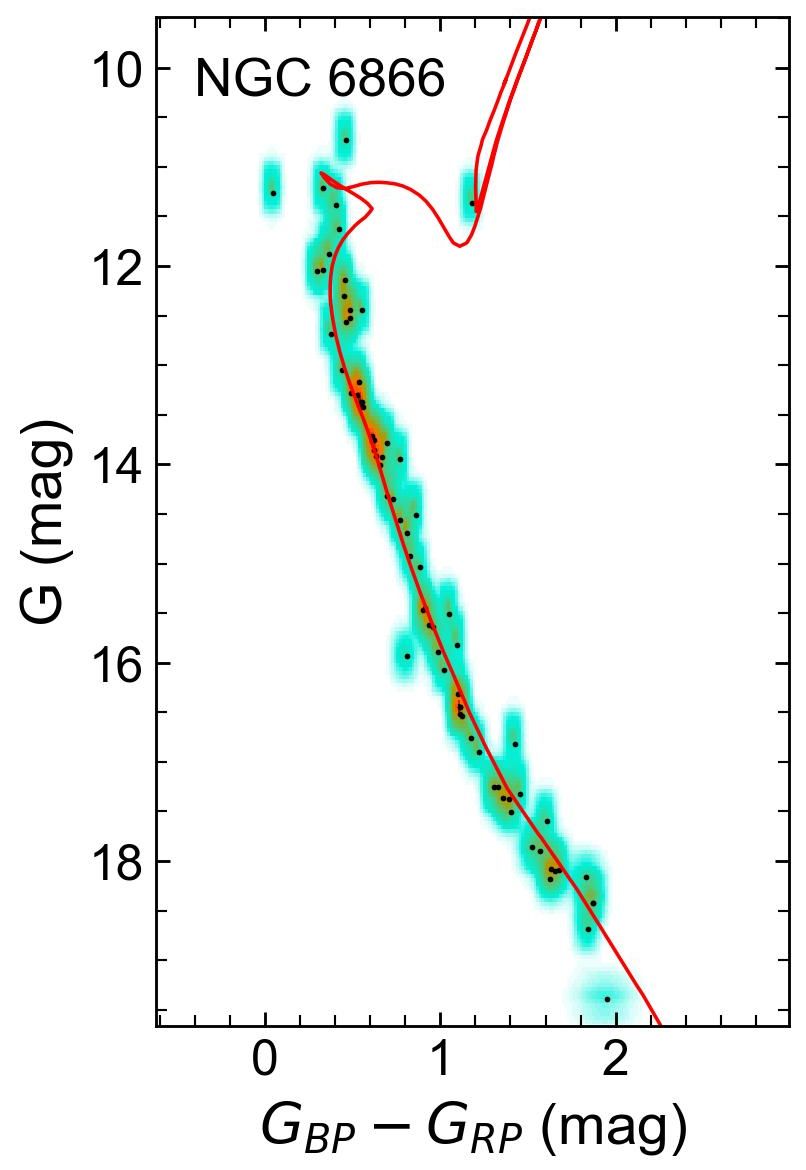}}\vspace{-1ex}
	\caption{GAIA CMDs of the four OCs. The symbols are the same as in Fig.~\ref{f5_cmdvub}.}
\label{f7_cmdgaia}
\end{figure}

Spectroscopic metal abundances are available for the programme clusters by \citet{don20} - see Table~\ref{t9_litcompbig}, reference~13. Thus, we adopt these measured values as input for \textit{fitCMD} to reduce the number of free parameters. 
The obtained best-fitting astrophysical parameters from \textit{fitCMD} - the reddening (colour excess), the distance modulus / distance in pc, and the Age in Gyr) for all five investigated CMDs are listed in Table~\ref{t3_fitcmd} and the observed CMDs (Hess diagrams) are shown in Figs.~\ref{f5_cmdvub}-\ref{f7_cmdgaia}. The B12 isochrones reproduce well the main-sequence, turn-off and red giant/red clump regions in the individual CMDs. 

\renewcommand{\tabcolsep}{5.2mm}
\renewcommand{\arraystretch}{1.2}
\begin{table*}
\centering
\caption{The derived astrophysical parameters using \textit{fitCMD}. CE is the colour excess (reddening).}
\label{t3_fitcmd}
\begin{tabular}{lAAAA}
	\hline
	NGC 6791  &    \mcl{CE}   & \mcl{$(V-M_{V})_{o}$} &  \mcl{d (pc)} & \mcl{Age (Gyr)}  \\
	\hline
	(U-B)     &  0.308&0.042  &   12.40&0.21   &  3023&294  &  5.80&0.60  \\
	(B-V)     &  0.174&0.031  &   13.16&0.20   &  4294&392  &  5.80&0.90  \\
	(V-I)     &  0.076&0.030  &   13.40&0.15   &  4777&327  &  7.00&0.90  \\
	(R-I)     &  0.036&0.027  &   13.41&0.14   &  4808&319  &  6.80&0.90  \\
	$(G_{BP}-G_{RP})$ &  0.179&0.024  &   13.21&0.13   &  4390&257  &  7.20&0.90  \\
	\hline
	NGC 6811  &     \mcl{}    &      \mcl{}    &   \mcl{}   &    \mcl{}   \\
	\hline
	(U-B)     &  0.030&0.011  &   10.29&0.23   &  1142&122  &  1.45&0.20  \\	
	(B-V)     &  0.010&0.061  &    9.93&0.33   &   970&147  &  1.20&0.20  \\
	(V-I)     &  0.001&0.077  &   10.03&0.34   &  1014&157  &  1.15&0.20  \\
	(R-I)     &  0.042&0.035  &    9.84&0.31   &   931&133  &  1.25&0.20  \\
	$(G_{BP}-G_{RP})$    &  0.023&0.083  &   10.01&0.30   &  1004&138  &  1.20&0.20  \\
	\hline 
	NGC 6819  &     \mcl{}    &      \mcl{}    &   \mcl{}   &    \mcl{}   \\
	\hline
	(U-B)     &  0.095&0.014  &   11.85&0.24   &  2343&264  &  2.85&0.30  \\
	(B-V)     &  0.111&0.024  &   12.04&0.33   &  2557&384  &  2.60&0.40  \\
	(V-I)     &  0.111&0.002  &   11.72&0.35   &  2212&358  &  3.10&0.30  \\
	(R-I)     &  0.089&0.001  &   11.57&0.45   &  2057&434  &  3.20&0.40  \\
	$(G_{BP}-G_{RP})$     &  0.169&0.011  &   11.91&0.26   &  2407&292  &  2.90&0.30  \\
	\hline
	NGC 6866  &     \mcl{}    &      \mcl{}    &    \mcl{}  &    \mcl{}   \\
	\hline
	(U-B)     &  0.066&0.029  &   10.67&0.34   &  1360&214  &  0.90&0.20  \\
	(B-V)     &  0.061&0.073  &   10.38&0.38   &  1189&208  &  1.00&0.20  \\
	(V-I)     &  0.087&0.079  &   10.38&0.35   &  1190&190  &  1.10&0.20  \\
	(R-I)     &  0.066&0.039  &   10.34&0.37   &  1169&198  &  0.95&0.20  \\
	$(G_{BP}-G_{RP})$     &  0.105&0.101  &   10.51&0.35   &  1266&206  &  1.00&0.20  \\
	\hline
\end{tabular}
\end{table*}

We note that Gaia~EDR3 $G$-magnitudes require a slight correction depending on the magnitude, colour, and astrometric solution \footnote{https://www.cosmos.esa.int/web/gaia/edr3-known-issues}. These corrections might be up to only 0.01\,mag in the colour range of the identified cluster members. However, most of our members show a five-parameter astrometric solution, which is generally unaffected. Only few stars would require a minimal correction of up to 2\,mmag. Thus, we have not applied a correction, because of the negligible influence on the derived cluster parameters, in particular if considering their errors.  

Furthermore, we have also determined the astrophysical parameters metallicity, the reddening, the distance modulus, and the age of our sample OCs by using the differential grid (DG) technique developed by \cite{poh10}, and improved by \cite{net13,net22}. This method was already applied to about 90 open clusters and provides well-scaled results for the metallicity at reasonable acccuracy \cite[see e.g.][]{net16}.  Thus, it also allows to verify the spectroscopic metallicities, which were adopted as input for the \textit{fitCMD} approach.For this, the photometric data of main-sequence cluster stars were transformed to luminosities and mean effective temperatures, the latter based on up to five colour indices using 2MASS \citep{skr06}, Gaia~EDR3, and our photometric data. These were compared to zero-age main-sequence (ZAMS) normalised B12 PARSEC isochrones. For details we refer to the paper by \cite{net22} and references therein. The derived astrophysical parameters are presented in Table~\ref{t4_difgrid} and the fits are shown in Fig.~\ref{f3_app_isoch}. 
The results agree well with the ones obtained by \textit{fitCMD} and the spectroscopic metallicities. We note that this method was not applicable to NGC~6791, because of the very old cluster age and the resulting small luminosity range of the main-sequence down to solar mass stars. 

\renewcommand{\tabcolsep}{2.6mm}
\renewcommand{\arraystretch}{1.2}
\begin{table*}
	\caption{Derived astrophysical parameters using the differential grids.}
	\label{t4_difgrid}
	\begin{tabular}{cccccccc}
		\hline
		& Z & [Fe/H] & E(B-V) &$(V-M_{V})_{o}$ & d~(pc) & $\log Age$ &  Age~(Gyr) \\
		\hline
		NGC 6811 & 0.014$\pm$0.002 &-0.03$\pm$0.06 &0.02 &9.90   & 955 & 9.10 & 1.26 \\
		NGC 6819 & 0.017$\pm$0.004 &0.06$\pm$0.11  &0.14 &11.80  & 2291 & 9.45 & 2.82 \\
		NGC 6866 & 0.017$\pm$0.005 &0.06$\pm$0.13  &0.09 &10.40  & 1202 & 9.00 & 1.00 \\
		\hline
	\end{tabular}
\end{table*}

\section{Structural parameters and Mass Function}
\label{sect:5}
\begin{figure}
	\centering{\includegraphics[width=0.475\columnwidth]{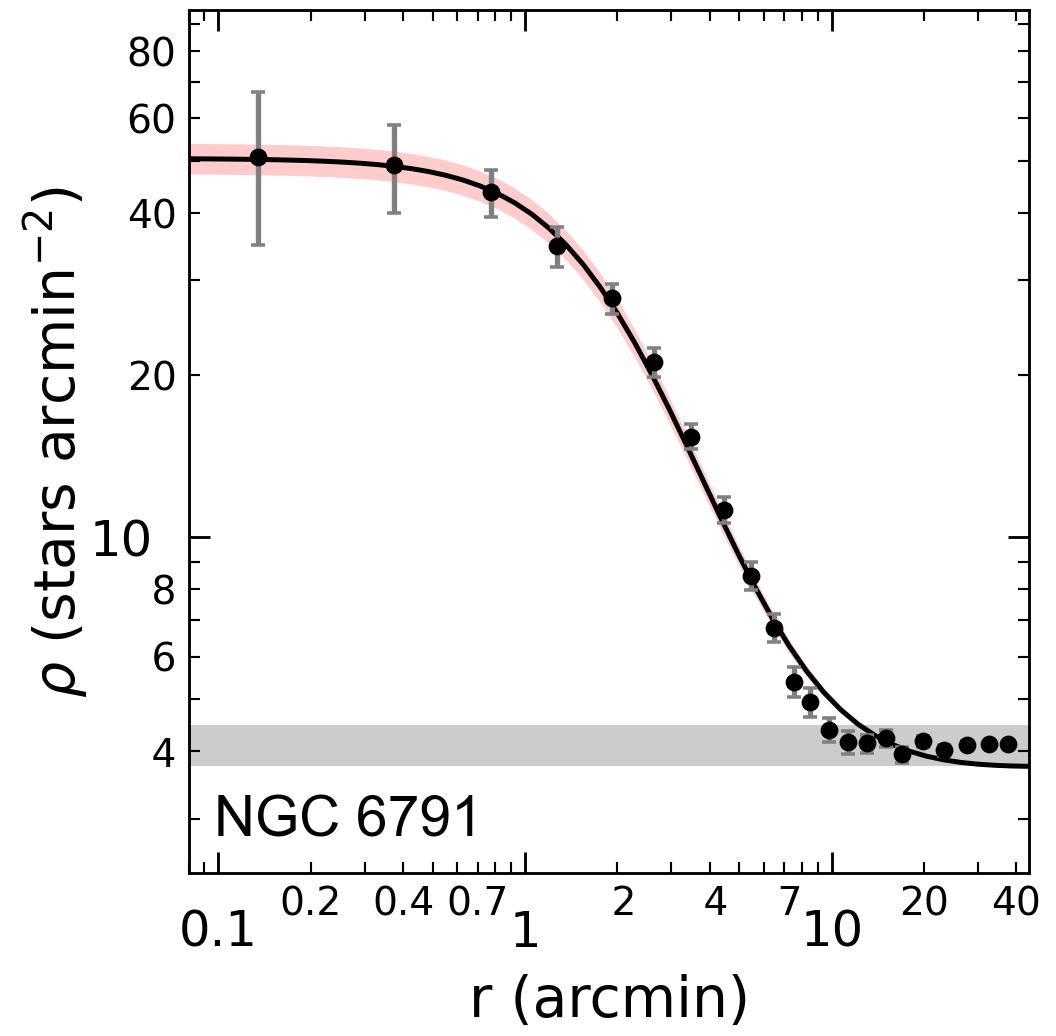} \hspace*{2ex}
		\includegraphics[width=0.48\columnwidth]{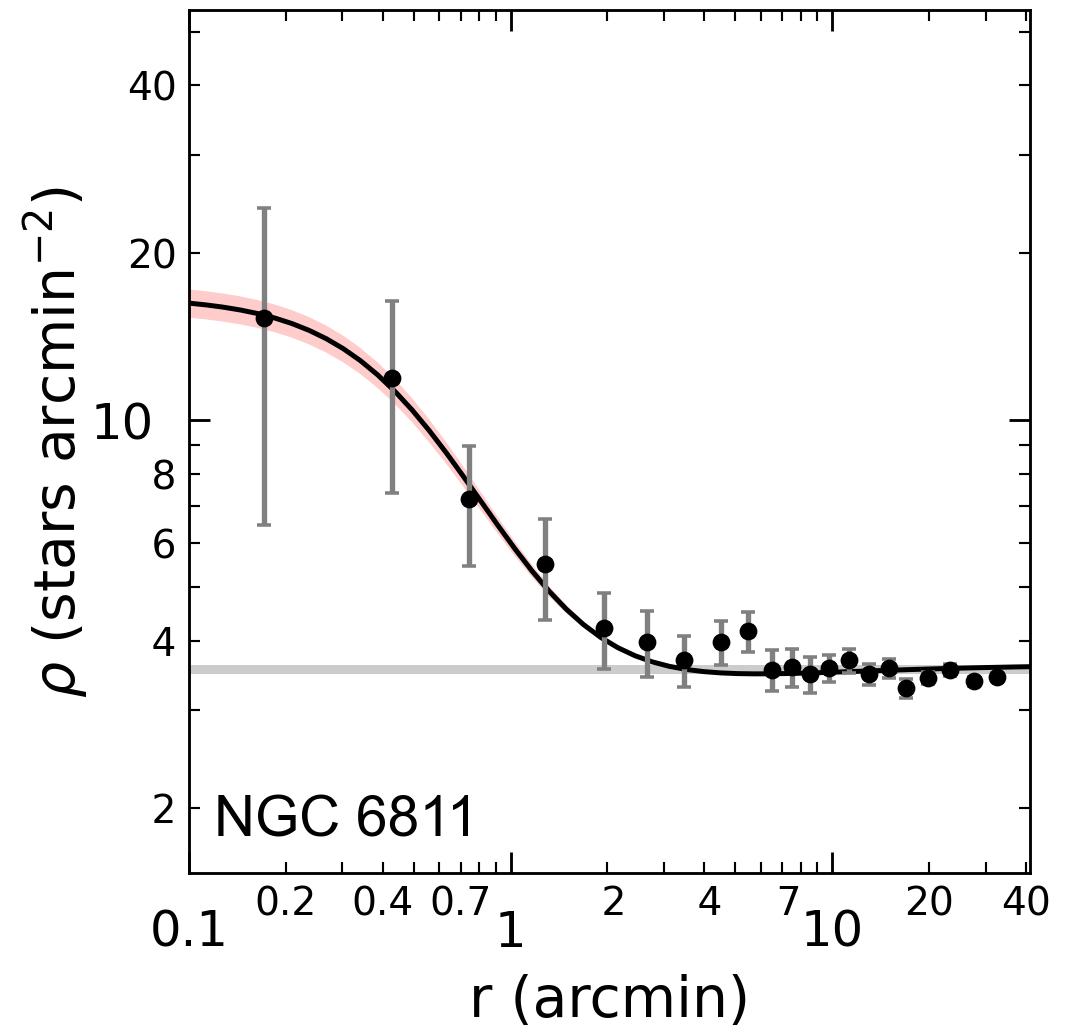}\\[2ex]
		\includegraphics[width=0.47\columnwidth]{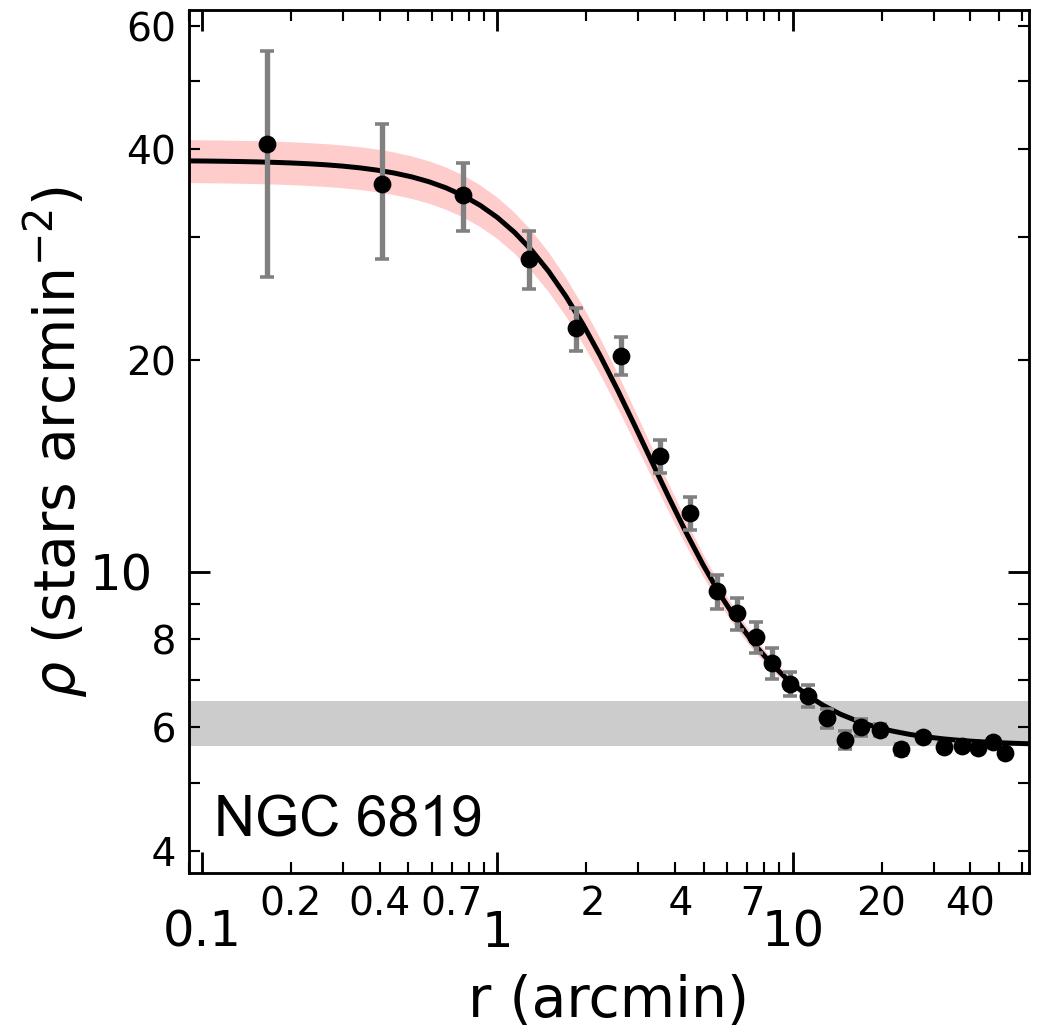}\hspace*{2ex}
		\includegraphics[width=0.48\columnwidth]{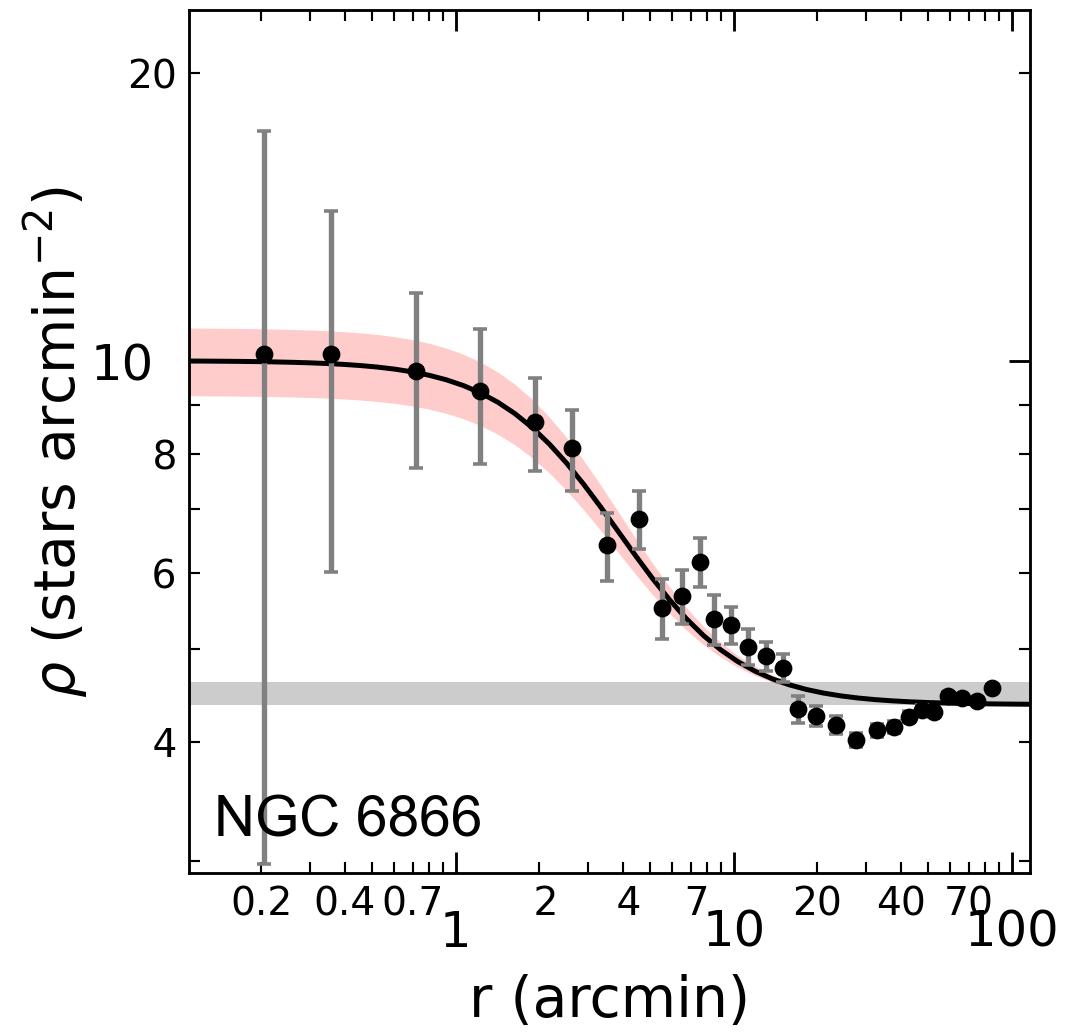}}\vspace*{-1ex}
	\caption{The radial density profiles of the four OCs. The solid lines show the best-fitting King profile and the horizontal gray area indicates the stellar background level measured in the comparison field. The $1\sigma$ King fit uncertainty is shown by the shaded range.}
\label{f8_rdp}
\end{figure}

The structural parameters of the four OCs were derived using stellar radial density profiles (RDPs). For this, we selected member stars that are brighter than 20\,mag in the G band and applied the relation  $\sigma(R) = \sigma_{bg} + \sigma_0/(1+(R/R_{core})^2$ by \cite{King1966}\footnote{Here, $\sigma_{bg}$, $\sigma_0$ and $R_{core}$ are the residual background and the central densities of stars, and the core radius, respectively.}. We obtained the cluster radii $(R_{RDP})$  by comparing the RDP level with the background (see Fig.~\ref{f8_rdp}). The low star content in the central parts of some OCs (NGC~6811 and NGC~6866) are responsible for the large uncertainties within  $R<1'$ of the RDPs.

Table~\ref{t5_struct} lists the derived structural parameters. The results for $R_{core}$ and $R_{RDP}$ in pc are almost consistent with the ones by \cite{buk2011}, who presents data for all four objects. \cite{Tarr2022}, on the other hand, list somewhat larger $R_{core}$ values for two objects in common (NGC~6811 and NGC~6866), and \cite{zho22} obtained for these by far the largest radii. 

For simplicity, the tidal radii $R_{t}$ in Table~\ref{t5_struct} were estimated using $R_{t}~(pc)=1.54~R_{RDP}$ given by \cite{pis07}. Notice that \cite{Bonatto2005} indicate $R_{t}~(pc)=1.4-1.9~R_{RDP}$ based on bright OCs. 
\cite{gao20} and \cite{Tarr2022}, on the other hand, use the three-parameter function ($R$, $R_{core}$, $R_{t}$) by \cite{King1962}. Our $R_{t}$ estimate for NGC~6791 is in reasonable agreement with literature, but for NGC~6811 and NGC~6866 our radii are much smaller. The differences for $R_{t}$ and $R_{core}$ can be explained as follows: the three-parameter \cite{King1962} model describes well the outer parts of a cluster, while the two-parameter \cite{King1966} model describes the central region of the clusters \citep{Bonatto2005}. Additionally, the tidal radius can only be derived well for globular clusters or populous OCs at higher Galactic latitude, but for sparse OCs close to the galactic plane an accurate determination of $R_{t}$ becomes difficult. 

Note that the results by \cite{gao20}, \cite{Tarr2022} and \cite{zho22} are based on Gaia data, but the membership selection differs. \cite{gao20} and \cite{zho22} use the GMM model with $P > 80\%$ and UPMASK with  $P > 70\%$, respectively. \cite{Tarr2022} consider the clustering algorithm named hierarchical density based spatial clustering of applications with noise (HDBSCAN) in its python implementation with using 0.5 as a probability cut-off for the membership. \cite{pla11} on the other hand use the astrometric data from Lick and Kitt Peak National observatory by adopting $P > 80\%$, while \cite{buk2011} and \cite{Gunes2012} apply a \cite{King1966} fit to 2MASS $JHK_{s}$ data and a photometric member selection. Furthermore, \cite{pis07} and \cite{kha13} adopt stars with kinematic/photometric membership probabilities higher than $60\%$ and apply a fit using the model by \cite{King1962}. All these factors results in different cluster sizes and do not allow a proper direct comparison, in particular for small samples in common.

\renewcommand{\tabcolsep}{1.8mm}
\renewcommand{\arraystretch}{1.2}
\begin{table*}	
	{\footnotesize
		\begin{center}
			\caption{Structural parameters of the four objects.}    
			\label{t5_struct}
		  	\resizebox{1.0\textwidth}{!}{ 
			\begin{tabular}{lcAAAA|AAAAccc}
				\hline
				\mcc{Cluster} & $(1')$   & \mcl{$\sigma_{0K}$} & \mcl{$\sigma_{bg}$} & \mcl{$R_{core}$}
				& \mcll{$R_{RDP}$} & \mcl{$\sigma_{0K}$} & \mcl{$\sigma_{bg}$} & \mcl{$R_{core}$}
				& \mcl{$R_{RDP}$} & \mcl{$R_{t}$} & CC \\
				& ($pc$) & \mcl{($*\,\prime^{-2}$)}&\mcl{($*\,\prime^{-2}$)}
				& \mcl{($\prime$)}& \mcll{($\prime$)} &\mcl{($*\,pc^{-2}$)}& \mcl{($*\,pc^{-2}$)}& \mcl{($pc$)}
				& \mcl{($pc$)} & \mcl{($pc$)} & \\
				\mcc{($1$)} & ($2$)    & \mcl{($3$)} & \mcl{($4$)} & \mcl{($5$)} & \mcll{($6$)}
				& \mcl{($7$)} & \mcl{($8$)} & \mcl{($9$)} & \mcl{($10$)}&\mcl{($11$)} &($12$) \\
				\hline
				NGC\,6791 & 1.27 & 3.75&0.36 & 50.6&3.6 & 1.97&0.08 & 11.27&0.45 & 4.8&0.4  &64.4&4.6 & 2.51&0.11 & 14.33&0.57 & &22.07& 0.998\\
				NGC\,6811 & 0.29 & 3.44&0.07 & 13.5&0.4 & 0.50&0.02 &  5.46&0.28 & 1.0&0.02 & 3.9&0.1 & 0.15&0.01 &  1.59&0.08 & &2.45&0.993\\
				NGC\,6819 & 0.69 & 5.65&0.34 & 32.9&2.7 & 2.00&0.11 & 13.05&0.58 & 3.9&0.2  &22.7&1.8 & 1.38&0.07 &  8.99&0.40 & &13.84&0.993\\
				NGC\,6866 & 0.36 & 4.38&0.11 &  5.6&0.6 & 3.15&0.25 & 15.02&0.57 & 1.6&0.04 & 2.0&0.2 & 1.15&0.09 &  5.48&0.21 & &8.44&0.971\\
				\hline
				\hline        					
				NGC\,6791 &      &   \mcl{}  &  \mcl{}  &  \mcl{}   &   \mcll{}  &  \mcl{}  &  \mcl{} & 2.62&0.23 & 24.61&2.61 & & & 1 \\
				
				&      &   \mcl{}  &  \mcl{}  &  \mcl{}   &   \mcll{}  &  \mcl{}  &  \mcl{} & 3.91&0.68 &\mcl{} & &29.54$\pm$5.42 &4 \\
				&      &   \mcl{}  &  \mcl{}  &  \mcl{}   &   \mcll{}  &  \mcl{}  &  \mcl{} & \mcl{3.80}  &\mcl{} & &27.00 &5 \\
				&      &   \mcl{}  &  \mcl{}  &  \mcl{}   &   \mcll{}  &  \mcl{}  &  \mcl{} & 4.40&0.30 &\mcl{} & &20.30$\pm$2.70 &6 \\	
				\hline
				NGC\,6811 &      &   \mcl{}  &  \mcl{}  &  \mcl{}   &   \mcll{}  &  \mcl{}  &  \mcl{} & 0.36&0.04 &  2.89&0.42 &  & & 1   \\
				
				&      &   \mcl{}  &  \mcl{}  &  \mcl{}   &   \mcll{}  &  \mcl{}  &  \mcl{} & \mcl{1.70} &\mcl{} & & 5.90 & 2   \\
				&      &   \mcl{}  &  \mcl{}  &  \mcl{}   &   \mcll{}  &  \mcl{}  &  \mcl{} & 1.81&0.39 &\mcl{} & & 6.78$\pm$1.22 & 4   \\
				&      &   \mcl{}  &  \mcl{}  &  \mcl{}   &   \mcll{}  &  \mcl{}  &  \mcl{} & 1.63&0.27 &\mcl{} & & 20.85$\pm$2.00 & 7   \\
				&      &   \mcl{}  &  \mcl{}  &  \mcl{}   &   \mcll{}  &  \mcl{}  &  \mcl{} &\mcl{4.30} &\mcl{} & & 8.80 & 8   \\
				\hline
				NGC\,6819 &      &   \mcl{}  &  \mcl{}  &  \mcl{}   &   \mcll{}  &  \mcl{}  &  \mcl{} & 2.20&0.20 & 17.67&1.77 & & & 1   \\
				&      &   \mcl{}  &  \mcl{}  &  \mcl{}   &   \mcll{}  &  \mcl{}  &  \mcl{} & 1.76&0.12& \mcl{} & &13.32$\pm$0.96 & 4   \\
				\hline
				NGC\,6866 &      &   \mcl{}  &  \mcl{}  &  \mcl{}   &   \mcll{}  &  \mcl{}  &  \mcl{} & 0.59&0.09 &  2.94&0.46 & & &1     \\
				&      &   \mcl{}  &  \mcl{}  &  \mcl{}   &   \mcll{}  &  \mcl{}  &  \mcl{} & 0.61&0.11 &\mcl{} & &6.50$\pm$0.96 & 4   \\
				&      &   \mcl{}  &  \mcl{}  &  \mcl{}   &   \mcll{}  &  \mcl{}  &  \mcl{} & 1.97&0.46&9.15&0.27 & & &3  \\ 
				&      &   \mcl{}  &  \mcl{}  &  \mcl{}   &   \mcll{}  &  \mcl{}  &  \mcl{} & 1.96&0.36 &\mcl{} & &25.66$\pm$4.70  & 7   \\
				&      &   \mcl{}  &  \mcl{}  &  \mcl{}   &   \mcll{}  &  \mcl{}  &  \mcl{} & \mcl{4.70} &\mcl{} & & 12.90 & 8   \\  
				
				\hline
			\end{tabular}
		}
		\end{center}
}
\flushleft
Notes: $R_{t}$ in Col.~11 represents the tidal radius and Col.~12 shows the correlation coefficient of the RDP fit. ($*\,\prime^{-2}$) and ($*\,pc^{-2}$) in Cols.~3-4 and 7-8 represent $stars~arcmin^{-2}$ and $stars~pc^{-2}$, respectively. For the radii in pc we adopt the distance based on Gaia data in Table~\ref{t3_fitcmd}. The very first panel lists our results, the subsequent ones literature data - References: 1: \cite{buk2011}, 2: \cite{pis07}, 3:\cite{Gunes2012}, 4: \cite{kha13},  5:\cite{pla11},  6: \cite{gao20}, 7: \cite{Tarr2022}, 8: \cite{zho22}
\end{table*}

\begin{figure}\label{f9_massfunc}
	\centering{\includegraphics[width=0.99\columnwidth]{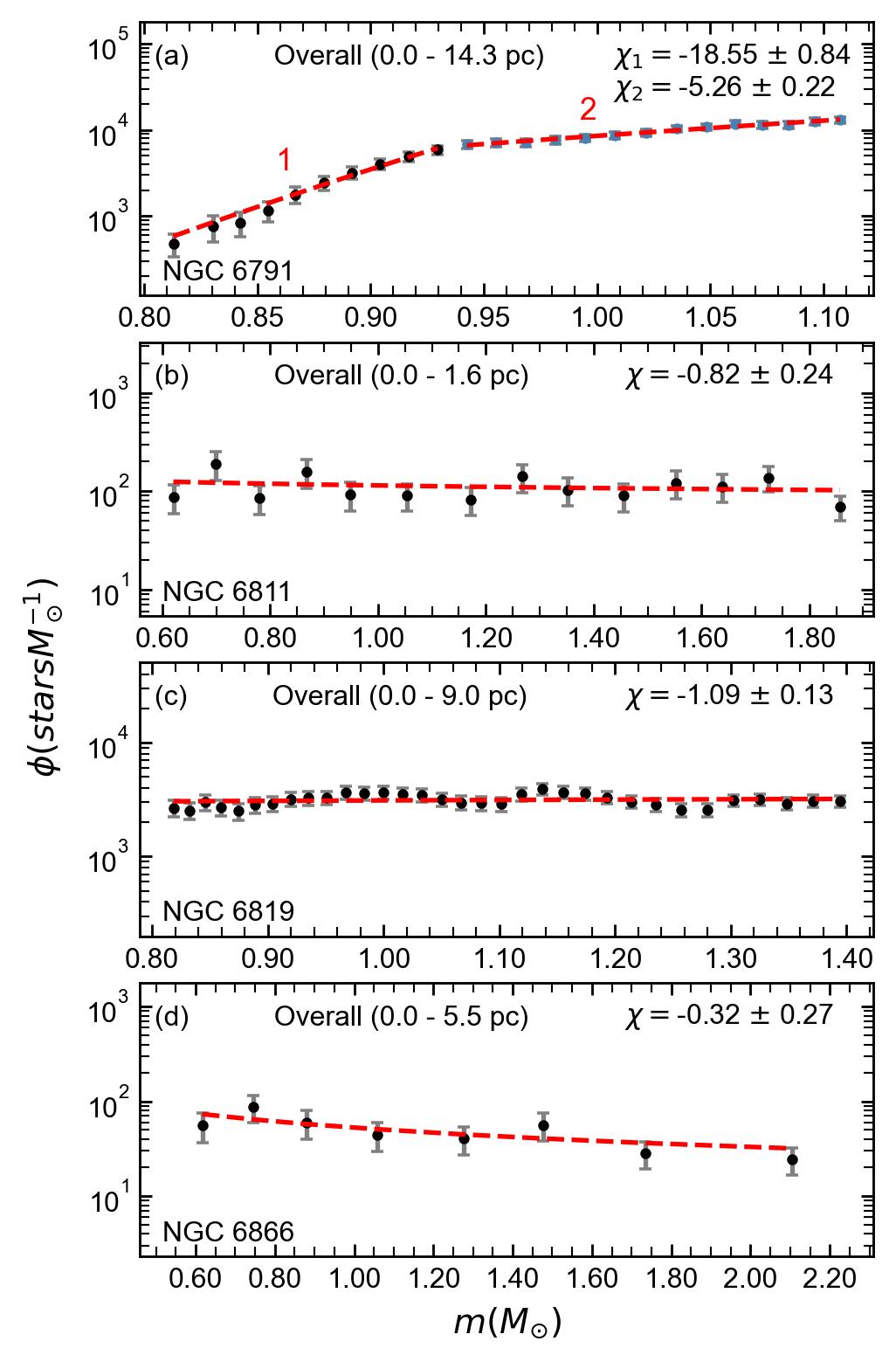}}\vspace*{-2ex}
	\caption{$\phi(m)$ versus $m~(M_{\odot})$ for the overall regions of the four OCs. 
		Here $\phi(m)= dN/dm$ (stars ~$m_\odot^{-1}$).}
\end{figure}

The mass function relation between $\phi(m)(stars ~M_\odot^{-1})$ and  m($M_\odot$) for the programme objects is presented in Fig.~\ref{f9_massfunc}, the obtained masses with the help of the B12 isochrones and \textit{fitCMD} are listed in Table~\ref{t6_massinfo}. With the exception of the last two columns, which refer to the full simulation of \textit{fitCMD} (considering stars down to $>0.08~M_\odot$), the data represent the simulation results of the actually observed cluster. 

The MF slopes are the observed ones, i.e., those computed directly from the observed CMDs. The theoretical IMF is used only for the purpose of estimating the completeness-corrected mass (by estimating the difference in the number of stars actually detected at a given magnitude with respect to the expected one). Therefore, all the other parameters are unaffected by this procedure.

The overall MF of NGC~6791 (Fig.~\ref{f9_massfunc}) shows a break followed by a flatter slope. This cluster is the most distant one in our sample, thus the mass range below $m \sim 0.93M_{\odot}$ is clearly affected by incompleteness. Therefore, we use the MF slope based on stars $m > 0.93M_{\odot}$ as representative for the overall cluster. The MFs of NGC\,6811, NGC\,6819 and NGC\,6866 show on the other hand rather flat slopes $(\chi=-0.82\pm0.24)$, $(\chi=-1.09\pm0.13)$ and $(\chi=-0.32\pm0.27)$, respectively.

\begin{figure}
	\centering{
		\includegraphics[width=0.48\columnwidth]{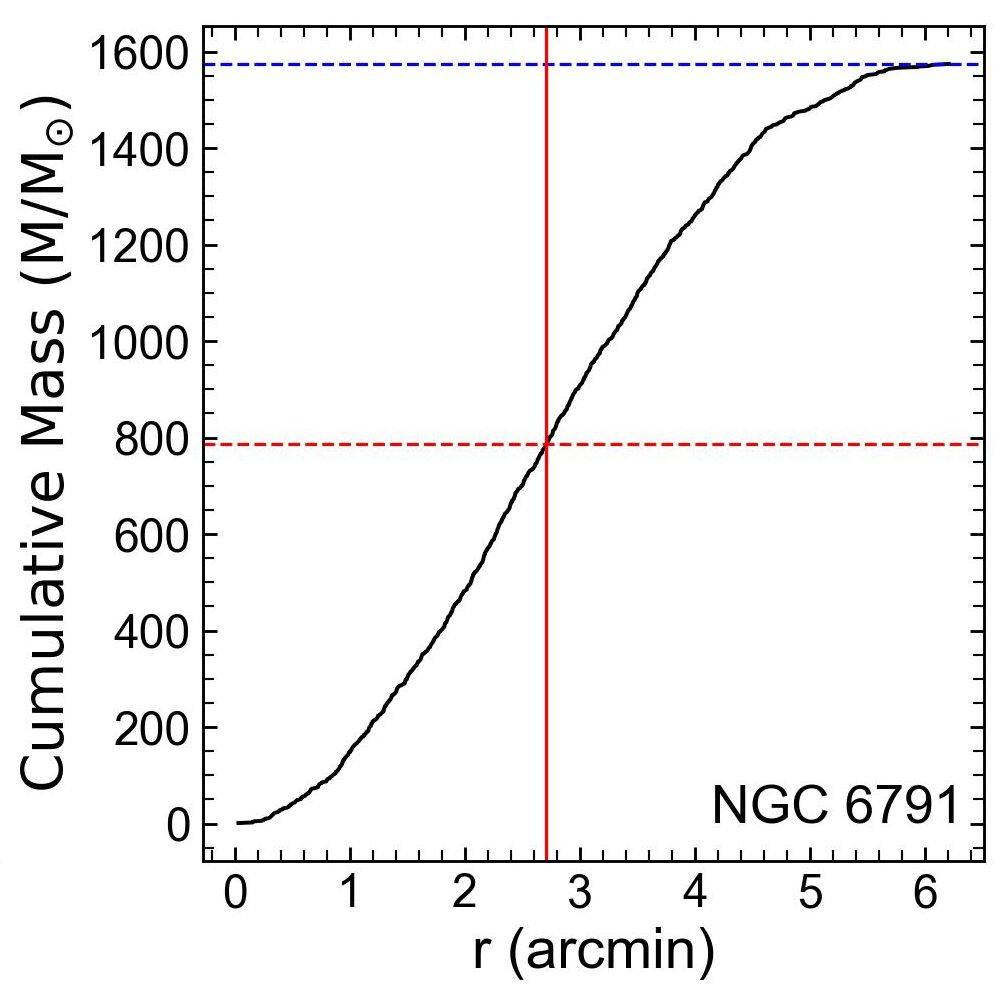}\hspace*{2ex}
		\includegraphics[width=0.47\columnwidth]{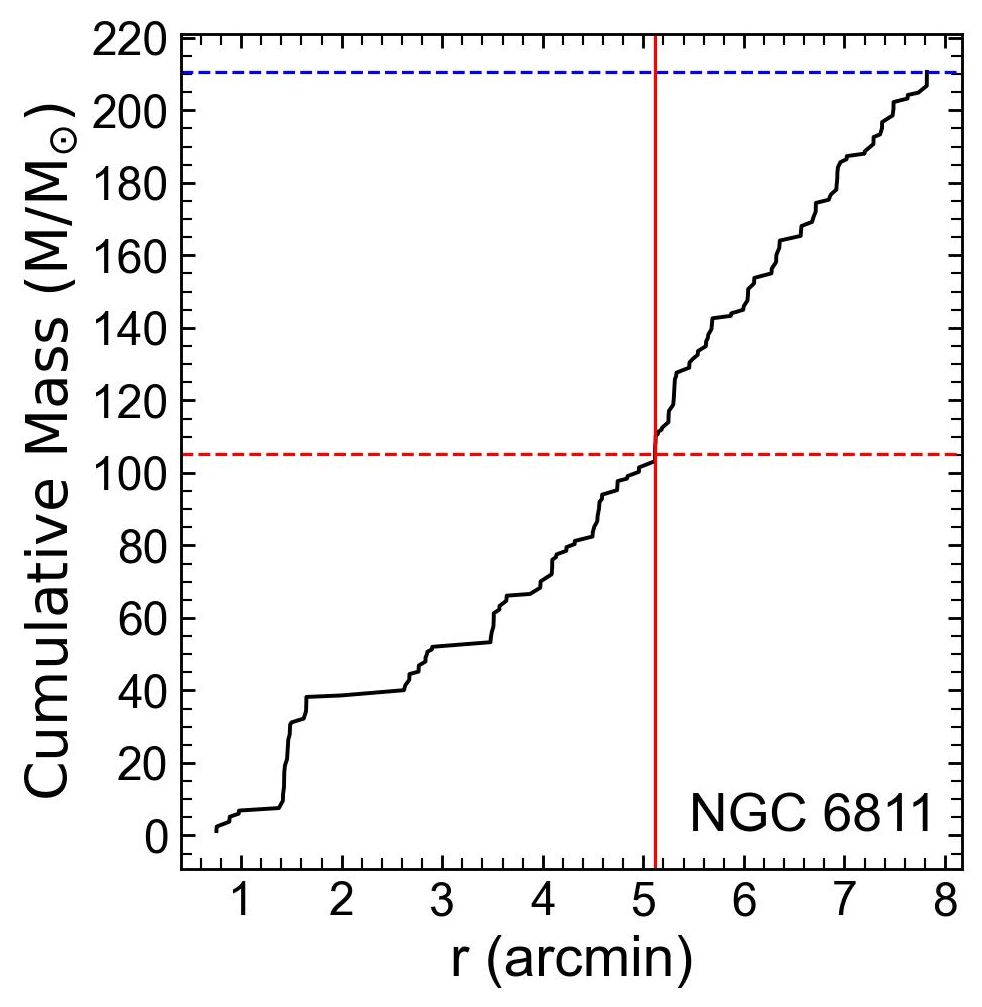}\\[2ex] 
		\includegraphics[width=0.48\columnwidth]{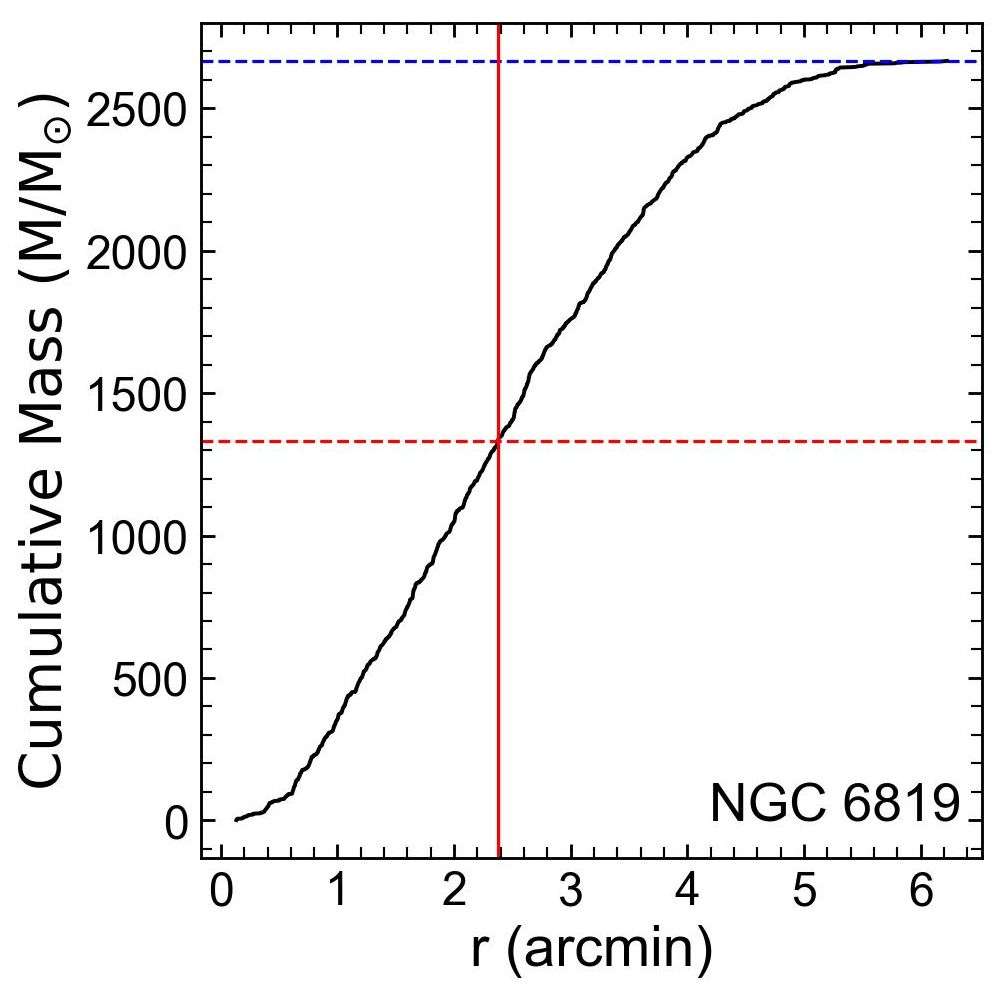}\hspace*{2ex}
		\includegraphics[width=0.47\columnwidth]{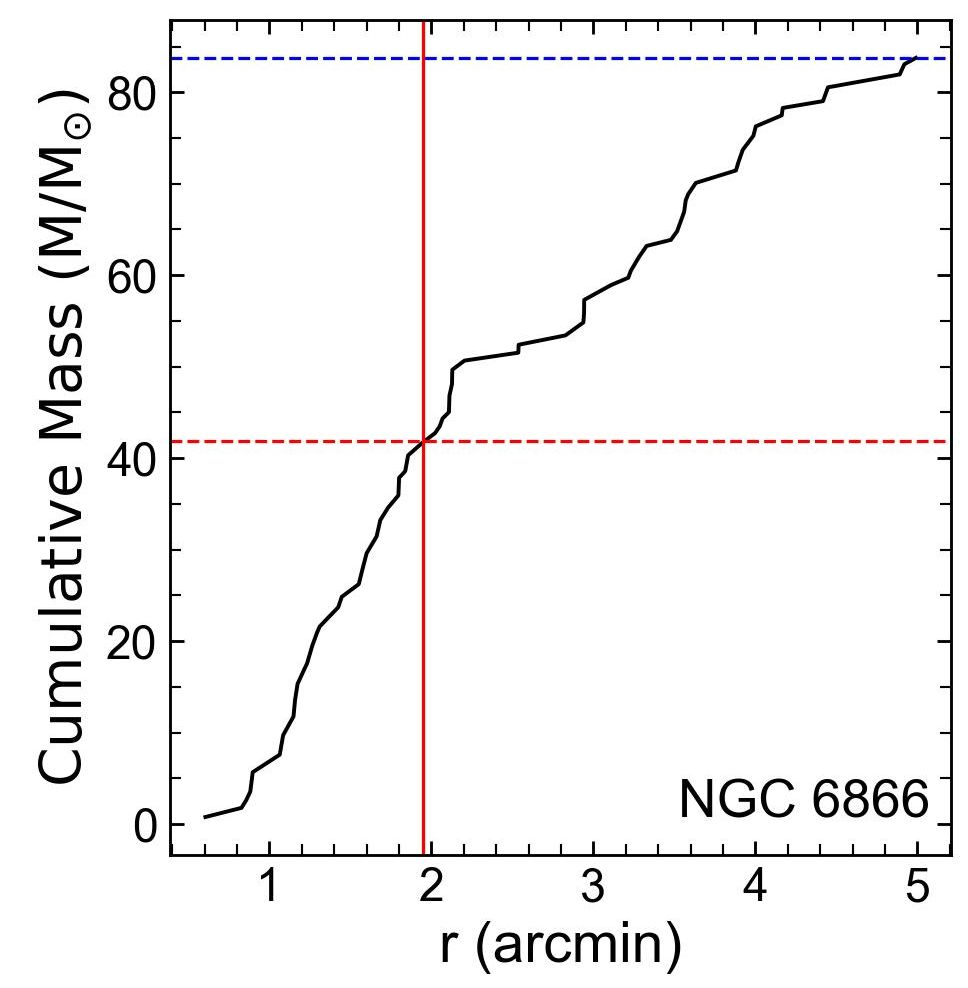}}\vspace*{-1ex}
	\caption{The cumulative mass versus radius of the four OCs. The vertical lines represent the half-mass radii correspond to the half-masses (horizontal dotted lines). The blue dotted lines show the total masses.}
	\label{f2_rhm}	
\end{figure}

\renewcommand{\tabcolsep}{3.1mm}
\renewcommand{\arraystretch}{1.3}
\begin{table*}
	\begin{center}
		\caption{Mass information for the overall regions of the four OCs.}
		\label{t6_massinfo}
		\begin{tabular}{lccAAAAll}
			\hline
			Cluster  & & Mass range & \mcl{MF slope} & \mcl{Observed mass} & \mcl{Mean Mass} & \mcl{Members}& Total mass&Members \\[-1.0ex]
			& & $M_{\odot}$ & \mcl{$\chi$}  & \mcl{$m_{obs}(M_{\odot})$} &\mcl{$(M_{\odot})$} &  \mcl{N1} &$m_{tot}(M_{\odot})$ &N2 \\
			\hline
			NGC\,6791  &1& 0.81 $-$ 0.93 &-18.55&0.84 &  298.2& 9.1  & 0.90&0.001&  333&10  & 60500 & 186362\\
			\quad\quad &2& 0.94 $-$ 1.11 & -5.26&0.22 & 1759.2&17.9  & 1.03&0.004& 1700&14  & 60500 & 186362 \\
			NGC\,6811  & & 0.62 $-$ 1.86 & -0.82&0.24 &  198.1&18.9  & 1.23&0.021&  161&14  & 2480  & 6502 \\
			NGC\,6819  & & 0.82 $-$ 1.39 & -1.09&0.13 & 2111.9&40.0  & 1.11&0.004& 1901&33  & 23100 & 68595 \\
			NGC\,6866  & & 0.62 $-$ 2.10 & -0.32&0.27 &  112.6&21.0  & 1.28&0.131&   88&12  & 556   & 1436 \\
			\hline
		\end{tabular}
	\end{center}
\end{table*}

\section{Dynamical parameters}\label{sect:6}
The relaxation time $t_{rlx}$ is obtained from the relation by \cite{Spitzer1971}
\begin{equation}
	t_{rlx}=\frac{8.9\times10^5\sqrt{N}\times{R_{h}}^{3/2}}{log(0.4N)\times \sqrt{m}}
\end{equation}
where $m$ and  $N$ refers to the mean mass and the number of the cluster members, respectively, adopted from Table~\ref{t6_massinfo}. $R_{h}$ is defined as the radius from the core that contains half the total mass of the cluster. This parameter was inferred from Fig.~\ref{f2_rhm}, showing the cumulative mass in dependence of the radius. The half-mass radii of NGC~6791, NGC~6811, NGC~6819 and NGC~6866 are 2$^{\prime}$.71, 5$^{\prime}$.41, 2$^{\prime}$.37, and  1$^{\prime}$.96, respectively. A conversion into pc is listed in Table~\ref{t7_dynamic}.

As an indicator of the dynamical evolution, the evolutionary parameter is estimated from the relation $\tau = Age/t_{rlx}$ by using the age obtained from the Gaia CMDs given in Table~\ref{t3_fitcmd}.

The disruption times of the four OCs have been determined from the equation by \citet{Binney2008}, as given in \cite{Converse2011}.
\begin{equation}
	t_{dis}=250~Myr\left(\frac{M}{300M_{\odot}}\right)^{1\slash2}\times\left(\frac{R_{h}}{2~pc}\right)^{-3\slash2}
\end{equation}
The Jacobi tidal radius $R_{J}$ and the Galactic mass $M_{G}$ inside a   Galactocentric radius $R_{GC}$ are estimated from the equations given by \cite{Kim2000},
\begin{equation}
\label{eq_jacobi}
	R_{J}=\left(\frac{M} {2M_{G}}\right)^{1/3}\times R_{GC}	
\end{equation}

\begin{equation}
	M_{G}=2\times10^{8} M_{\odot} \left(\frac{R_{GC}} {30 pc}\right)^{1.2}	
\end{equation}

where $M$ is the observed cluster mass taken from Table~\ref{t6_massinfo} and the Galactocentric distances of the clusters are given in Table~\ref{t8_overall}. All the derived dynamical parameters are provided in Table~\ref{t7_dynamic}.

\renewcommand{\tabcolsep}{4.5mm}
\renewcommand{\arraystretch}{1.2}
\begin{table*}
	\begin{center}
		\caption{The dynamical parameters of the four OCs.}
		\label{t7_dynamic}
		{\small
			\begin{tabular}{lAAAAAA}
				\hline
				Cluster  & \mcl{$M_G$}&\mcl{$R_J$}&\mcl{$R_{h}$}&\mcl{$t_{rlx}$}&\mcl{$t_{dis}$}&\mcl{$\tau$}\\
				  &\mcl{$10^{9}\,M_{\odot}$}&\mcl{pc}&\mcl{pc}&\mcl{Myr}&\mcl{Myr}&\mcl{}\\
				\hline
				NGC\,6791 &  160.27&0.11  &  13.91&0.06  &  3.44&0.07  &  81.4&2.7  &  268&9   &   88&11 \\
				NGC\,6811 &  164.80&0.06  &   6.82&0.09  &  1.49&0.03  &  11.0&0.8  &  294&20  &  109&20 \\
				NGC\,6819 &  160.31&0.08  &  14.79&0.05  &  1.64&0.10  &  26.8&1.1  &  895&38  &  108&12 \\
				NGC\,6866 &  164.67&0.01  &   5.64&0.12  &  0.71&0.05  &   3.0&0.4  &  683&77  &  330&77 \\
				\hline
			\end{tabular}
		}
	\end{center}
\end{table*}

\subsection{Indicators of Dynamical Evolution}

As an indicator for the internal dynamical evolution, the mass segregation degree (small/mild/large) provides a proxy if higher or lower mass stars are dominant in the cluster. 
In the sense, the MF slopes ($\chi$) also reflect the ratio between massive and low-mass stars. Low-mass stars are transferred from the core to the halo, and are then lost to the field. 
Mass segregation is directly related to $t_{rlx}$ and $\tau$, where $t_{rlx}$ gives the time it takes for a star to move from one end of the cluster to the other. 
The smaller $t_{rlx}$, the sooner the star can leave the cluster. Thus, with time the cluster loses stars and dynamically evolves. A high $\tau$ value of an OC implies 
an advanced dynamical evolution, showing the degree to which it has lost its low-mass stars to the field. There is a negative relationship between $t_{rlx}$ and $\tau$ -  
higher $\tau$ and lower $t_{rlx}$ values indicate advanced mass segregation.

As to external dynamical evolution, the three-parameter \cite{King1962} model describes well the outer parts of a cluster and provides the tidal radius $R_{t}$. 
Internal relaxation and the stripping of stars from the cluster by the Galactic tidal field depend on $R_{t}$. $R_{J}$ (Eq.~\ref{eq_jacobi}) 
is also the distance from the cluster center at which the external gravitation of the Galaxy has more influence on the cluster stars than the cluster itself \citep{vonHoerner1957}.  
Since the OCs are mostly in nearly circular orbits, $R_{J}$ should be close to $R_{t} = R_{RDP}$ (tidally filling).

As noted by A18, OCs are exposed to significant mass loss processes towards their final disruption in case the $R_{t}$ radii are larger than the $R_{J}$ radii. 
In the opposite case, stars within the Roche lobe are gravitationally bound to the OCs, and therefore such OCs keep their stellar contents within their $R_{J}$ radii. 
The cluster members outside the $R_{J}$ radii are more influenced by the external potential of the Galaxy. Clusters that show comparatively equal radii are in the transitional 
phase towards their final disruption stage.  

According to \cite{pia17a,pia17b}, the concentration parameter $c = \log (R_{t}/R_{core})$ is the efficiency of heating on cluster stars caused by dynamic interactions in the center of the clusters.  
This parameter is almost negatively correlated with $R_{h}/R_{t}$ and increases with age as a result of their dynamic evolution.

According to \cite{heg2003}, \cite{pia09}, A20 and A21, $R_{h}/R_{t}$ is as an indicator of the tidal influence of the Galaxy on the dynamical evolution of the OCs.  
Lower $R_{h}/R_{t}$ ratios indicate a more compact cluster, which is less subject to tidal stripping/disruption caused by Galactic gravitational forces. $R_{core}/R_{h}$ is a measure of the compactness of a cluster in its inner regions.  
Note that a small half-mass radius indicates a dense core relative to the overall size, and that clusters with low $R_{core}/R_{h}$ ratios are also less subject to tidal disruption. 

The ratio $R_{h}/R_{J}$ is a useful measure for the degree of tidal filling that an OC experiences in the tidal field of the Galaxy. In the sense it is defined as the Roche volume 
filling factor, and characterizes the impact of the tidal field - lower $R_{h}/R_{J}$ ratios indicate a weaker tidal field impact. We note that the tidal field strength itself weakens with increasing Galactocentric distance.

\section{Kinematics and Orbital Parameters}
\label{sect:7}
Based on Gaia EDR3 radial velocities ($V_R$) for the bright giants in NGC~6791 ($N= 46$), NGC~6811 ($N = 5$), NGC~6819 ($N = 121$), and NGC~6866 ($N = 2$) we derived weighted cluster averages (Table~\ref{t8_overall}). The heliocentric velocities ($U$, $V$, $W$) in the right-hand system have been obtained using the radial velocities, the median proper motion components, the cluster distances and the algorithm by \cite{joh87}. For the distance we adopt the result based on the Gaia EDR3 CMD (Table~\ref{t3_fitcmd}). These space velocities were transformed to the components $U'$, $V'$, $W'$ by correcting for the solar motion $(U, V, W)_{\odot} = (+11.10, +12.24, +7.25)$ km\,s$^{-1}$ \citep{sch10} with respect to the local standard of rest (LSR). For this we adopt $R_{\odot}=8.2\pm0.1$ kpc \citep{bg16} and $V_{LSR}$ = 239\,km\,s$^{-1}$ \citep{bru11}. The heliocentric cartesian distances ($x'$, $y'$, $z'$) in kpc and LSR-velocity components ($U'$, $V'$, $W'$) have been converted to the Galactic Rest of Frame (GSR) i.e., ($x$, $y$, $z$) and ($V_{x}$, $V_{y}$, $V_{z}$) using the equations by \cite{kep07}. The azimuthal velocity  ($V_{\Phi}$) in km\,s$^{-1}$ is estimated using 
\begin{equation*} 
	V_{\Phi} =  \frac{x V_{y} - y V_{x}}{R}
\end{equation*}
We note that $V_{\Phi}<0$ means prograde.  From the "MWPotential2014" code in the galpy-code library \footnote[1]{http://github.com/jobovy/galpy} by \cite{bov15}, peri- and apo-galactic distances $(R_{min},~R_{max})$ and the maximum height above the Galactic plane (z$_{max}$) in kpc are obtained.
MWPotential2014 is an axisymmetric Galactic potential which includes a spherical Galactic bulge, a Miyamoto-Nagai disc, and a halo with a Navarro-Frenk-White profile. For details about the parameters and properties of the Galactic components we refer to \cite{bov15}.

For the orbital eccentricity (ecc) we adopt
\begin{equation*} 
	ecc = \frac{R_{max}-R_{min}}{R_{max}+R_{min}}
\end{equation*}

The current mean Galactocentric radius $R_{m}=(R_{min}+R_{max})/2$ is also known as the guiding or mean orbital
radius. The orbits have been integrated using the obtained kinematic parameters for the ages of the four OCs within the Galactic potential. Furthermore, their orbital angular momentum components  $J_{z}$ (kpc km\,s$^{-1}$) are calculated from the equation by \cite{kep07}. All these parameters are listed in Table~\ref{t8_overall}.

\begin{figure}
	\centering{
		\includegraphics[width=0.47\columnwidth]{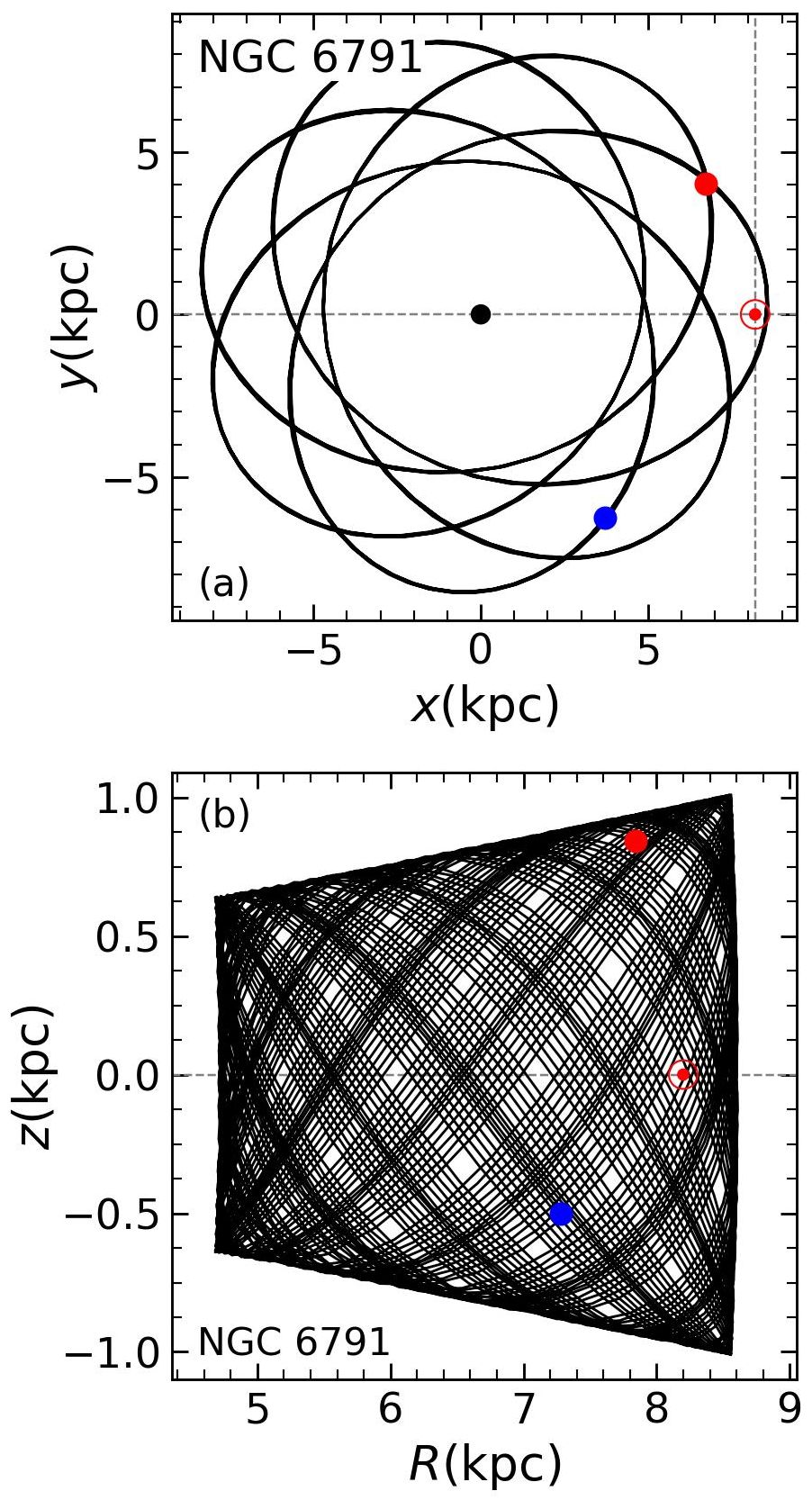}\hspace*{1ex}
		\includegraphics[width=0.47\columnwidth]{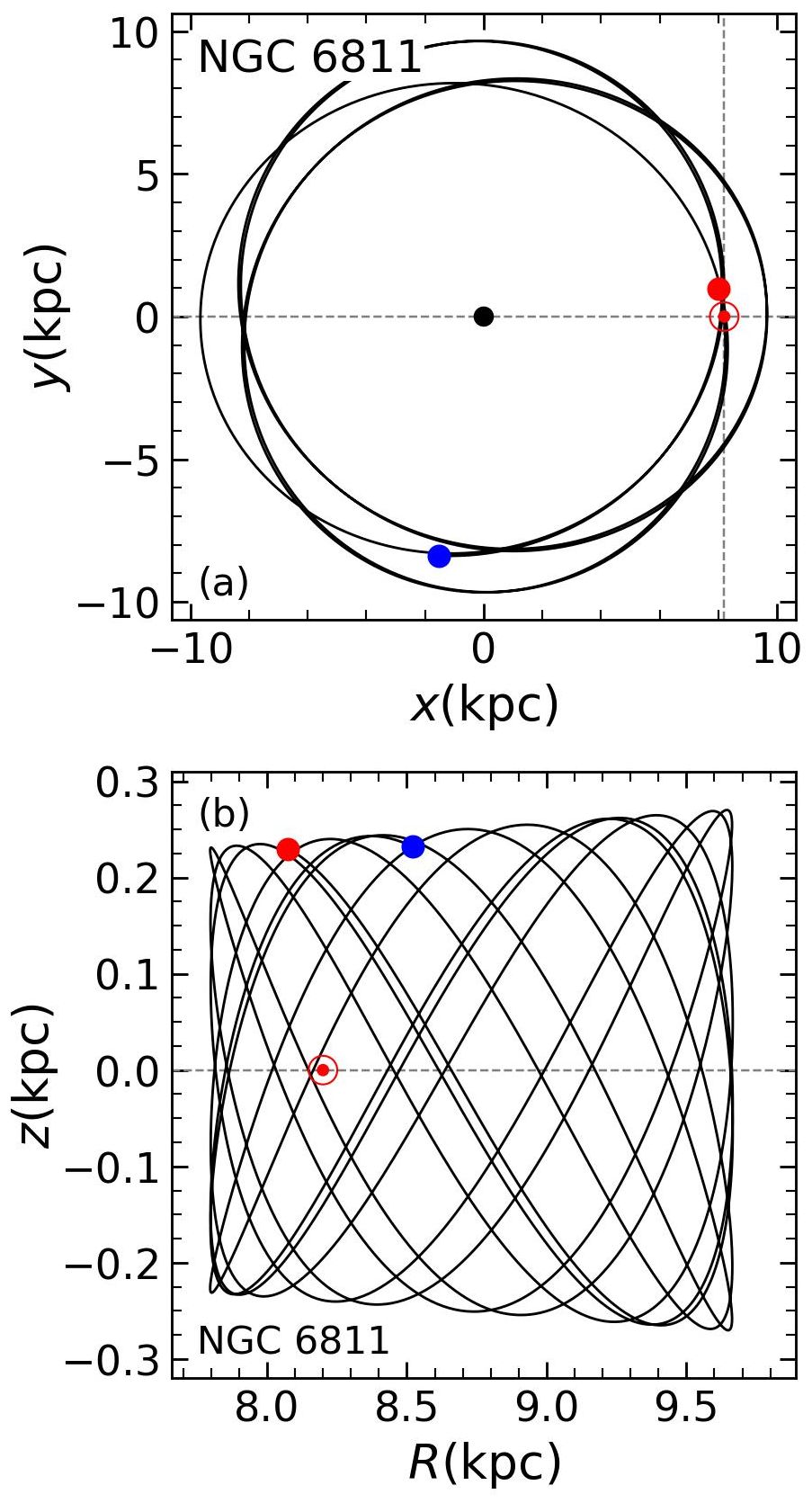}\\[1ex]
		\includegraphics[width=0.47\columnwidth]{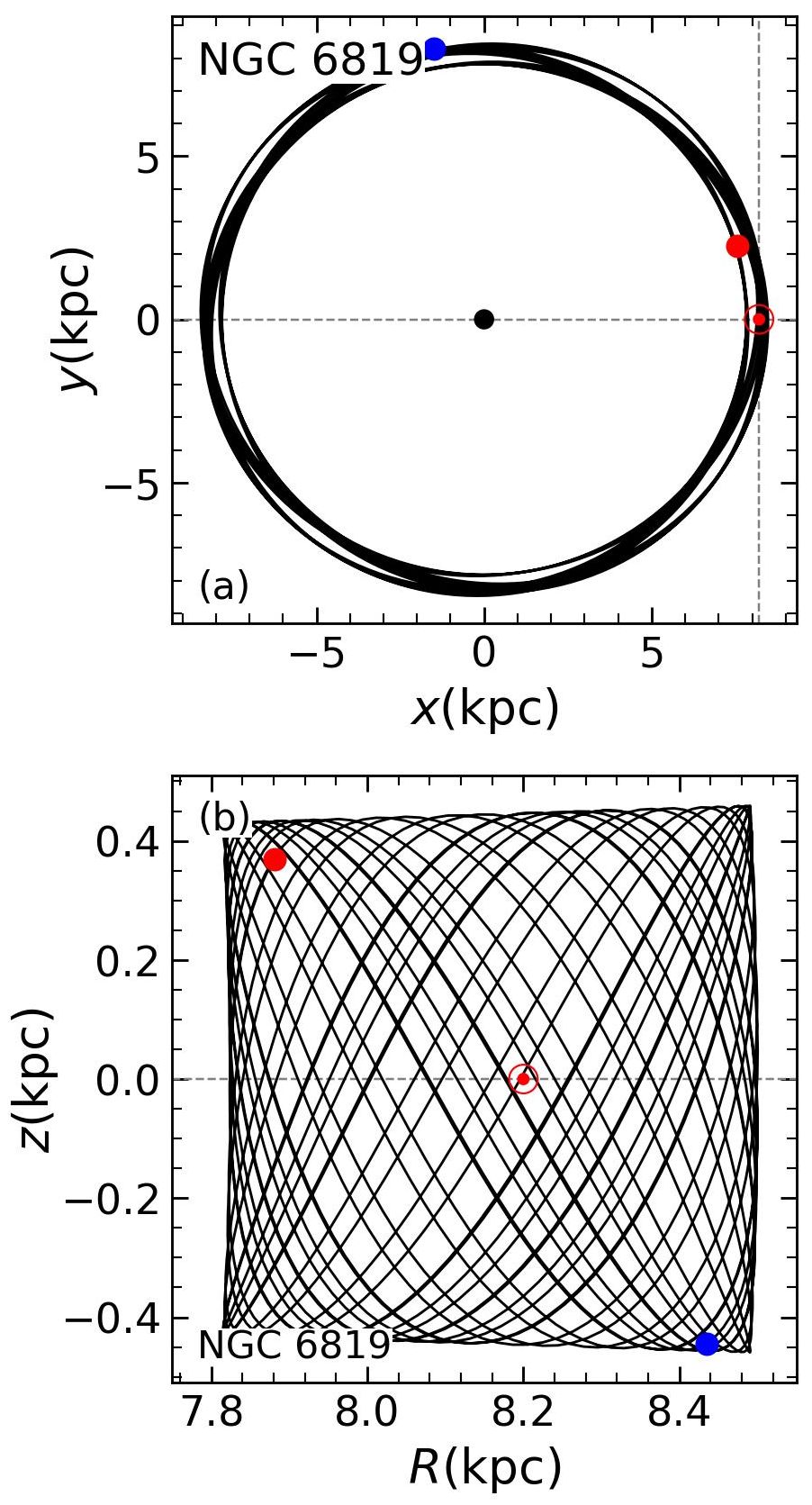}\hspace*{1ex}
		\includegraphics[width=0.47\columnwidth]{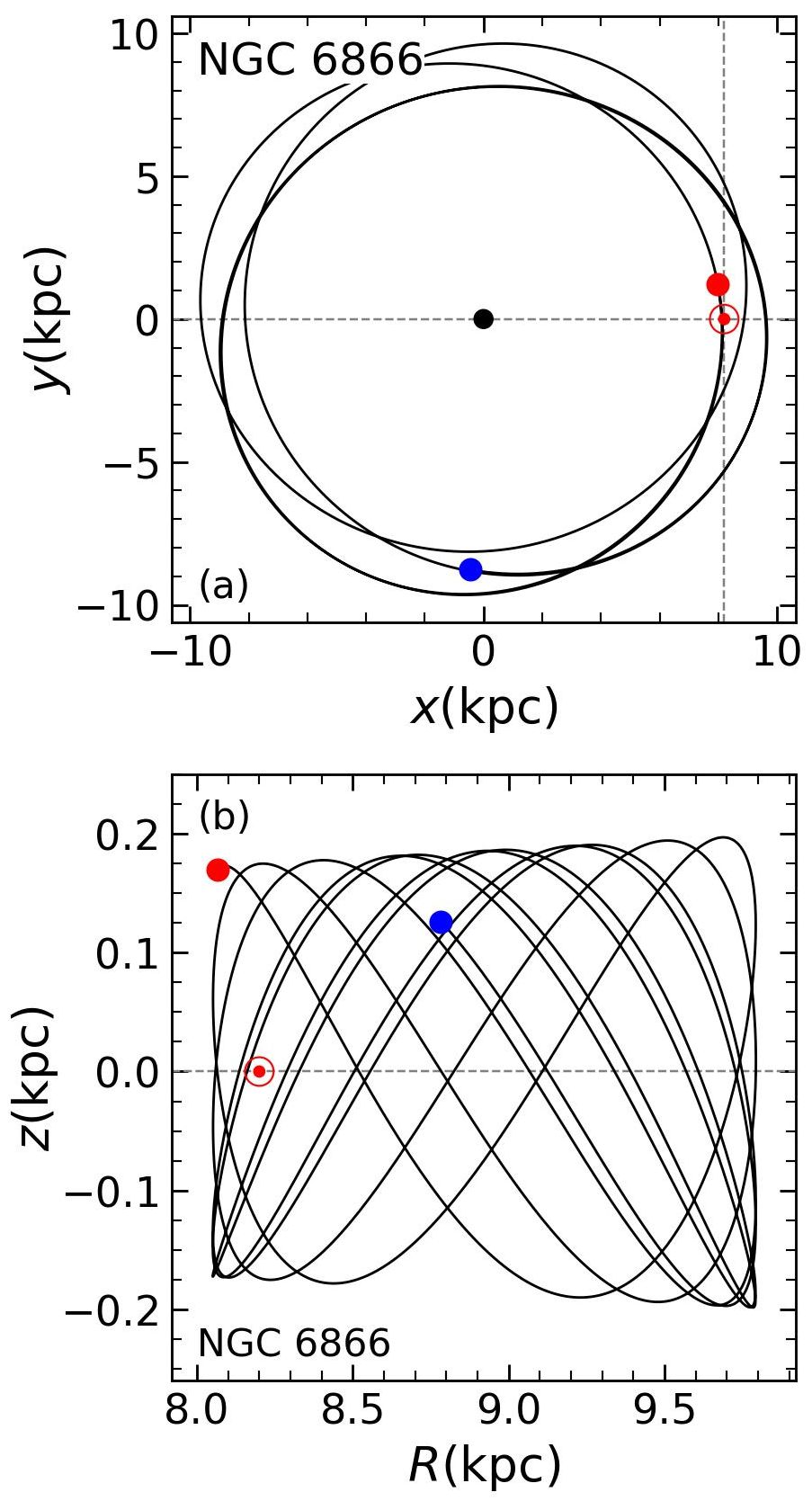}\vspace*{-1ex}
	}
	\caption {Galactic orbits of the four OCs based on the galpy-code package and "MWPotential2014". The trajectories represent the paths traveled by the OCs during their age. The filled blue/red dots show their initial/present day positions. The red circle shows the location of the Sun.}
\label{f10_orbit}
\end{figure}

The rotational velocities, the eccentricities, and the orbital angular momentum values indicate that the four OCs have Galactic thin disc properties. Figure~\ref{f10_orbit} shows the Galactic orbits of the objects. The x-y~(kpc) plane is known as projected on to the Galactic plane, whereas z-R (kpc) is the meridional plane. They follow a circular path around the Galactic center with eccentricities in the range of 0.04$-$0.29 and are orbiting near the Galactic disc. Therefore, they might be affected by the tidal forces of the disc. According to their revolution periods T~(Myr) around the Galactic center, NGC~6791 has made 31 revolutions around the Galactic center, and the remaining objects show 5 to 14 revolutions (see Table~\ref{t8_overall}). 

The orbits in the z-R plane show boxy-like type properties, thus the four OCs move in the meridional planes within the confined spaces and are oscillating along the z-axis.  The orbit of NGC~6791 is confined in the range of $\sim 7.2 < R_{GC} \leq 7.8$ kpc, therefore it is interacting with the inner region of the Galaxy. The confined spaces of the others are as follows: $\sim 8.0 < R_{GC} \leq 8.6 $ kpc for NGC~6811,  $\sim 7.82 < R_{GC} \leq 8.00$ kpc for NGC~6819, $\sim 8.0 < R_{GC} \leq 8.9$ kpc for NGC~6866. 

Their initial and present day positions in our Galaxy are shown in Fig.~\ref{f10_orbit} with filled blue and red dots, respectively. For this, the times were adopted as zero (initial) and the cluster age (present day). Their closest approaches to the current solar position are determined as $(d~(kpc),~t (Gyr))$ = (0.46, 4.86) for NGC~6791), (0.99, 1.19) for NGC~6811, (0.55, 2.74) for NGC~6819, (0.97, 0.97) for NGC~6866, respectively.

The derived radial / rotational velocities and the orbital parameters of the objects in Table~\ref{t8_overall} are in reasonable agreement with literature values \citep{car22,Tarr2021}. We note that this paper and \cite{Tarr2021} use the same Galactic components for the Galactic potential, but \cite{car22} also consider the bar and spiral arms in addition to the three main components.

\newcommand{\hpm}{$\,\pm\,$}
\renewcommand{\tabcolsep}{1.5mm}
\renewcommand{\arraystretch}{1.2}
\newcommand{\hgray}{\rowcolor{DarkGray!40}}
\begin{table}
	\begin{center}
		\caption{Velocities and orbital parameters of the objects.}
		\label{t8_overall}
		\begin{tabular}{lccccc}
			\hline
			& NGC\,6791 & NGC\,6811 & NGC\,6819 & NGC\,6866&Ref.\\
			\hline
			$V_R$         & -47.20\hpm0.01 & 5.40\hpm8.1 & 2.50\hpm0.01 & 9.5\hpm26.0&1\\ 
			& -47.75\hpm0.17 & 7.17\hpm0.13 & 2.80\hpm0.14 &12.44\hpm0.34&2\\ 
			& -46.49\hpm0.53 & 6.92\hpm0.16 & 2.96\hpm0.26& --&3\\ \hgray
			$U$           &    28.33  &    44.87  &    52.01  &    34.86  &1 \\
			$V$           &   -57.13  &    -2.13  &   -13.56  &     4.49  &1\\ \hgray
			$W$           &   -21.49  &    -3.50  &     8.15  &    -9.82  &1 \\
			$V_{\Phi}$    &  -185.80  &  -253.92  &  -245.56  &  -259.70  &1\\
			&  -189.04  &  -256.75  &   -246.78 &  -263.77  &2\\ 
			&   189.50  &   256.70  &   247.00  &     --    &3\\ \hgray
			ecc           &     0.29  &     0.11  &     0.04  &     0.10  &1 \\ \hgray
			&    0.28   &    0.12   &    0.05   &     0.11  &2 \\ \hgray
			&    0.35   &    0.17   &    0.13   &      --   &3 \\
			R$_{min}$     &     4.71  &     7.80  &     7.82  &     8.05  &1 \\
			&     4.83  &     7.91  &     7.94  &     8.19  &2 \\
			&     4.12  &     6.69  &     6.44  &     --    &3 \\ \hgray
			R$_{max}$     &     8.62  &     9.67  &     8.50  &     9.79  &1 \\ \hgray
			&     8.62  &     10.02 &     8.70  &     10.23 &2 \\ \hgray
			&     8.42  &     9.50  &     8.42  &     --    &3 \\
			R$_m$         &     6.67  &     8.74  &     8.16  &     8.92  &1 \\ \hgray
			z$_{max}$     &     1.01  &     0.27  &     0.46  &     0.20  &1 \\ \hgray
			&      0.95 &    0.31   &     0.52  &     0.23  &2 \\ \hgray
			&     1.26  &    0.30   &     0.55  &      --   &3 \\
			R$_{in}$      &     7.28  &     8.52  &     8.43  &     8.78  &1 \\ \hgray
			R$_{GC}$      &     7.84  &     8.07  &     7.88  &     8.07  &1 \\ \hgray
			&    7.92  &     8.20  &     8.02  &     8.20  &2 \\ \hgray
			&     7.94  &     8.20  &     8.03  &      --   &3 \\
			$J_z$         & -1465.32  & -2050.79  & -1937.37  & -2095.60  &1 \\ \hgray
			$T$           &   235     &   213     &   204     &   214     &1 \\
			$N_{Rev}$     &    30.7   &     5.6   &    14.2   &     4.7   &1 \\
			\hline
		\end{tabular}
	\end{center}
	\flushleft
	Notes: The weighted average radial velocities, ($V_R$) km\,s$^{-1}$, space velocity components and rotational velocity  ($U$, $V$, $W$, $V_{\Phi}$) km\,s$^{-1}$, eccentricity (ecc), peri- and apogalactic distances, initial and present day distances (R$_{max}$, R$_{min}$, R$_{m}$, z$_{max}$, R$_{in}$, R$_{GC}$) (kpc). ($J_{z}$) (kpc km s$^{-1}$) and $T(Myr)$ are the orbital angular momentum and the time of one revolution around the Galactic center, respectively. $N_{Rev}$ is the number of the revolutions over the age of the cluster. References: 1: This paper, 2: \cite{Tarr2021}, 3: \cite{car22}.
\end{table}

\section{Discussion and Conclusion}
\subsection{Comparison of astrophysical parameters}
\label{sect_comp-para}
In Sect.~\ref{sect:4} we derived the astrophysical parameters of the clusters based on various colours and methods (see also Tables \ref{t3_fitcmd} and \ref{t4_difgrid}. A comparison of our findings for $E(B-V)$, $d~(kpc)$, and $Age$~(Gyr) of the four OCs to literature is given in Table~\ref{t9_litcompbig} and Fig.~\ref{f12_litcomp}. For the Fig. \ref{f12_litcomp}(a) the individual colour excesses are converted to $E(B-V)$ using the reddening ratios given by \citet{bes98} and $E(B-V)=0.775E(G_{BP}-G_{RP})$ by \cite{brag18}.

Our individual results mostly lie within the error range based on the $(B-V)$ colour.  However, e.g. the reddening obtained for NGC~6791 from the $(R-I)$ and $(V-I)$ colours is somewhat lower. These colours also provide closer distances and a somewhat older age of NGC~6819.   
	
\begin{figure}
	\centering{\includegraphics[width=0.7\columnwidth]{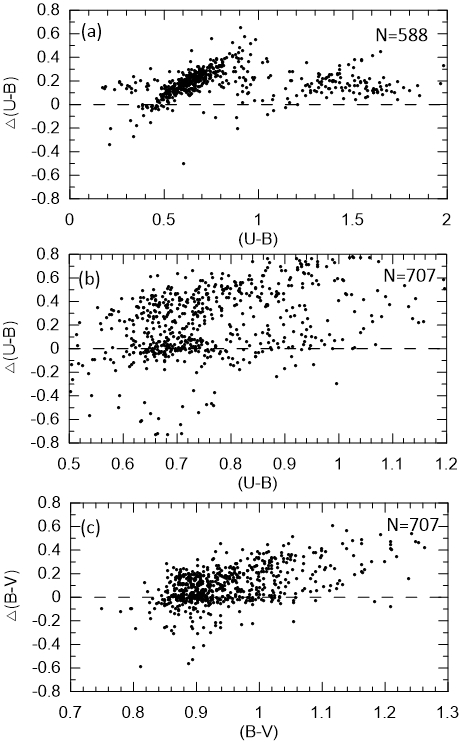}}
	\caption{The comparison of our CCD $UB$ / CCD $UBVI$ data for NGC~6791 to literature data: panel~(a) \citep{car13}, and panels~(b)-(c) \citep{kal95}. The difference $\Delta$ means our data minus literature.}
	\label{f11_ubcomp}
\end{figure}

\begin{figure*}
	\centering{\includegraphics[width=1.3\columnwidth]{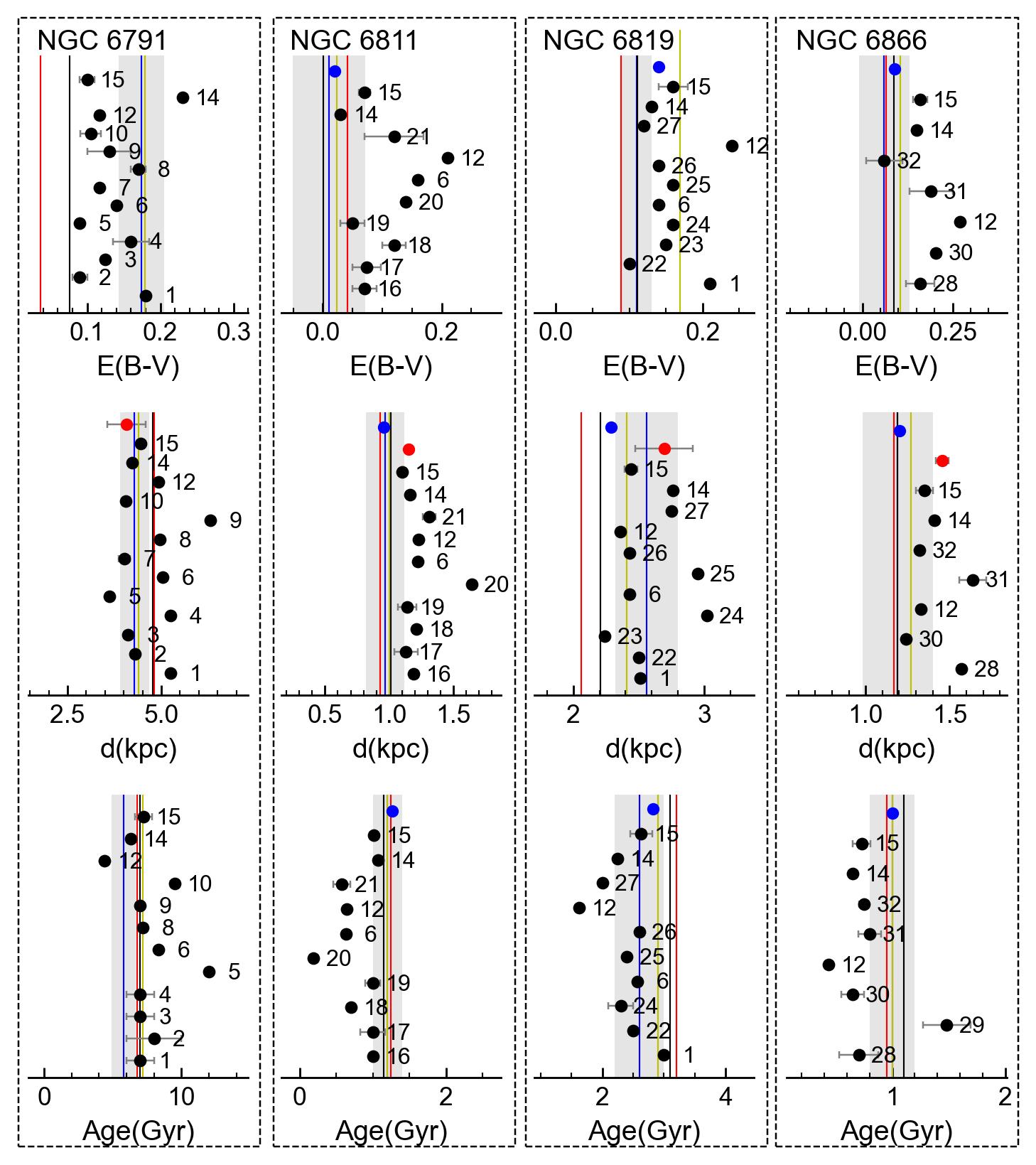}}
	\caption{Comparison of the astrophysical parameters  obtained in this work with literature (Table~\ref{t9_litcompbig}). The vertical line represent our findings based on the $(B-V)$ (blue),  $(R-I)$ (red),  $(V-I)$ (black), and $(G_{BP}-G_{RP})$ (yellow) colours, respectively. The shaded regions show the uncertainties of the parameters for $(B-V)$. The filled red and blue dots represent the distances from Gaia EDR3 parallaxes, and the astrophysical parameters of the DG method, respectively. In panel (a) the colour excess in the individual colours are converted to $E(B-V)$ according to Sect. \ref{sect_comp-para}}
\label{f12_litcomp}	
\end{figure*}

A broad age range (4.4--12~Gyr) can be found in the literature in particular for NGC~6791, though some results adopt solar metallicity isochrones for this metal-rich object. Also our results show some deviations, the results based on the $V/(U-B)$ and $V/(B-V)$ CMDs are about 1.0-1.4 Gyr younger than using the other colours. These two CMDs also provide a younger age compared to e.g. \cite{car13} and \cite{kal95} who indicate an age of 7.0 and 7.2\,Gyr from CCD~$UB$ and CCD~$UBV$ data. On the other hand, the results by these authors agree well with our finding using Gaia photometry. To investigate the age difference, Fig.~\ref{f11_ubcomp} compares our photometry for NGC~6791 to the CCD~$UB$ photometry by \cite{car13} and CCD~$UBV$ photometry by \cite{kal95}. The mean differences are up to $\Delta (U-B)=+0.80$ and $\Delta (B-V) = 0.50$, respectively. The $(U-B)$ colours by \cite{car13} and \cite{kal95} are systematically bluer than our $(U-B)$ and $(B-V)$ colours.  However, the agreement between the reddening results based on our $(B-V)$ and the Gaia colour gives a hint for a correct $(B-V)$ scale, also an additional comparison with the homogeneous photometry by \cite{stet03}, which gives $\Delta (B-V) = 0.006 \pm 0.097$ over the complete magnitude range. We are unfortunately unable to clarify the difference in $(U-B)$, a colour for which a standard transformation is challenging for CCDs anyway \citep[see e.g.][]{sung00}.

Our astrophysical parameters based on Gaia EDR3 are within the uncertainties almost compatible with the ones by \cite{cantat2020} (reference 14) or \cite{dia18} (reference 15 in Table~\ref{t9_litcompbig} and Fig.~\ref{f12_litcomp}), except e.g. for the reddening of NGC 6791. Our distances from the Gaia EDR3 parallaxes also lie in the given error range based on the $(B-V)$ colour.
Figure~\ref{f12_litcomp} indicates that some literature values present an agglomeration close to our results, but there are certainly also discrepancies that might stem from the usage of different membership techniques, isochrone sets, metal abundances, reddening determination and various photometric data, as discussed e.g. by \cite{moi10}. However, even homogeneous methods do not necessarily produce agreeing results for close OCs, mean intrinsic errors of about 0.2\,dex for the age or 0.35\, mag for the distance modulus are quite common \citep[see e.g.][]{net15}. In this context Gaia parallaxes might provide reasonable cluster distances for closer objects, whereas isochrone fits are favourable for more distant clusters \citep{mon20}.

As already noted in Sect.~\ref{sect:4}, the results based on the DG method are also consistent with the ones obtained by \textit{fitCMD} and agree well with available spectroscopic metallicities.

\renewcommand{\tabcolsep}{1.2mm}
\renewcommand{\arraystretch}{1.0}
\begin{table*}
	\caption{Literature comparison of the astrophysical parameters.}
	\label{t9_litcompbig}
	\resizebox{1.0\textwidth}{!}{ 
		\begin{tabular}{lllllllll}
			\hline
			E(B-V)  & $(V-M_{V})_{0}$& d~(kpc)  & Z  & [Fe/H] & Age~(Gyr)  & Isochrone  & Photometry    & Reference \\
			\hline
			NGC 6791       &   &  &   &   &    &    &        & \\
			\hline
			0.174$\pm$0.031 & 13.16$\pm$0.20 & 4.29$\pm$0.39& 0.030 & 0.30           & 5.80$\pm$0.90   & Bressan et al. (2012)  & CCD UBVRI & This paper \\
			0.18      & 13.60    & 5.25     & 0.04  & --  & 7.00$\pm$1.00  & Bertelli (2008)  & CCD UB  & 1 \\
			0.09$\pm$0.01    & 13.07$\pm$0.05      & 4.30    & 0.046  & 0.39$\pm$0.05              & 8.00$\pm$2.00 & Girardi et al.(2000)  & CCD BVI  & 2 \\
			0.125  & 13.07  & 4.11  & 0.04  & 0.40 & 7.00$\pm$1.00 &Bertelli et al.(2008)  & CCD UB  & 3   \\
			0.16$\pm$0.025 & 13.60$\pm$0.15 & 5.25         & --    & 0.45$\pm$0.04  &
			7.00$\pm$1.00        & Girardi et al. (2002)   & ubyCa$H_beta$               & 4 \\
			0.09           & 12.79          & 3.61         & 0.035 & 0.30       & 12.00             & VandenBerg (2002)       & CCD BVI         & 5\\
			0.14           & --         & 5.04         & --    & 0.38$\pm$0.01  & 8.32           & Marigo et al. (2008)            & 2MASS~JH$K_{s}$             & 6\\
			0.117          & 13.02$\pm$0.08 & 4.02$\pm$0.15& --    & --             & --             & Red Clump         & SDSS DR8/2MASS          & 7 \\
			0.17$\pm$0.01  & 13.48   & 4.97     & 0.039 &     & 7.20    & VandenBerg and Poll (1989)   & CCD UBVI       & 8\\
			0.13$\pm$0.03  & 14.00$\pm$0.02 & 6.31$\pm$0.06& 0.02  & --   & 7.00 & Ciardullo and Demarque (1977)   & Photoelectric UBV      & 9\\
			0.105$\pm$0.014& 13.04$\pm$0.08 & 4.06$\pm$0.02&   & 0.42$\pm$0.07  & 9.50$\pm$0.30 & Delahaye and Pinsonneault (2015)& BVI$_{c}$-2MASS~JH$K_{s}$         & 10 \\
			--             & --         & --           & --    & 0.32$\pm$0.02  & --             & --                              & Spectroscopy              & 11\\
			0.117     & 13.50     & 4.93    & solar & solar    & 4.42     & Girardi et al. (2002)           & 2MASS JH$K_{s}$     & 12 \\
			--             & --             & --           & 0.030 & 0.35$\pm$004           & --             & --                              & Spectroscopy          & 13 \\
			0.23(Av=0.70)  & 13.13          & 4.23         & solar & solar          & 6.31           & Bressan et al. (2012)           & Gaia DR2                    & 14 \\
			0.10$\pm$0.01(Av=0.313$\pm$0.028)  & --   & 4.46$\pm$0.10  & -- & 0.399$\pm$0.024      & 7.23$\pm$0.62    & Bressan et al. (2012)           & Gaia DR2     & 15 \\[1.5ex]
			\hline
			NGC 6811   &    &    &     &     &    &      &      & \\
			\hline
			0.010$\pm$0.061&  9.93$\pm$0.33 & 0.97$\pm$0.15& 0.014 & --0.04   & 1.20$\pm$0.20  & Bressan et al. (2012)        & CCD UBVRI          & This paper \\
			0.07$\pm$0.02  & 10.37$\pm$0.03 & 1.19$\pm$0.02& 0.0147& 0.04    & 1.00$\pm$0.05  & Bressan et al. (2012)            & BVIc                        & 16 \\
			0.074$\pm$0.024& 10.26$\pm$0.18 & 1.13$\pm$0.09& 0.012 & --0.19         & 1.00$\pm$0.17  & Yale-Yonsei/Padova              & CCD UBVRI                   & 17 \\
			0.12$\pm$0.02  & 10.42$\pm$0.03 & 1.21$\pm$0.02& --    & --       & 0.70      & Castellani et al. (1992)        & photoelectric UBVRI         & 18 \\
			0.05$\pm$0.02  & 10.29$\pm$0.14 & 1.14$\pm$0.07& 0.012 & 0.04$\pm$0.01  & 1.00$\pm$0.10  & Bressan et al. (2013)           & BV of Janes et al. (2013)     & 19 \\
			0.14           & --             & 1.64         & --    & --             & 0.19           & Girardi et al. (2003)           & uvby                        & 20 \\
			0.16           & --             & 1.22         & 0.025 & --0.02         & 0.63           & Marigo et al. (2008)            & 2MASS~JH$K_{s}$              & 6 \\
			& --             & --           & 0.014    & --0.05$\pm$0.02         & --             & --                              & Spectroscopy                 & 13 \\
			0.21           & 10.51          & 1.23         & solar & solar     & 0.64           & Girardi et al. (2002)           & 2MASS~JH$K_{s}$             & 12 \\
			0.12$\pm$0.05  & 10.59$\pm$0.09 & 1.31$\pm$0.05& 0.019 & --             & 0.58$\pm$0.12& Girardi et al. (2000)           & CCD BV           & 21 \\
			0.03 (Av= 0.09)& 10.33          & 1.16         & solar & solar      & 1.07           & Bressan et al. (2012)           & Gaia DR2                    & 14 \\
			0.07$\pm$0.01(Av=0.213$\pm$0.033)  & --   & 1.10$\pm$0.01 & -- & 0.032$\pm$0.015  & 1.01$\pm$0.05    & Bressan et al. (2012)           & Gaia DR2     & 15 \\[1.5ex]
			\hline
			NGC 6819       &   &    &       &    &         &       &     & \\
			\hline
			0.11$\pm$0.02  & 12.04$\pm$0.33 & 2.56$\pm$0.38& 0.017 & 0.05  & 2.60$\pm$0.40 & Bressan et al. (2012)     & CCD UBVRI           & This paper \\
			0.21           & 12.00        & 2.51         & 0.04  & --    & 3.00 & Bertelli et al. (2008)                 & CCD UB                      & 1 \\
			0.10           & 12.30$\pm$0.12 & 2.50          & --    & --             & 2.50            & Hyades main sequence            & CCD BVR                     & 22 \\
			0.15           & 11.75$\pm$0.09 & 2.24         & --    & --             & --             & --                              & --                          & 23 \\
			0.16$\pm$0.007 & 12.40$\pm$0.12 & 3.02     & --    & --0.06$\pm$0.04& 2.30$\pm$0.20   & Demarque et al. (2004)          & uvby                        & 24 \\
			0.14           & --        & 2.43     & -- & 0.07$\pm$0.01 & 2.57  & Marigo et al. (2008)            & 2MASS~JH$K_{s}$                & 6\\
			0.16           & 12.35          & 2.95     & --    & 0.00/-0.10   & 2.40    & VandenBerg et al. (1998)        & BV                          & 25 \\
			0.14           & 11.93$\pm$0.10 & 2.43    & --    & 0.09      & 2.60       & Dotter et al. (2008)            & CCD VI                      & 26 \\
			& --             & --           & 0.014 & 0.05$\pm$0.03      & --             & --                              & Spectroscopy                 & 13 \\
			0.24           & 11.94          & 2.36         & solar & solar   & 1.62           & Girardi et al. (2002)           & 2MASS~JH$K_{s}$                 & 12 \\
			0.12           & 12.20           & 2.75         & 0.02  & 0.09           & 2.00    & Bressan et al. (1993)           & CCD VI                      & 27\\
			0.13 (Av= 0.40) & 12.21          & 2.76         & solar & solar          & 2.24           & Bressan et al. (2012)           & Gaia DR2      & 14 \\
			0.16$\pm$0.02(Av=0.487$\pm$0.051)  & --   & 2.44$\pm$0.05 & -- & 0.093$\pm$0.006& 2.63$\pm$0.18   & Bressan et al. (2012)           & Gaia DR2     & 15\\[1.5ex]
			\hline
			NGC 6866       &     &    &       &     &    &    &  & \\
			\hline
			0.06$\pm$0.07  & 10.38$\pm$0.38 & 1.19$\pm$0.21& 0.016 & 0.02 & 1.00$\pm$0.20 & Bressan et al. (2012)           & CCD UBVRI              & This paper \\
			0.16$\pm$0.04  & 10.98$\pm$0.24 & 1.57$\pm$0.02& 0.014 & --       & 0.70$\pm$0.17 & Bressan et al. (2012)           & CCD UBVRI                & 28\\
			& --             & --           & --    & 0.016          & 1.48$\pm$0.21  & Paxton et al. (2013)            & KIC 8263801 RG star& 29 \\
			0.204$\pm$0.002& 10.47$\pm$0.02 & 1.24$\pm$0.01& solar & solar          & 0.65$\pm$0.10  & Bertelli et al. (2008)          & Kepler Input Catalogue      & 30 \\
			0.27           & 10.70           & 1.33         & solar & solar   & 0.44           & Girardi et al. (2002)           & 2MASS~JH$K_{s}$                & 12 \\
			& --             & --           & 0.016 & 0.01$\pm$0.01           & --             & --                              & Spectroscopy                 & 13 \\
			0.19$\pm$0.06  & 11.08$\pm$0.11 & 1.64$\pm$0.08& solar & solar          & 0.80$\pm$0.10  & Girardi et al. (2002)       & 2MASS~JH$K_{s}$       & 31 \\
			0.06$\pm$0.05  & 10.61$\pm$0.02 & 1.32$\pm$0.01& 0.015 & --0.10         & 0.75$\pm$0.04  & Marigo et al. (2008)            & CCD UBVRI      & 32 \\
			0.15 (Av= 0.48)& 10.74          & 1.41         & solar & solar          & 0.65           & Bressan et al. (2012)           & Gaia DR2        & 14 \\
			0.16$\pm$0.02(Av=0.498$\pm$0.074)  & --   & 1.35$\pm$0.05 & -- & 0.047$\pm$0.018& 0.73$\pm$0.08  & Bressan et al. (2012)       & Gaia DR2     & 15 \\
			\hline
		\end{tabular}}
\flushleft		
Notes: Our results are those based on $(B-V)$. The colour excesses of different photometric systems were converted to $E(B-V)$.
	\\ [1.7ex]
	\renewcommand{\tabcolsep}{1mm}
	\renewcommand{\arraystretch}{0.7}
	\scriptsize
	\begin{tabular}{@{}r l m{3mm} r l m{3mm} r l}
		1 & Carraro, G. et al. 2013, AJ, 146,128          & & 12 & Kharchenko, N. V. et al. 2013, A\&A, 558, 53          & & 23 & Abedigamba, O.P. et al.  2016, NewAst, 46, 90 \\
		2 & Carraro, G. et al. 2006, AJ, 643, 1151        & & 13 & Donor, J. et al. 2020, AJ, 159, 199                    & & 24 & Anthony-Twarog, B. et al. 2014, AJ, 148, 51  \\
		3 & Buzzoni, A. et al. 2012, ApJ, 749, 35         & & 14 & Cantat-Gaudin, T. et al.  2020, A\&A, 640, 1          & & 25 & Rosvick, J.M. and Vandenberg, D.A., 1998, AJ, 115, 1516  \\
		4 & Anthony-Twarog, B. et al. 2007, AJ,133, 1585  & & 15 & Dias et al. 2021, MNRAS, 504, 356                     & & 26 & Yang, S.C. et al. 2013, ApJ, 762, 1  \\
		5 & Stetson, P. et al. 2003, PASP, 115,3          & & 16 & Sandquist,E.L. et al. 2016, ApJ, 831, 11              & & 27 & Bragaglia, A. and Tosi, M. 2006,  AJ, 131, 1544,  \\
		6 & Frinchaboy, P.M. et al. 2013, ApJL, 777, 1    & & 17 & Janes, K.J. et al. 2013, AJ, 145, 7,14                & & 28 & Janes, K. J.  et al. 2014, AJ, 147, 139  \\
		7 & Gao, Xin-hua et al. 2012, ChA\&A, 36,1        & & 18 & Glushkova, E.V. et al. 1999, AstL, 25, 86,15          & & 29 & Tang, Y. et al. 2018, ApJ, 866, 59,  \\
		8 & Kaluzny, J. et al. 1995, A\&AS, 114,1         & & 19 & Molenda-Zakowicz, J. et al. 2014, MNRAS, 445, 2446  & & 30 & Balona, L.A. et al.  2013 MNRAS, 429, 1466  \\
		9 & Harris, W. et al. 1981, AJ, 86, 1332          & & 20 & Pena, J.H. et al. 2011, RMxAA, 47, 309                & & 31 & G\"une\c{s}, O. et al. 2012,  NewAstr, 17,720  \\
		10 & An, D. et al. 2015, ApJ, 811,46               & & 21 & Luo, Y.P. et al. 2009, NewAst, 14, 584,               & & 32 &Akkaya, \.I et al. 2015, NewAst , 34, 195  \\
		11 & Boesgaard, A.M. et al. 2015, ApJ, 799, 202    & & 22 & Kalirai, J.S. et al. 2001, AJ, 122, 266               & & \\
	\end{tabular}
\end{table*}

\subsection{Dynamical evolution}

\begin{figure}
	\centering{\includegraphics[width=0.98\columnwidth]{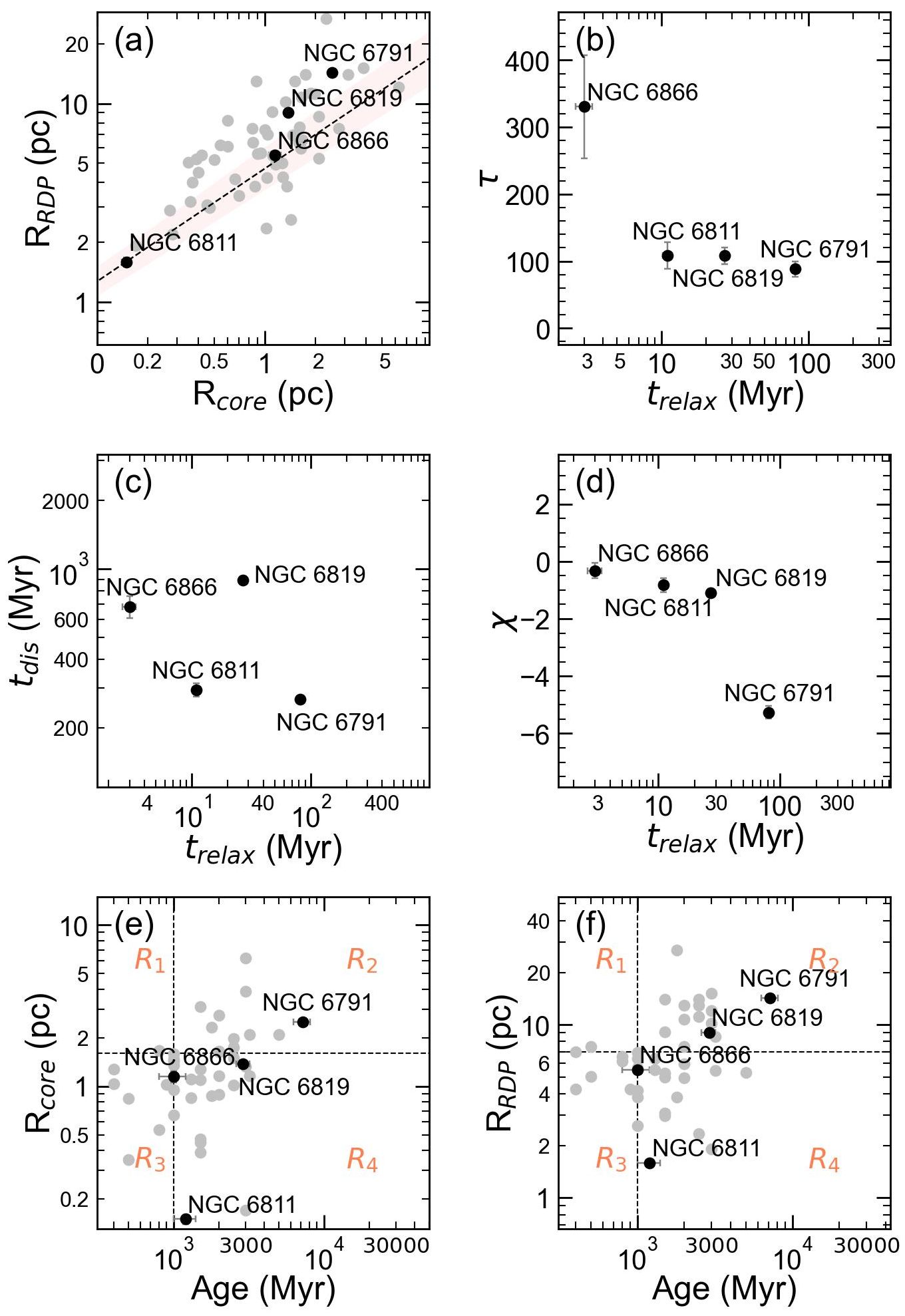}}\vspace*{-1ex}
	\caption {$R_{RDP}$ versus $R_{core}$  (panel~a),  $\tau$ versus $t_{rlx}$ (panel~b), $t_{diss}$ versus $t_{rlx}$ (panel c), $\chi$ versus $t_{rlx}$ (panel~d), and $(R_{core},~R_{RDP})$ versus Age (panels~e-f). The relation and its 1$\sigma$ uncertainty as shaded area in panel~(a), the regions $R1-R4$ in panels~(e)-(f), and the comparison objects as filled grey dots are from \citet{Gunes2017}.}
\label{f13_keplerdyn1}	
\end{figure}

There appears to be an increasing trend between $R_{core}$ and $R_{RDP}$. \cite{Gunes2017} obtained the relation $R_{RDP} =(4.69\pm0.35)R_{core}^{(0.56\pm0.11)}$ and all our objects follow this relation as seen in Fig.~\ref{f13_keplerdyn1}(a).

\begin{figure}
	\centering{\includegraphics[width=0.98\columnwidth]{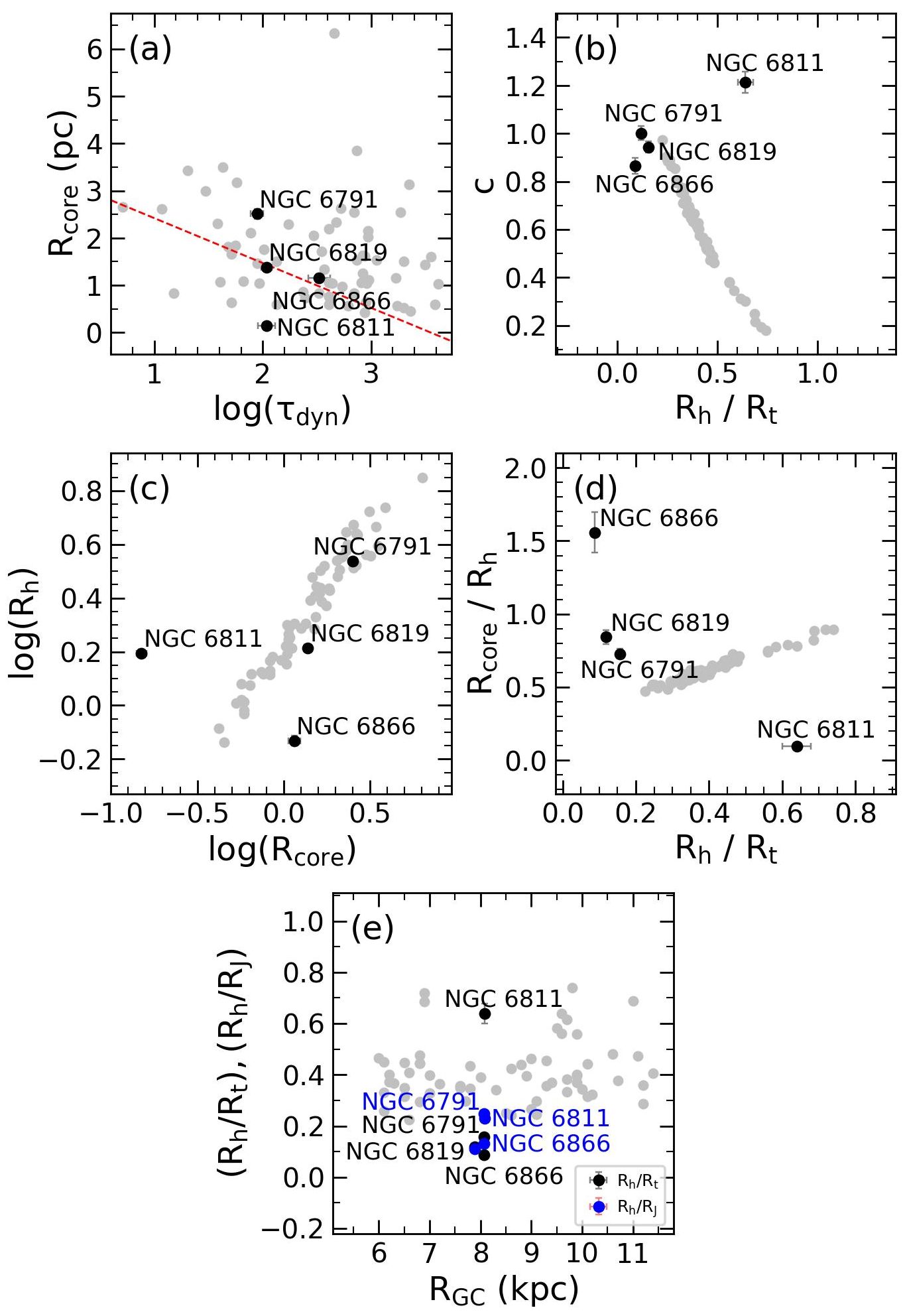}}\vspace*{-1ex}
	\caption {$R_{core}$ versus $\log(\tau)$ (panel~a), $c$ versus $R_{h}/R_{t}$ (panel~b), $\log (R_{h})$ versus $\log (R_{core})$ (panel~c), $R_{core}/R_{h}$ versus $R_{h}/R_{t}$ (panel~d), 
	$R_{h}/R_{t}$ and $R_{h}/R_{J}$  versus $R_{GC}$ (panel~e). The grey filled dots and the dashed line in panels~(a)-(e) are from A20 and A21. Black and blue filled dots in panel~(e) represent the ratios with King-profile $R_{t}$ and Jacobi $R_{J}$ tidal radii, respectively.}
\label{f14_keplerdyn2}	
\end{figure}

The relaxation times ($t_{rlx}$) of the four OCs are much smaller than $\tau$ (see panel~b of Fig.~\ref{f13_keplerdyn1}). Therefore, they can be considered as dynamically relaxed. 
On the other hand, the disruption times of the four OCs are higher than their relaxation times (see Fig.~\ref{f13_keplerdyn1}(c) and Table~\ref{t7_dynamic}). This necessarily implies that the four OCs break up in the initial phase of rapid expansion because of the death of their most massive (bright) stars \citep{Converse2011}. 
 
Following the works by \cite{Camargo2009, Gunes2017, Cakmak2021}, the four OCs are plotted in a $(R_{core},~R_{RDP})$ versus Age diagram in Fig.~\ref{f13_keplerdyn1}~(e)-(f). This relationship is related to survival and dissociation rates of the OCs \citep{Camargo2009}. The horizontal and vertical dotted lines in these panels define the regions $R1$ to $R4$ used by \citet{Gunes2017} to separate small and large sized OCs.

The very steep negative MF of NGC~6791 $(\chi=-5.26)$ for stars $m > 0.93m_{\odot}$ and the value of $\tau = 88$ indicates advanced dynamical evolution. While its massive stars move into inner regions, its low mass stars are transferred to its outer regions. Therefore, the $\chi$ value is highly negative relative to its large $R_{RDP}=14$ pc.  However, there is an incompatibility between $\chi$ and  $\tau = 88$ of this OC. In this context, there may be a large number of primordial massive stars in this cluster. From panels (e)-(f) in Fig.~\ref{f13_keplerdyn1}, the location in the $R2$ region indicates an expansion, which is caused by possible stellar black-holes and binaries, and mass segregation in its core region.

The overall negative/steep MF of NGC~6819 $(\chi=-1.09)$ indicates mild scale mass segregation due to the relatively large values of $(t_{rlx},~\tau)=(27~Myr, 108)$. The outer parts of NGC 6819 show an expansion with time because of the transport of low-mass stars into the halo. This is the reason why it occupies the $R2$ region in Fig.~\ref{f13_keplerdyn1}~(f). Due to its sparse structure and mass segregation, it has a small core, resulting in the position in the $R4$ region of Fig.~\ref{f13_keplerdyn1}~(e). Its steep $\chi$ points out that the high mass stars slightly outnumber the low-mass stars in its central part and that low-mass stars are transferred to its outer part. It seems to lose few stars to the field, owing to the internal and external perturbations.

NGC~6811 with a negative/flat overall MF $(\chi=-0.82)$ and NGC~6866 with a positive/flat MF $(\chi=-0.32)$ present signs of small scale mass segregation, in the sense that their low mass stars slightly outnumber their high mass stars. Their large $\tau$ values are linked to their advanced dynamical evolution. The position of NGC~6811 in the $R4$ regions indicates a possible shrinkage of the core/cluster radii. This means that it shrunk in size and mass with time as it lost its star content because of the presence of massive GMCs, and tidal effects from disc and Bulge crossings as external perturbations.  Instead of shrinking in size and mass with time, NGC~6811 may also have a primordial origin in conjunction with high molecular gas density in the Galactic directions \citep{Camargo2009, van91}. Note that NGC~6866 lies on the border of the $R2$ and $R4$ regions. This object with a small mass will undergoe a dynamic evolution towards the $R4$ region as a result of mass segregation and core collapse.

To present the internal and external dynamical evolutionary context for our sample OCs, we have plotted in Fig.~\ref{f14_keplerdyn2} the relations $R_{core}$ versus $\log(\tau)$ (panel~a), $c$ versus $R_{h}/R_{t}$ (panel~b), $\log (R_{h})$ versus $\log R_{core}$ (panel~c), $R_{core}/R_{h}$ versus $R_{h}/R_{t}$ (panel~d), $R_{h}/R_{t}$ and $R_{h}/R_{J}$  versus $R_{GC}$ (panel~e).

The four OCs show a decreasing trend between $R_{core}$ and $\log (\tau)$. As shown by A20 and A21 (their data included as grey points in Fig.~\ref{f14_keplerdyn2}), this trend points out that our OCs are dynamically evolved and that they are losing their star content to the field due to internal/external dynamical processes.

Based on the concentration parameter $c$ one can infer that the four objects are more compact clusters (panel~b of Fig.~\ref{f14_keplerdyn2}). Three OCs are close to the OC data by A20 and A21, but NGC~6811 stands far away. Furthermore, except NGC~6866, the objects follow the relation $R_{h} > R_{core}$ as shown in panel~(c) of the figure. The deviating OC has a small mass (113~$M_{\odot}$), and its $R_{h}$ determination may also be affected by the small number of member stars and their sparse distribution. A similar behaviour can be noticed also e.g. for NGC~6573 studied by A18. However, we note that the data by A20 and A21 shown in Fig.~\ref{f14_keplerdyn2} are based on the simple relation $R_{h} = 1.3~a$ by using Plummers' $a$ parameter \citep{plummer1911} from the profile fit.

The $R_{h}/R_{t}$ ratios of NGC~6791 and NGC~6819 fall in the range of 0.10-0.23 according to the figures~2 in chapter 33 by \cite{heg2003}. In this context they are tidally filled OCs. Hence, they are exposed to more intense tidal effects due to their large tidal radii.  With the relatively high $R_{core}/R_{h}$ ratios, these OCs expand the point of being tidally filling, and they undergo two-body relaxation and mass segregation in their core regions. Their steep negative MFs and relatively large $\tau$ values support this. 
These findings are also consistent with their expanding core/cluster radii due to the likely presence of stellar black holes and binaries (Fig.~\ref{f13_keplerdyn1}e-f) and their relatively large $t_{rlx}$ and $t_{dis}$ values (see panel~c of Fig.~\ref{f13_keplerdyn1}).

NGC~6866, which shows $R_{core}/R_{h}=1.62$, and NGC~6811 with $R_{h}/R_{t} = 0.61$ deviate from the trend presented in Fig.~\ref{f14_keplerdyn2}(d). NGC~6811 is also beyond the tidally limit value  $R_{h}/R_{t} = 0.40$ given by \cite{heg2003}. Note that Collinder 110 with low $c$ and large $R_{GC}$ has a high $R_{h}/R_{t}$ as is evident from fig.14(d) of A20. The high $R_{h}/R_{t}$ ratio of NGC 6811 suggests that it feels less the tidal effects by the Galactic gravitational field without being tidally disrupted. Furthermore, its low value $R_{core}/R_{h} = 0.10$ as an indicator of its compactness implies that its central part is in advanced evolutionary stage without being tidally disrupted. The internal processes such as mass segregation can move its stars into to the outskirts of this OC. The large $\tau$ value and the consequent loss of low mass results in a small cluster size and mass. 
The high $R_{core}/R_{h}$ ratio of NGC~6866 can be explained by that its core region is exposed to mass segregation and the ow ratio $R_{h}/R_{t}=0.08$ indicates that it is less subject to tidal effects. Note that low $R_{h}/R_{t}$ imply its survival against tidal disruption.		
	
All these features of NGC~6811 and NGC~6866 are also supported by their contracting cores due to mass segregation and core-collapse (Fig.~\ref{f13_keplerdyn1}e-f) and their small $t_{rlx}$ and $t_{diss}$ values (see panel~c of Fig.~\ref{f13_keplerdyn1}).
The differences between the OCs that occupy almost the same age and $R_{GC}$ depend on the initial conditions at cluster formation and their environments \citep{sch2006,Angelo2018}. In the sense NGC~6811 might have been born as a small size cluster.

The $R_{h}/R_{J}$ ratios of the four OCs in Fig.~\ref{f14_keplerdyn2}(e) show no trend with $R_{GC}$. They are potentially exposed to stronger Galactic tidal field because of their locations close to the solar circle. Our sample OCs do not show a ratio $R_{h}/R_{J}>0.5$, such clusters would be disrupted with small dissolution time depending on the strength of tidal perturbations. The $R_{h}/R_{J}$ ratios of NGC~6791 and NGC~6811 fall in the range $[0.20, 0.35]$ given by A18, indicating that they are tidally affected. The dynamical evolution of NGC~6819 and NGC~6866 with a more compact $R_{h}/R_{J}$ ratio is primarily driven by their internal relaxation.

The theoretical masses of the four OCs are obtained from the relation of \cite{King1962} with the help of the results for $R_{min}$ and $R_{J}$:
\begin{equation*} 
M_{theo}=3.5M_{G}\left(\frac{R_{J}}{R_{min}}\right)^3
\end{equation*}
Here, $M_{G}$, $R_{J}$, and $R_{min}$ are taken from Tables~\ref{t7_dynamic} and \ref{t8_overall}, respectively. With these values, their theoretical masses have been calculated as 14449~$M_{\odot}$ (NGC~6791), 386~$M_{\odot}$ (NGC~6811), 3796~$M_{\odot}$ (NGC~6819), and 198~$M_{\odot}$ (NGC~6866), respectively. By taking into account their total masses (Table~\ref{t6_massinfo}), for instance, the total mass of NGC\,6791 (60500~$M_{\odot}$) decreases to $\sim14449_{\odot}$ due to $R_{min}=4.71$~kpc, which corresponds to 76\% of its initial total mass. Similarly, NGCC~6811, NGC~6819 and NGC~6866 appear that they lost 84\%, 83\%, and 64\% of their initial total masses, respectively.

The fact that NGC~6791 indicates a tidal radius much larger than the Jacobi radius might be linked to significant mass loss that occurs since its formation about 7\,Gyr ago. This can be also concluded for NGC~6819 and NGC~6866 with $R_{t}/R_{J}\gtrsim 1$.  NGC~6811, on the other hand, with $R_{t}/R_{J}=0.36$ keeps it stellar content within its Jacobi radius.

In summary from Figs.~\ref{f13_keplerdyn1}-\ref{f14_keplerdyn2}, the degree of mass segregation and the strength of two-body relaxation played some role in shaping the inner parts of the OCs. Their position in the first Galactic quadrant, different initial formation conditions, strength of Galactic tidal perturbations from spiral arms and Galactic disc/Bulge, Galactic tidal field, and encounters with GMCs are responsible for their mass losses. These old OCs survived despite the internal/external dynamical processes and 5 to 30 revolutions around the Galactic center.

\subsection{Radial migration and cluster orbits}

We estimate the birth radii ($R_{birth}$) of the objects following \cite{net22} by using their result for the current metallicity gradient based on young open clusters, the model by \cite{min18} for the time evolution of the Galactic Interstellar Medium (ISM) metallicity gradient, the spectroscopic metallicities listed in Table~\ref{t10_migrate}, and distances and ages obtained from the Gaia EDR3 CMD (Table~\ref{t3_fitcmd}). 
For details we refer to \cite{net22}, the results are given in Table~\ref{t10_migrate}.

\renewcommand{\tabcolsep}{1.4mm}
\renewcommand{\arraystretch}{1.2}
\begin{table}
	\caption{The migration distances $(d_{mig} = R_{guide}-R_{birth})$. }
	\label{t10_migrate}
	\begin{tabular}{rrrrrrrc}
		\hline
		Cluster  &  [Fe/H] &Age& $R_{min}$ &$R_{max}$ &$R_{guide}$ &$R_{birth}$ &$d_{mig}$\\
		&    dex   &Gyr &    kpc    &    kpc   &    kpc &     kpc    &   kpc\\
		\hline
		NGC 6791 & 0.35 & 7.20&4.71 & 8.62 & 6.67 &--0.45 & 7.12 \\
		NGC 6811 &--0.05 & 1.20&7.80 & 9.67 & 8.74 & 8.34 & 0.40 \\
		NGC 6819 & 0.05 & 2.90&7.82 & 8.50 & 8.16 & 5.79 &  2.37 \\
		NGC 6866 & 0.01 & 1.00&8.05 & 9.79 & 8.92 & 7.41 &  1.51 \\
		\hline
	\end{tabular}  
\end{table}

We note that in this paper we adopt a solar distance of $R_{GC} = 8.2$\,kpc and a slightly different determination of the guiding radius ($R_{guide}$) than used by \cite{net22}. However, the derived $R_{birth}$ distances are in good agreement with their results, suggesting that NGC~6791 has its origin in or close to the Galactic center.  We estimate that the cluster could have radially migrated by about 7~kpc. The migration distances $(d_{mig} = R_{guide}-R_{birth})$ of the objects generally increase with age, thus they follow the trend of the cluster population as shown by \cite{net22}.

NGC~6791, an old metal-rich OC (7--8 Gyr, $[Fe/H] = 0.35-0.47$), is an intriguing system and therefore target of numerous works that deal with its astrophysical parameters and origin  \citep{pet98,ant07,car07,cas17,don20}. 
  
\cite{twa2011} evaluate that NGC~6791 might show indication of an external origin because of its position in the solar circle, the high metal abundance, and the quite eccentric orbit. According to \cite{jil12},  a strong bar and spiral arm effects are responsible for its migration  from the inner disc $(R=3-5~ kpc)$ to the current position.
	
\cite{car22} list for NGC~6791 an eccentricity of 0.35, which is close to our result (see Table~\ref{t8_overall}). The large eccentricity could indicate that the cluster was formed in inner Galactic regions, as its metal content suggests, but its orbit has been perturbed in such a way that it acquires a higher eccentricity, thus spending a significant fraction of time at larger radii. Anyway, it is not uncommon that the oldest clusters show larger eccentricities \citep[see e.g.][]{Tarr2021}.

The large eccentricity $(ecc = 0.59)$ obtained by \cite{car06} was also interpreted as a core of a large system, which is exposed to strong tidal stripping. Another scenario suggested by them is that  this OC was formed in the inner side of the Galaxy, close to the metal-rich bulge.

\cite{lind17} suggest that NGC~6791 is either an original member of the thick disc or a former member of the Galactic bulge, because of the high metal $([Fe/H]=0.28-0.34)$ and high-$\alpha$ $(0.08-0.10)$ abundances which are obtained for five members from APOGEE DR13 data.
	
\cite{mart18} report that NGC~6791 formed in the inner thin disc or in the
bulge, and then migrated to its current location. Also \cite{vil18} conclude based on the location of NGC~6791 $(z, R_{GC})= (1, 8)$~kpc that is spatially member of the Galactic disc. From the spectroscopic findings $[Fe/H]=0.313\pm0.005$ and $[\alpha/Fe]=+0.06\pm0.05$ for the giant sample, they support a scenario of a Galactic bulge origin with radial migration to its current position.
 
Our migration and eccentricity values ($d_{mig}=7.12$~kpc and $ecc=0.29$) for NGC~6791 provide a support in favor of the radial migration scenarios. 

NGC~6811 seems to be still very close to its birth position and shows by far the lowest migration rate (migration distance in dependence of the age). For the remaining objects (NGC~6819 and NGC~6866) we estimate  migration rates of about 0.8-1.5 kpc/Gyr. These values are quite consistent with the results by \cite{net22}. They note that objects up to about 2\,Gyr show mean migration rates of about 1\,kpc/Gyr and that there is a decrease of the migration rate with age, because older clusters also tend to be dynamically hotter objects. Their study shows a significant scatter of the migration distances, which they attribute among others to the merge of objects at different Galactic locations. However, the individual dynamical stage of the OCs might contribute to the scatter as well. An analysis of the overall mean migration distances in dependence on Galactic location and dynamical stage will certainly require a much larger sample in respect of both, OC metallicities and detailed dynamical knowledge for these objects.

\subsection{Concluding remarks}

In this paper we provide a detailed study of the open clusters NGC~6791, NGC~6811, NGC~6819 and NGC~6866 in respect of the ``classical'' astrophysical parameters like the age and the distance, but also for the mass and mass-function, cluster radii, orbital parameters, dynamical evolution, and radial migration.

Such information is of general importance to trace the evolution of the OC population in the Galactic context. The studied objects are covered also by the Kepler prime field, thus the results can be furthermore used in the context of variable stars - e.g. for studies related to their evolution or dependency on the various OC parameters.

\section*{Acknowledgments}
This work was supported by Scientific Research Projects Coordination Unit of Istanbul University. Project Number: FBA-2017-23599. We thank our referee for her/his valuable suggestions/comments. We thank M.~Angelo and J.~Santos for the valuable discussion on Fig.~15. O.~Günes is thanked for his interpretation on the dynamical evolution. G.~Carraro is thanked for providing the CCD $UB$ data of NGC~6791 for the comparison. We thank R.~Carrera for providing the orbital parameters for Table~8 and his comments on the origin of NGC 6791. The open cluster data is based upon the observations carried out at the Observatorio Astronómico Nacional on the Sierra San Pedro Mártir (OAN-SPM), Baja California, México. This paper has made use of results from the European Space Agency (ESA) space mission Gaia, the data from which were processed by the Gaia Data Processing and Analysis Consortium (DPAC). Funding for the DPAC has been provided by national institutions, in particular the institutions participating in the Gaia Multilateral Agreement. The Gaia mission website is http: //www.cosmos.esa.int/gaia. This paper has also made use of the WEBDA database, operated at Department of Theoretical Physics and Astrophysics of the Masaryk University, Brno. This publication also makes use of SIMBAD database-VizieR (http://vizier.u-strasbg.fr/viz-bin/VizieR?-source=II/246.).

\section*{Data Availability}
The photometric data of the four OCs can be requested from Raul Michel (rmm@astro.unam.mx).


\clearpage

\appendix\section{Photometric Errors and Differential Grids Fits}
\renewcommand{\tabcolsep}{2.6mm}
\renewcommand{\arraystretch}{1.1}
\begin{table}
	\centering
	\caption{The mean photometric errors of $V$, $(R-I)$, $(V-I)$, $(B-V)$ and $(U-B)$ for NGC~ 6791, NGC~6811, NGC~6819 and NGC~6866 in terms of $V$ mag.}
	\label{t1_app_err}
	\begin{tabular}{cccccc}
		\hline
		\hline
		&  & &    NGC~6791  & &  \\
		\hline
		V &$\sigma_{V}$&$\sigma_{R-I}$&$\sigma_{V-I}$ &$\sigma_{B-V}$ &$\sigma_{U-B}$ \\
		\hline
		11 - 12 & 0.005 & 0.007 & 0.007 & 0.006 & 0.009 \\
		12 - 13 & 0.004 & 0.008 & 0.007 & 0.006 & 0.008 \\
		13 - 14 & 0.005 & 0.009 & 0.008 & 0.009 & 0.014 \\
		14 - 15 & 0.005 & 0.007 & 0.007 & 0.008 & 0.019 \\
		15 - 16 & 0.006 & 0.009 & 0.008 & 0.011 & 0.027 \\
		16 - 17 & 0.009 & 0.013 & 0.013 & 0.017 & 0.050 \\
		17 - 18 & 0.014 & 0.018 & 0.019 & 0.030 & 0.082 \\
		18 - 19 & 0.027 & 0.031 & 0.036 & 0.056 & 0.122 \\
		19 - 20 & 0.056 & 0.050 & 0.067 & 0.108 &   --  \\
		20 - 21 & 0.104 & 0.055 & 0.107 &   --  &   --  \\
		\hline
		\hline
		&  & &    NGC 6811  & &  \\
		\hline
		V &$\sigma_{V}$&$\sigma_{R-I}$&$\sigma_{V-I}$ &$\sigma_{B-V}$ &$\sigma_{U-B}$ \\
		\hline
		10 - 11 & 0.031 & 0.044 & 0.032 & 0.036 & 0.024 \\
		11 - 12 & 0.015 & 0.030 & 0.017 & 0.019 & 0.013 \\
		12 - 13 & 0.010 & 0.022 & 0.012 & 0.015 & 0.012 \\
		13 - 14 & 0.008 & 0.018 & 0.010 & 0.012 & 0.011 \\
		14 - 15 & 0.006 & 0.012 & 0.009 & 0.011 & 0.012 \\
		15 - 16 & 0.007 & 0.009 & 0.009 & 0.010 & 0.012 \\
		16 - 17 & 0.010 & 0.014 & 0.014 & 0.015 & 0.023 \\
		17 - 18 & 0.018 & 0.020 & 0.022 & 0.027 & 0.040 \\
		18 - 19 & 0.035 & 0.036 & 0.042 & 0.051 & 0.053 \\
		\hline
		\hline
		&   & & NGC 6819 &      &  \\
		\hline
		V &$\sigma_{V}$&$\sigma_{R-I}$&$\sigma_{V-I}$ &$\sigma_{B-V}$ &$\sigma_{U-B}$ \\
		\hline
		10 - 11 & 0.024 & 0.009 & 0.008 & 0.027 &  --   \\
		11 - 12 & 0.010 & 0.033 & 0.031 & 0.028 & 0.030 \\
		12 - 13 & 0.013 & 0.031 & 0.028 & 0.021 & 0.019 \\
		13 - 14 & 0.009 & 0.022 & 0.019 & 0.015 & 0.017 \\
		14 - 15 & 0.014 & 0.014 & 0.017 & 0.017 & 0.013 \\
		15 - 16 & 0.011 & 0.012 & 0.014 & 0.014 & 0.013 \\
		16 - 17 & 0.012 & 0.016 & 0.015 & 0.015 & 0.018 \\
		17 - 18 & 0.014 & 0.022 & 0.020 & 0.020 & 0.029 \\
		18 - 19 & 0.021 & 0.034 & 0.031 & 0.033 & 0.046 \\
		19 - 20 & 0.029 & 0.045 & 0.041 & 0.045 & 0.056 \\
		\hline
		\hline
		&   & &NGC 6866 &      &  \\
		\hline
		V &$\sigma_{V}$&$\sigma_{R-I}$&$\sigma_{V-I}$ &$\sigma_{B-V}$ &$\sigma_{U-B}$ \\
		\hline
		10 - 11 & 0.020 & 0.016 & 0.035 & 0.031 & 0.027 \\
		11 - 12 & 0.027 & 0.019 & 0.030 & 0.042 & 0.039 \\
		12 - 13 & 0.019 & 0.012 & 0.021 & 0.029 & 0.025 \\
		13 - 14 & 0.013 & 0.010 & 0.015 & 0.019 & 0.017 \\
		14 - 15 & 0.008 & 0.009 & 0.011 & 0.013 & 0.013 \\
		15 - 16 & 0.008 & 0.011 & 0.011 & 0.011 & 0.016 \\
		16 - 17 & 0.010 & 0.012 & 0.013 & 0.014 & 0.025 \\
		17 - 18 & 0.013 & 0.016 & 0.017 & 0.019 & 0.036 \\
		18 - 19 & 0.020 & 0.024 & 0.026 & 0.030 & 0.050 \\
		\hline
	\end{tabular}%
\end{table}%

\begin{figure}
	\centering{\hspace*{0.5ex}
		\includegraphics[width=0.47\columnwidth]{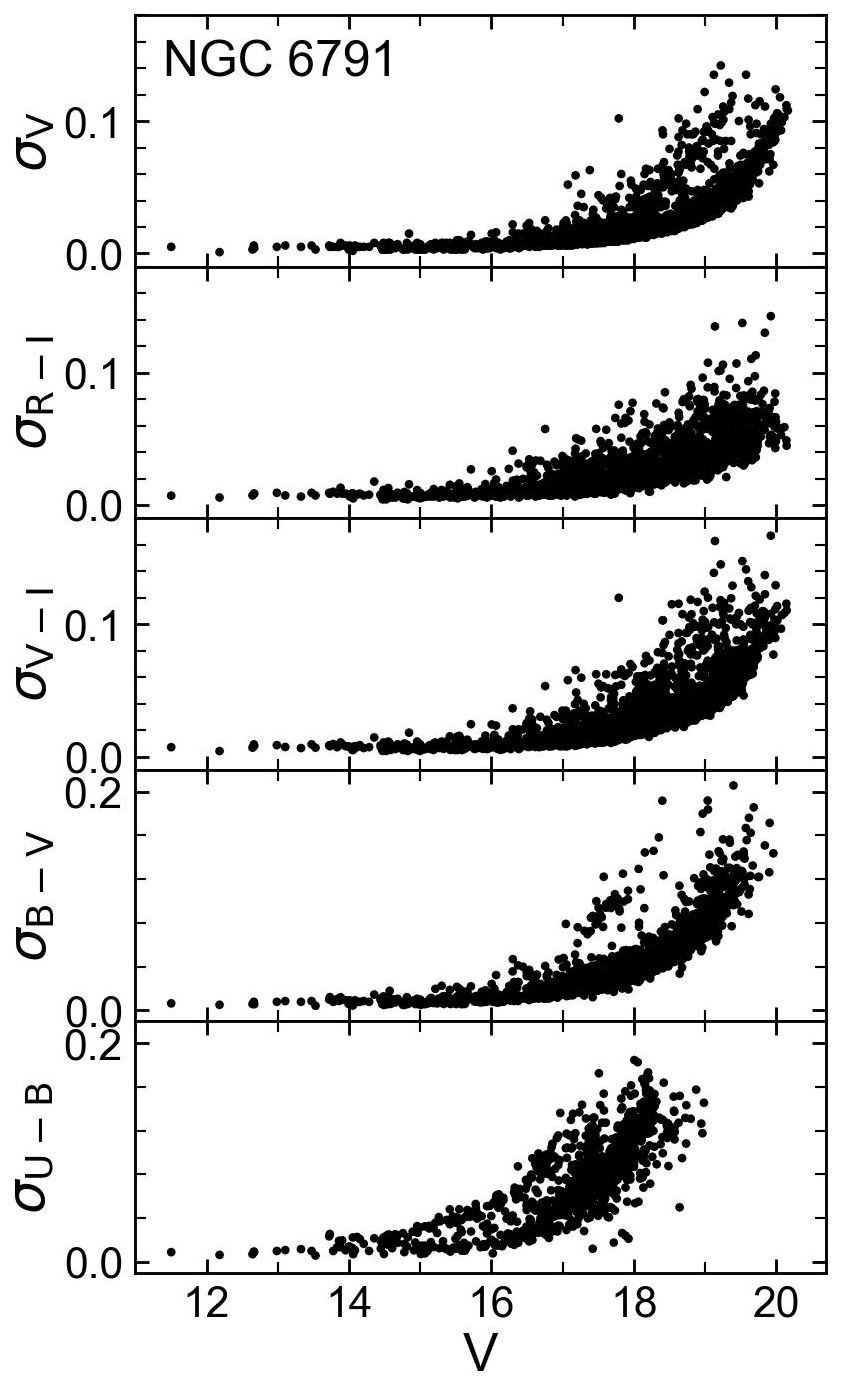}\hspace*{1.5ex}
		\includegraphics[width=0.49\columnwidth]{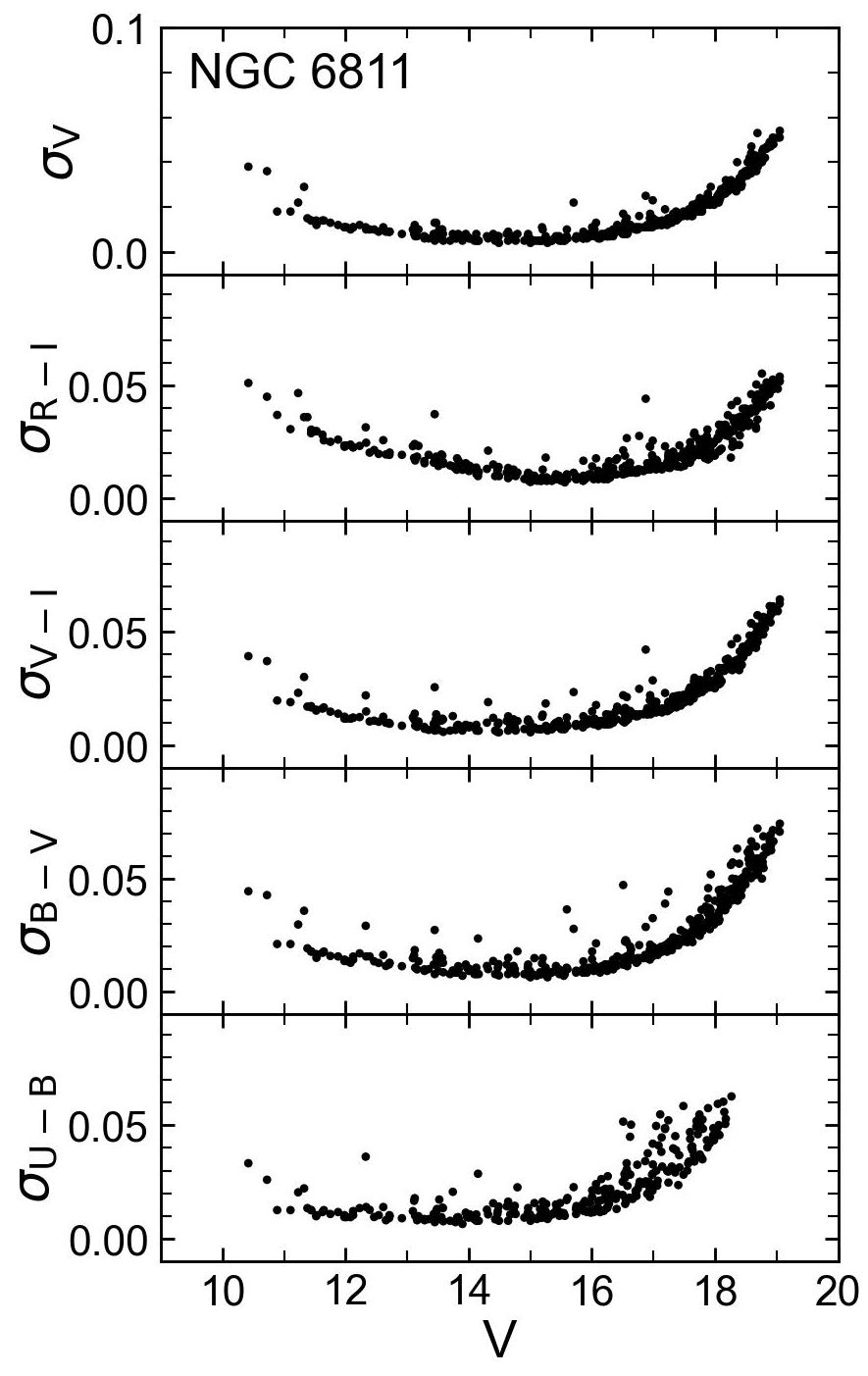}\\[1ex]
		\includegraphics[width=0.49\columnwidth]{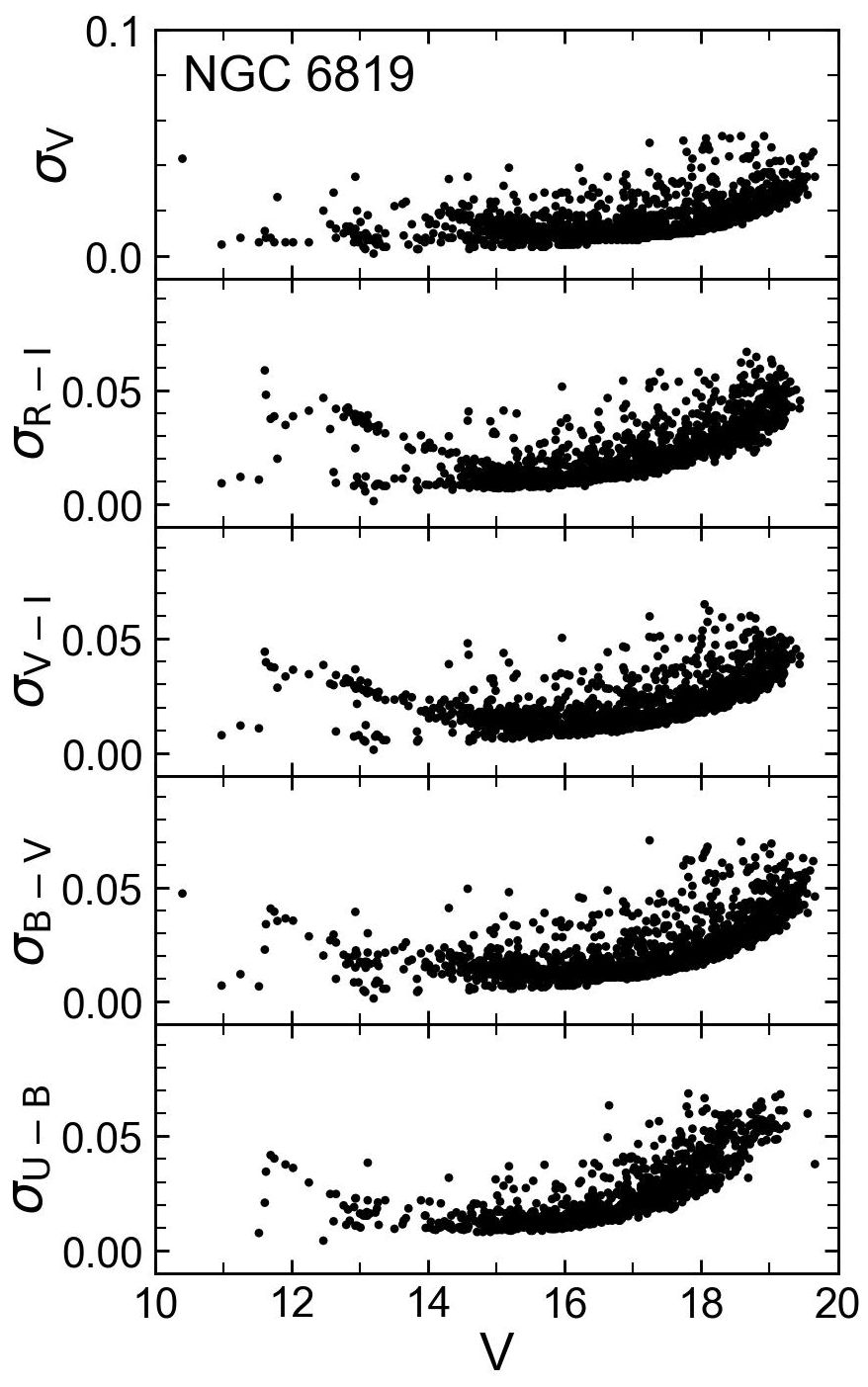}\hspace*{0.5ex}
		\includegraphics[width=0.48\columnwidth]{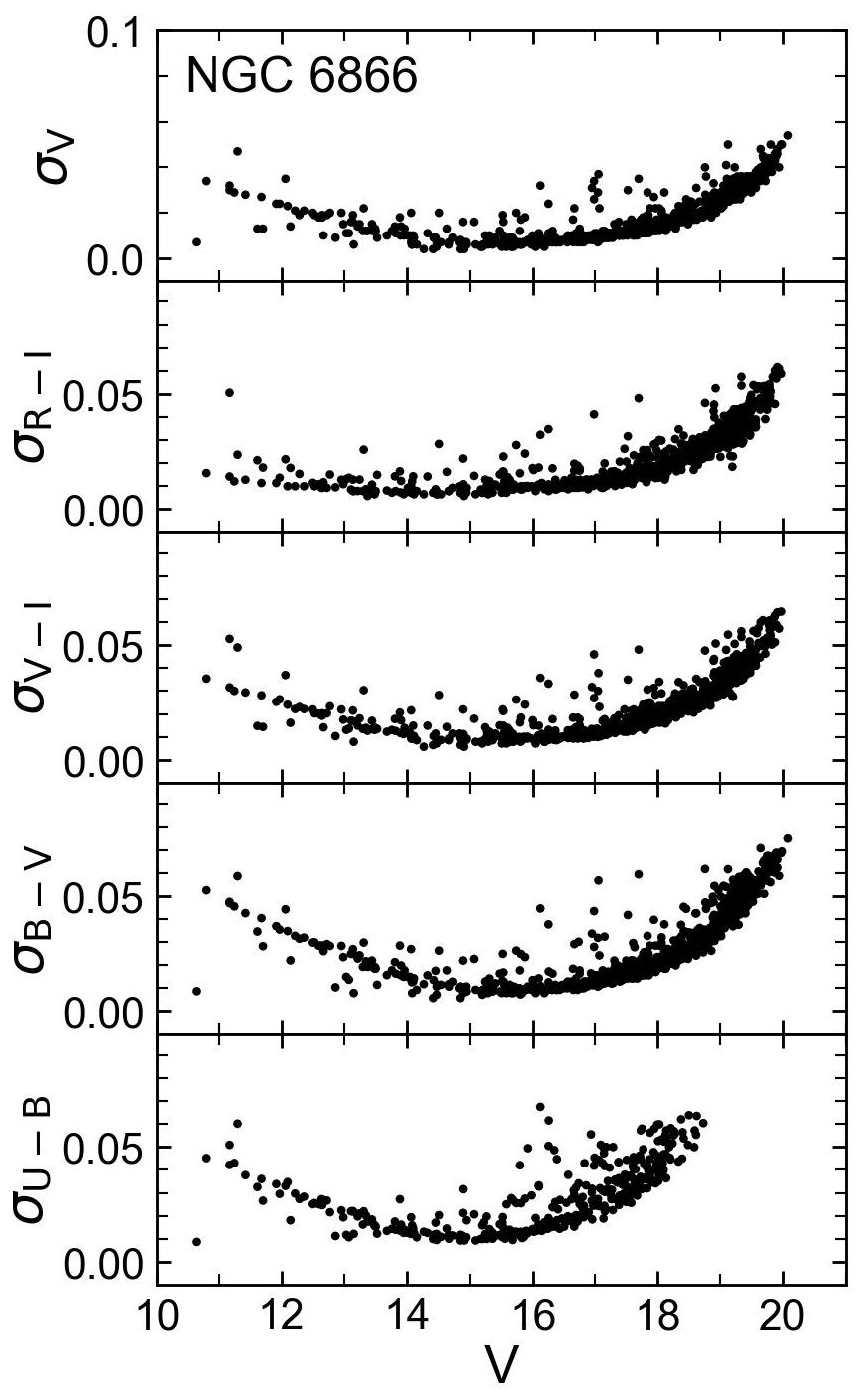}	}\vspace{-1ex}
	\caption{The distribution of the photometric errors of V, (R-I), (V-I), (B-V) and (U-B) against V mag for NGC 6791, NGC 6811, NGC 6819 and NGC 6866.}
	\label{f1_app_pherr}
\end{figure}

\begin{figure}
	\centering{
		\includegraphics[width=0.48\columnwidth]{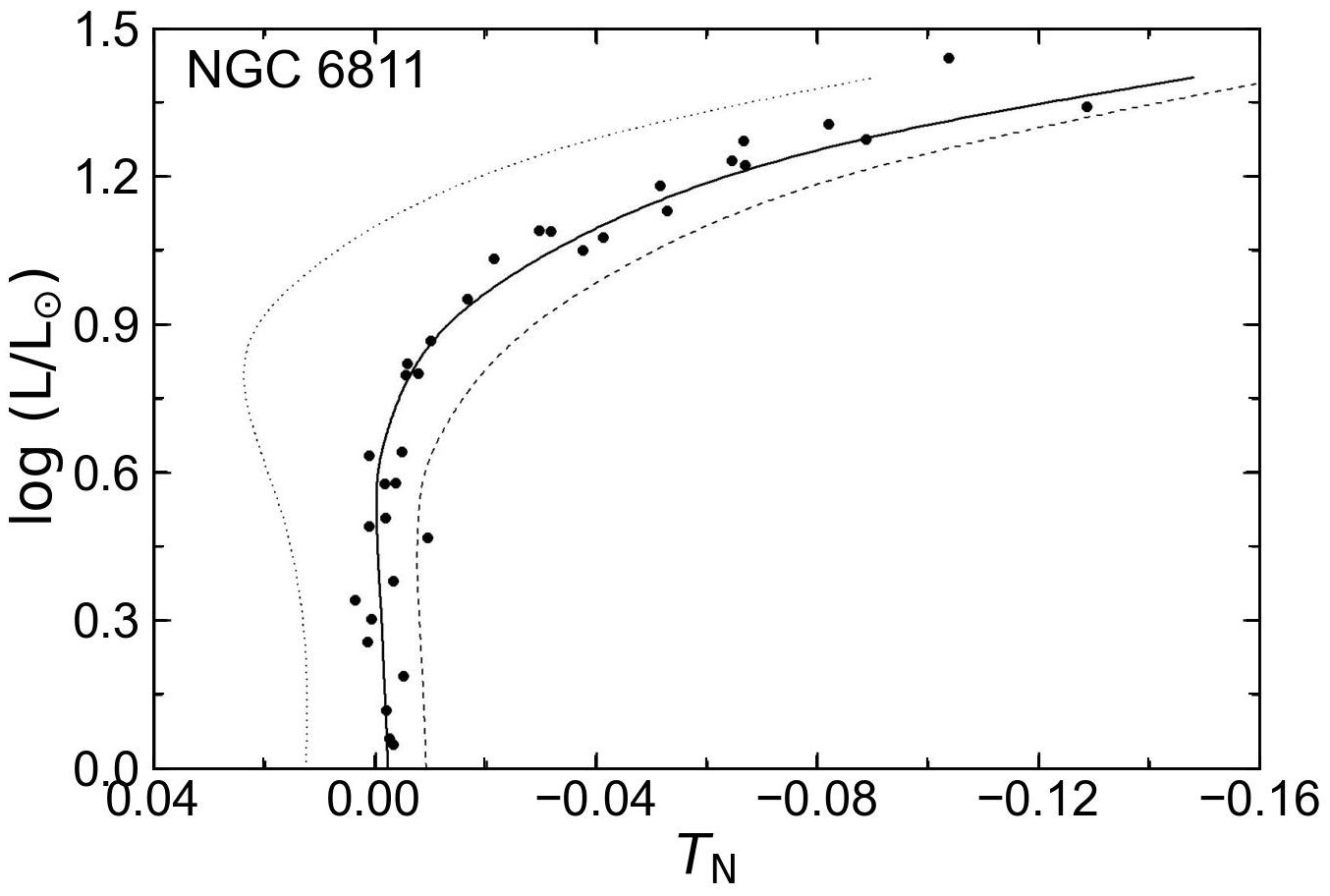}\hspace*{2ex} 
		\includegraphics[width=0.48\columnwidth]{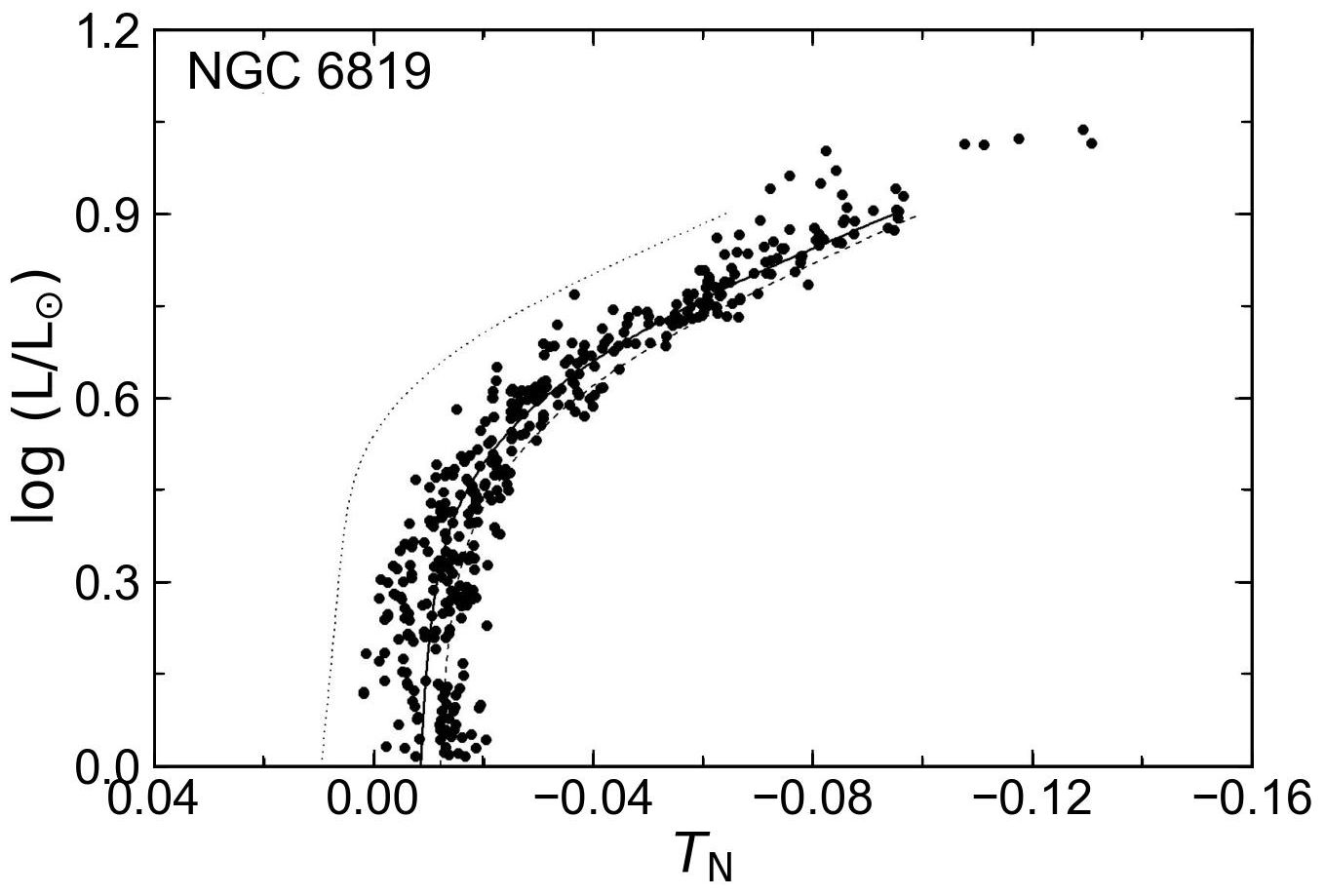}\\[1ex]
		\includegraphics[width=0.48\columnwidth]{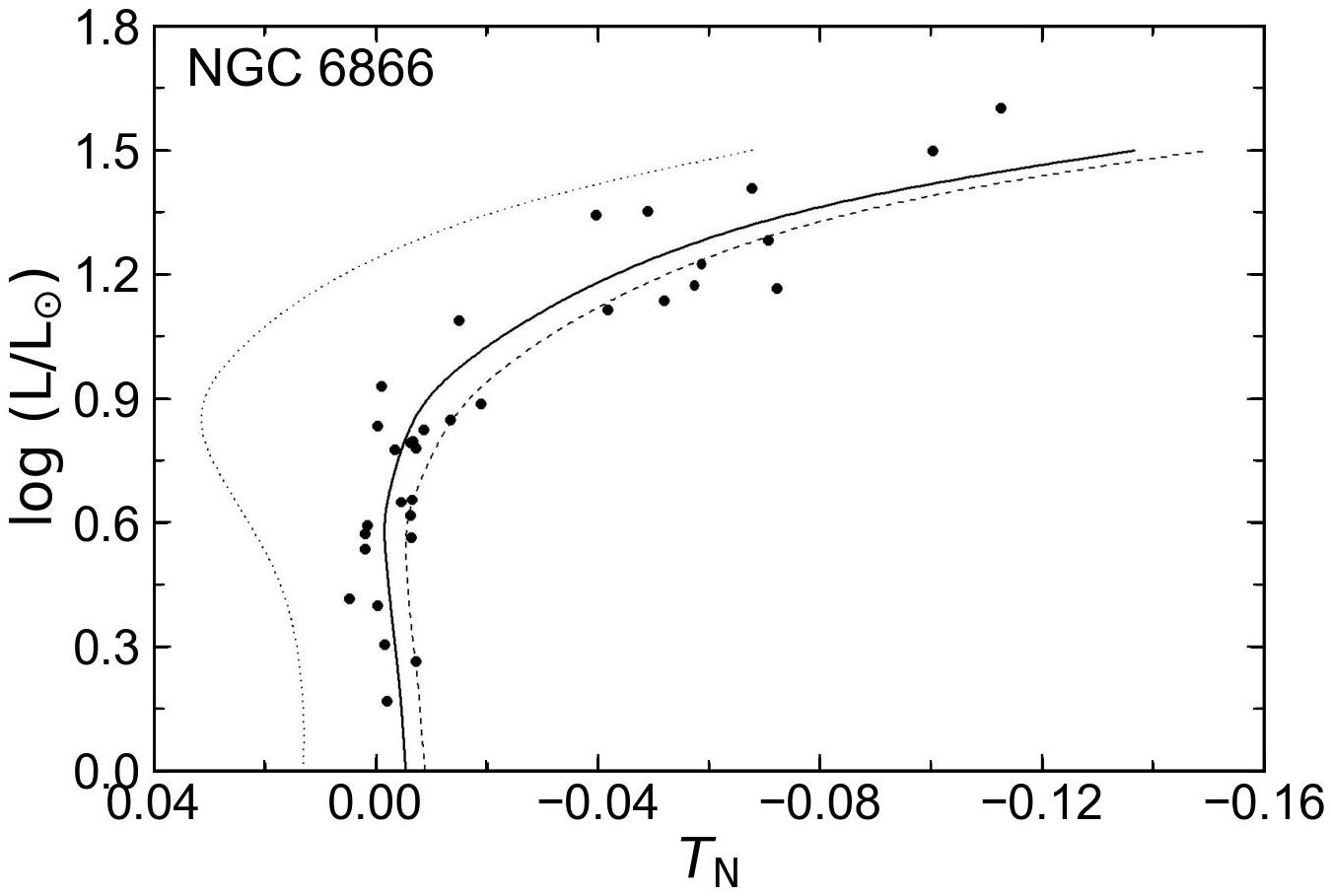}}\vspace{-2ex}
	\caption{The thick line represents the isochrone for the parameters given in Table~4, the dotted line for $Z=0.006$, and the dashed line for $Z=0.02$. $T_N$ is the temperature difference in dex between the star and the ZAMS at solar metallicity using the mean temperature based on up to five colours.}
	\label{f3_app_isoch}
\end{figure}

\bsp	
\label{lastpage}

\begin{thebibliography}{99}	
	\bibitem[Abedigamba et al. (2016)]{ab16} Abedigamba O.P., et al., 2016, \na, 46, 90 
	
	\bibitem[Akkaya et al.\ (2010)]{akk10} Akkaya \'I., Schuster W.J., Michel R., Chavarr\'ia-K C., Moitinho A., V\'azquez R., Karata\c{s} Y., 2010, \rmxaa, 46, 385  
	
	\bibitem[Akkaya Oralhan et al. (2015)]{akk15} Akkaya Oralhan \'I., Karata\c{s} Y., Schuster W.J., Michel R., Chavarr\'ia C., 2015, \na, 34, 195
	
	\bibitem[Akkaya Oralhan et al. (2019)]{akk19} Akkaya Oralhan I., Michel R., Schuster W.J., Karata\c{s} Y.,  Karslı Y., Chavarr\'ia C., 2019, \japa, 40,33
	
	\bibitem[An et al. (2015)]{an15} An D., et al., 2015, \apj, 811, 46
	
	\bibitem[Anders et al. (2017)]{and07} Anders F., et al., 2017, \aap, 600, 70
	
	\bibitem[Angelo et al. (2018)]{Angelo2018} Angelo M.S., Piatti A.E., Dias W.S., Maia F.F.S., 2018,\mnras, 477, 3600 (A18)
	
	\bibitem[Angelo et al. (2020)]{Angelo2020} Angelo M.S., Santos Jr J.F.C., Corradi W.J.B., 2020,\mnras, 493, 3473 (A20)
	
	\bibitem[Angelo et al. (2021)]{Angelo2021} Angelo M.S., Corradi W.J.B, Santos Jr J.F.C., Maia F.F.S., Ferreira F.A., 2021, \mnras, 500, 4338 (A21)
	
	\bibitem[Anthony-Twarog et al.(2007)] {ant07} Anthony-Twarog B.J., Twarog B.A., Lindsay M., 2007, \aj, 133, 1585 
	
	\bibitem[Anthony-Twarog et al.(2014)] {ant14} Anthony-Twarog B., et al., 2014, \aj, 148, 51 
	
	\bibitem[Bailer-Jones et al. (2018)]{Bailer2018} Bailer-Jones C.A.L., Rybizki J., Fouesneau M., Mantelet G., Andrae R., 2018, \aj, 156, 58
	
	\bibitem[Bailer-Jones et al.\ (2021)]{Bailer2021} Bailer-Jones C.A.L., Rybizki J., Fouesneau M., Demleitner M., Andrae R., 2021, \aj, 161, 147
	
	\bibitem[Balaguer-Nunez et al.\ (1998)]{bal98} Balaguer-Nunez L., Tian K.P., Zhao J.L., 1998, \aaps, 133, 387
	
	\bibitem[Balona et al.\ (2013)]{bal13} Balona L.A., et al., 2013, \mnras, 429, 1466 
	
	\bibitem[Baumgardt et al.\ (2010)]{bau2010} Baumgardt H., Parmentier G., Gieles M., Vesperini E., 2010, \mnras, 401, 1832 	
		
	\bibitem[Bertelli et al.\ (2008)]{ber08} Bertelli, G., Girardi, L., Marigo, P., \& Nasi, E., 2008, \aap, 484, 815
	
 	\bibitem[Bessell et al.\ (1998)]{bes98} Bessell,M.S., Castelli, F., Plez, B., 1998, \aap, 333, 231
 
	\bibitem[Binney \& Tremaine (2008)]{Binney2008} Binney J., Tremaine S., Galactic Dynamics Princeton Univ Press Princeton, NJ;2008.
	
	\bibitem[Bland-Hawthorn et al. (2016)]{bg16} Bland-Hawthorn J., Ortwin Gerhard O., 2016, \araa, 54, 529
	
	\bibitem[Boesgaard et al.(2015)]{boes15} Boesgaard A.M., et al., 2015, \apj, 799, 202
	
	\bibitem[Bonatto et al.\ (2005)]{Bonatto2005} Bonatto C., Bica E., Santos Jr J.F.C., 2005, \aap, 433, 917
	
	\bibitem[Bonatto \& Bica (2007)]{Bonatto2007} Bonatto C., Bica E., 2007, \mnras, 377, 1301
	
	\bibitem[Bonatto \& Bica (2007)]{Bonatto2011} Bonatto C., Bica E., 2011, \mnras, 415, 313
		
	\bibitem[Bonatto (2019)]{Bonatto2019} Bonatto C., 2019, \mnras, 483, 2758
	
	\bibitem[Bovy (2015)]{bov15} Bovy J., galpy: A python Library for Galactic Dynamics., 2015, \apjs,  216, 29
	
	\bibitem[Bragaglia \& Tosi (2006)]{bt2006} Bragaglia A., Tosi M., 2006, \aj, 131, 1544 

	\bibitem[Bragaglia et al. (2018)]{brag18} Bragaglia A., Fu X., Mucciarelli A., Andreuzzi G., Donati P., 2018, \aap, 619, A176
	
	\bibitem[Bressan et al. (2012)]{bre12} Bressan A., Marigo P., Girardi L., Salasnich B., Dal Cero C., Rubele S., Nanni A., 2012, \mnras, 427, 127
	
	\bibitem[Bressan et al. (1993)]{bre13} Bressan, A., Fagotto, F., Bertelli, G., and Chiosi, C.  , 1993, \aaps, 100, 647 
	
	\bibitem[Brunthaler et al.\ (2011)]{bru11} Brunthaler A., Reid M.J., Menten K.M., et.al., 2011, \an, 332, No.5, 461
	
	\bibitem[\protect\citeauthoryear{Bukowiecki et al.}{2011}]{buk2011} Bukowiecki L., Maciejewski G., Konorski P., Strobel A., 2011, \actaa, 61, 231
	
	\bibitem[Buzzoni et al \ (2012)]{buz12} Buzzoni, A., Bertone, E., Carraro, G., Buson, L., 2012, \apj, 749, 35
		
	
	\bibitem[Cakmak et al.\ (2021)]{Cakmak2021} Cakmak H., Gunes O., Karatas Y., Bonatto C., 2021, \an, 324, 975
	
	\bibitem[Camargo et al.(2009)]{Camargo2009} Camargo D., Bonatto C., Bica E., 2009, \aa, 508, 211
	
	\bibitem[Cantat-Gaudin et al. (2018)]{cantat2018} Cantat-Gaudin T., Jordi C., Vallenari A., et al., 2018, \aap, 618, 93
	
	\bibitem[Cantat-Gaudin et al. (2020)]{cantat2020} Cantat-Gaudin T., Anders F., Castro-Ginard A., et al., 2020, \aap, 640, 1
	
	\bibitem[Carraro et al.\ (2006)]{car06} Carraro G., Villanova S., Demarque P., et al., 2006, \aj, 643, 1151
	
	
	\bibitem[Carraro et al.\ (2013)]{car13} Carraro G., et al., 2013, \aj, 146, 128  
	
	\bibitem[Carretta et al.(2007)]{car07}	Carretta E.,  Bragaglia A., Gratton R.G., 2007, \aap, 473, 129
	
	\bibitem[Carrera et al.(2022)]{car22} Carrera R.,  Casamiquela, L., Carbajo-Hijarrubia, J., et al., 2022, \aap, 658, A14		
	
	\bibitem[Casamiquela et al.(2017a)]{cas17} Casamiquela L., et al., 2017, \mnras, 470, 4363 
	
	\bibitem[Converse \& Stahler (2011)]{Converse2011} Converse J.M., Stahler S.W. 2011, \mnras, 410, 2787
	
	\bibitem[Dias et al.\ (2018)]{dia18} Dias W.S., Monteiro H., Lepine J.R.D., prates R., Gneiding C.D., Sacchi M. \ 2018, \mnras, 481, 3887
	
	\bibitem[Dias et al.\ (2021)]{dia21} Dias W.S., Monteiro H., Moitinho, A., Lepine J.R.D., Carraro, G., Paunzen, E., Alessi, B., Villela, I., 2021, \mnras, 504, 356
	
	\bibitem[Donor et al.\ (2020)]{don20} Donor J. et al., 2020, \aj, 159, 199
	
	\bibitem[Frinchaboy et al.\ (2020)]{fri20} Frinchaboy P.M. et al. 2013, ApJL, 777, 1    
	
	\bibitem[Gaia Collaboration et al. \ (2021)]{gaia3} Gaia Collaboration,   
	Brown A.G.A,  Vallenari A., Prusti T., de Bruijne J.H.J.  et al., 2021, \aap, 649, A1 	
	
	\bibitem[Gao \& Chen (2012)]{gao12} Gao, Xin-hua, Chen L., 2012, \chaa, 36, 1
	\bibitem[Gao (2020)]{gao20} Gao, Xin-hua., 2020, Astrophys.Space.Sci., 365, 24
	
	\bibitem[Gieles et al. (2007)]{Gieles2007} Gieles M., Athanassoula E., Portegies-Zwart S., 2007, \mnras, 376, 809
	
	\bibitem[\protect\citeauthoryear{Girardi et al.}{2002}]{gir02} Girardi L., Bertelli G., Bressan A., Chiosi C., Groenewegen M.A.T., Marigo P., Salasnich B., Weiss A., 2002, \aap, 391, 195
	
	\bibitem[Glushkova et al. (2012)]{glu1999} Glushkova E.V., et al., 1999, AstL, 25, 86,15
	
	\bibitem[G\"{u}ne\c{s} et al. (2012)]{Gunes2012} G\"une\c{s} O., Karata\c{s} Y., Bonatto C., 2012, New Astronomy, 17, 720
	
	\bibitem[G\"{u}ne\c{s} et al. (2017)]{Gunes2017} G\"une\c{s} O., Karata\c{s} Y., Bonatto C., 2017, Astronomische Nachritten (AN),  338, 464.
	
	\bibitem[Harris et al.(1981)]{har81} Harris W., et al., 1981, \aj, 86, 1332
	
	\bibitem[Heggie \& Hut (2003)]{heg2003} Heggie D. and  Hut P., 2003, the gravitational million-body problem: A multidisciplinary approach to star cluster dynamics, Cambridge Univ. Press, chapter 33.
	
	\bibitem[Janes et al.(2013)]{jan13} Janes K.J., et al., 2013, \aj, 145, 7,14
	\bibitem[Janes et al.(2014)]{jan14} Janes K.J., et al., 2014, \aj, 147, 139 
	\bibitem[Jilkova et al.(2012)]{jil12} Jilkova L., Carraro G., Jungwiert B., Minchev I. 2012, \aap, 541, A64
	
	\bibitem[Johnson \& Soderblom (1987)]{joh87} Johnson D.R.H., Soderblom D.R., 1987, \aj, 93, 864
	
	\bibitem[\protect\citeauthoryear{Kharchenko et al.}{2013}]{kha13} Kharchenko N.V., Piskunov A.E., Schilbach E., R{\"o}ser S., Scholz R.D., 2013, \aap, 558, 53
	
	\bibitem[Kalirai et al. (2001)]{kal01} Kalirai J.S. et al. 2001, \aj, 122, 266
	
	\bibitem[Kaluzny \& Rucinski (1995)]{kal95} Kaluzny J., Rucinski S.M., 1995, \aaps, 114, 1
	
	\bibitem[Kepley et al.  (2007)]{kep07} Kepley A., Morrison H.L., Helmi A, Kinman, T.D., 2007, \aj, 134, 1579
	
	\bibitem[Kim et al. (2000)]{Kim2000} Kim S.S., Figer D.F., Lee H.M., Morris M.,2000, \apj, 545, 301
	
	\bibitem[King (1962)]{King1962} King I., 1962, \aj, 67, 471
	
	\bibitem[King (1966)]{King1966} King I., 1966, \aj, 71, 64
	
	\bibitem[Lamers \& Gieles (2006)]{Lamers2006} Lamers H.J.G.L.M., Gieles M., 2006,  \aap, 455, 17
	
     \bibitem[\protect\citeauthoryear{Landolt}{2009}]{lan09} Landolt A.U., 2009, \aj, 137, 4186

	\bibitem[Larsen (2006)]{Larsen2006} Larsen S.S., 2006,  An ISHAPE Users Guide p 14. 
	
	\bibitem[Lindegren et al.\ (2021)]{lin21}  Lindegren L., Klioner S. A., Hernandez J., et al., 2021, \aap, 649, 9 
	
	\bibitem[Linden et al.\ (2017)]{lind17}  Linden S.T., Pryal M., Hayes C.R., et al., 2017, \apj, 842, 49 
			
	\bibitem[Luo et al.\ (2009)]{lu09} Luo Y.P. et al., 2009, NewAst, 14, 584 	
	
	\bibitem[Martinez-Medina et al. (2018)]{mart18} Martinez-Medina L.A., Gieles M., Pichardo B., Peimbert A., 2018, \mnras, 474, 32
	
	\bibitem[Mermilliod (1992)]{Mermilliod1992} Mermilliod J.C., 1992, Bull. Inform, 40, 115
	
	\bibitem[Minchev et al.\ (2018)]{min18} Minchev I., Anders F.,  Recio-Blanco A., et al. 2018, \mnras, 481, 1645
	
	\bibitem[\protect\citeauthoryear{Moitinho}{2010}]{moi10} Moitinho A., 2010, Star clusters: basic galactic building blocks, Proceedings IAU Symposium No.266, eds. R.de Grijs \& J.R.D. Lepine 
	
	\bibitem[Molenda-Zakowicz et al. \ (2014)]{mz14} Molenda-Zakowicz J. et al. 2014,\mnras, 445, 2446
		
	\bibitem[Monteiro et al.\ (2020)]{mon20} Monteiro H., Dias W.S., Moitinho A., Cantat-Gaudin T., Lepine J.R.D., Carraro G., Paunzen E., 2020, \mnras, 499, 1874

	\bibitem[Netopil et al. \ (2015)]{net15} Netopil M., Paunzen E., Carraro G., 2015, \aap, 582, A19	
		
	\bibitem[Netopil \& Paunzen \ (2013)]{net13} Netopil,  M., Paunzen,  E., 2013, \aap, 557, A10
	
	\bibitem[Netopil et al. \ (2016)]{net16} Netopil M., Paunzen E., Heiter U., Soubiran C., 2016, \aap, 585, A150
	
	\bibitem[Netopil et al.\ (2022)]{net22} Netopil M., Oralhan- Akkaya, \'I., Cakmak H., Michel R., Karatas Y. 2022, \mnras, 509, 421
	
	\bibitem[Pedregosa et al.\ (2011)]{ped11} Pedregosa F., et al., Scikit-learn: Machine Learning in Python, JMLR 12, pp. 2825-2830, 2011.
	
	\bibitem[Pena et al.\ (2011)]{pen11} Pena J.H., et al., 2011, \rmxaa, 47, 309 
	
	\bibitem[Peterson \& Green(1998)]{pet98} Peterson R.C., Green E.M., 1998, \apj, 502, 39 
	
	\bibitem[\protect\citeauthoryear{Piatti et al.}{2017a}]{pia17a} Piatti A.E., Dias W.S., Sampedro L.M., 2017a, \mnras, 466, 392	
	
	\bibitem[\protect\citeauthoryear{Piatti et al.}{2017b}]{pia17b} Piatti A.E., 2017b, \mnras, 465, 2748
	
	\bibitem[Piatti et al. (2019)]{pia09} Piatti A.E., Angelo M.S., Dias W., 2019, \mnras,  488, 4648
	
	\bibitem[Piskunov et al. (2007)]{pis07} Piskunov A.E., Schilbach E., Kharchenko N.V., Roser S., Scholz R.D., 2007, \aap,  468, 151
	
	\bibitem[Platais et al. (2011)]{pla11} Platais, I., Cudworth, K.M., Kozhurina-platais, V. et al., 2011, \apj~Letters, 733, L1	

	\bibitem[Plummer (1911)]{plummer1911} Plummer, H.C., 1911, \mnras, 71, 460	
	
	\bibitem[P\"ohnl and Paunzen \ (2010)]{poh10} P\"ohnl H., Paunzen E., 2010, \aap, 514, A81
	
	\bibitem[Rosvick and Vandenberg \ (2010)]{rv10} Rosvick J.M., Vandenberg D.A., 1998, \aj, 115, 1516
	
	\bibitem[Reid et al. (2019)]{rei19} Reid, M.J., Menten, K.M., Brunthaler, A., et al., 2019, \apj, 885, 131	

	\bibitem[Sandquist et al.\ (2016)]{san16} Sandquist E.L., et al., 2016, \apj, 831, 11
	
	\bibitem[Sariya et al.\ (2012)]{sar12} Sariya D.P., Yadav R.K.S., Yadav Bellini A., 2012, \aap, 543, 87
	
	\bibitem[Schilbach et al. (2006)]{sch2006} Schilbach E., Kharchenko N.V., Piskunov A.E., Roser S., Scholz R.D., 2006, \aap, 456, 523
	
	\bibitem[Sch{\"o}nrich and Binney (2010)]{sch10} Sch{\"o}nrich R., Binney J., Dehnen W., 2010, \mnras 403, 1829
	
	\bibitem[\protect\citeauthoryear{Schuster et al.}{2007}]{suc07} Schuster W.J., Michel R., Dias W., Tapia-Peralta T., V\'azquez R., Macfarland J., Chavarr\'{\i}a C., Santos C., Moitinho A., 2007, Galaxy Evolution Across the Hubble Time, eds.\ F.~Combes and J.~Palou$\breve{\rm s}$, Proceedings of the International Astronomical Union, IAU Symposium No.~235, (Cambridge, United Kingdom: Cambridge University Press), p.~331
	
	\bibitem[\protect\citeauthoryear{Skrutskie et al.}{2006}]{skr06} Skrutskie M.F., Cutri R., Stiening R., Weinberg M.D., Schneider S.E., Carpenter J.M., Beichman C., Capps R., 2006, \aj, 131, 1163
	
	\bibitem[Spitzer \& Hart (1971)]{Spitzer1971} Spitzer L., Hart M. ,1971, \aj, 164, 399
	
	\bibitem[\protect\citeauthoryear{Stetson}{1987}]{stet87} Stetson P.B., 1987, \pasp, 99, 191
	
	\bibitem[\protect\citeauthoryear{Stetson et al.}{2003}]{stet03} Stetson P.B., Bruntt H., Grundahl F., 2003, \pasp, 115, 413

	\bibitem[\protect\citeauthoryear{Sung \& Bessell}{2000}]{sung00} Sung H., Bessell M., 2000, \pasa, 17, 244
		
	\bibitem[Tang et al. (2018)]{tan18} Tang Y., et al., 2018, \apj, 866, 59
	
	\bibitem[Tapia et al.\ (2010)]{tap10} Tapia M.~T., Schuster W.~J., Michel R., Chavarr\'ia-K C., Dias W.~S., V\'azquez R., \& Moitinho A., 2010, \mnras, 401, 621
	
	\bibitem[Tarricq et al. (2021)]{Tarr2021} Tarricq Y., Soubiran C., Casamiquela L. et al., 2021, \aap, 647, A19
		
	\bibitem[Tarricq et al. (2022)]{Tarr2022} Tarricq Y., Soubiran C., Casamiquela L., Castro-Ginard A., Olivares J., Miret-Roig N., Galli P.A.B., 2022, \aap, 659, A59
	
	\bibitem[Twarog et al. (2011)]{twa2011} Twarog B. A., Carraro G., \& Anthony-Twarog B. J. 2011, \apj, 727, L7
	
	\bibitem[\protect\citeauthoryear{van Dokkum}{2001}]{van01} van Dokkum P.G., 2001, \pasp, 113, 1420 
	
	\bibitem[van den Berg et al. (1991)]{van91} van den Berg S., Morbey C., Pazder J., 1991, \apj, 375, 594	
	
	\bibitem[Villanova et al. (2018)]{vil18} Villanova S., Carraro G., Geisler D., Monaco L., Assmann P., 2018, \apj, 867, 34
	
	\bibitem[von Hoerner (1957)]{vonHoerner1957} von Hoerner S., 1957, \aj, 125, 451
	
	\bibitem[Wu et al. (2002)]{wu02} Wu Z.Y., Tian K.P., Balaguer-Nunez L., Jordi C., Zhao J.L., and Guibert J., 2002, \aap, 381, 464 
	
	\bibitem[Yang  et al. (2013)]{yan13} Yang S.C., et al., 2013, \apj, 762, 1
	
	\bibitem[Zhong et al. (2022)]{zho22} Zhong J., Chen L., Jiang Y.,  Qin S., and Hou J., 2022, \aj, 164,54
		
\end{thebibliography}
\end{document}